\documentclass[twocolumn]{aastex62}

\usepackage{amsmath}
\usepackage{graphicx}   % Include figure files
\usepackage{bm}         % bold math
\usepackage{natbib}
\usepackage{etoolbox}
\usepackage{rotating}
\usepackage{array}
\usepackage{booktabs}
\usepackage{amssymb}

\makeatletter
\patchcmd{\NAT@citex}
  {\@citea\NAT@hyper@{%
     \NAT@nmfmt{\NAT@nm}%
\hyper@natlinkbreak{\NAT@aysep\NAT@spacechar}{\@citeb\@extra@b@citeb}%
     \NAT@date}}
  {\@citea\NAT@nmfmt{\NAT@nm}%
   \NAT@aysep\NAT@spacechar\NAT@hyper@{\NAT@date}}{}{}

\patchcmd{\NAT@citex}
  {\@citea\NAT@hyper@{%
     \NAT@nmfmt{\NAT@nm}%
\hyper@natlinkbreak{\NAT@spacechar\NAT@@open\if*#1*\else#1\NAT@spacechar\fi}%
       {\@citeb\@extra@b@citeb}%
     \NAT@date}}
  {\@citea\NAT@nmfmt{\NAT@nm}%
\NAT@spacechar\NAT@@open\if*#1*\else#1\NAT@spacechar\fi\NAT@hyper@{\NAT@date}}
  {}{}
\makeatother

\makeatletter
\DeclareRobustCommand{\textsupsub}[2]{{%
  \m@th\ensuremath{%
    ^{\mbox{\fontsize\sf@size\z@#1}}%
    _{\mbox{\fontsize\sf@size\z@#2}}%
  }%
}}
\makeatother

\newcommand{\lsun}{\mbox{L$_\odot$}}
\newcommand{\msun}{\mbox{M$_\odot$}}

\newcommand{\kms}{\mbox{km\,s$^{-1}$}}

\newcommand{\ee}[1]{\mbox{${} \times 10^{#1}$}}% scientific number format
\newcommand{\jj}[2]{\mbox{$J = #1\rightarrow#2$}}
\newcommand{\jyb}{\mbox{Jy\,beam$^{-1}$}}
\newcommand{\cc}{\mbox{cm$^{-3}$}}
\newcommand{\h}{\mbox{$^{\text{h}}$}}
\newcommand{\m}{\mbox{$^{\text{m}}$}}

\newcommand{\dd}{\mbox{$^{\text{d}}$}}
\newcommand{\J}{\mbox{$J$}}
\newcommand{\K}{\mbox{$K$}}
\newcommand{\Ka}{\mbox{$K_\text{a}$}}
\newcommand{\Kc}{\mbox{$K_\text{c}$}}

\newcommand{\htcn}{\mbox{H$^{13}$CN}}
\newcommand{\methylformate}{\mbox{CH$_{3}$OCHO}}
\newcommand{\methylformatev}{\mbox{CH$_{3}$OCHO\,$v=1$}}
\newcommand{\methanol}{\mbox{CH$_{3}$OH}}
\newcommand{\tmethanol}{\mbox{$^{13}$CH$_{3}$OH}}
\newcommand{\dmethanol}{\mbox{CH$_{2}$DOH}}
\newcommand{\methanold}{\mbox{CH$_{3}$OD}}

\newcommand{\dimethylether}{\mbox{CH$_{3}$OCH$_{3}$}}
\newcommand{\acetone}{\mbox{CH$_{3}$COCH$_{3}$}}
\newcommand{\ethanol}{\mbox{C$_{2}$H$_{5}$OH}}
\newcommand{\acetaldehyde}{\mbox{CH$_{3}$CHO}}
\newcommand{\ethylcyanide}{\mbox{CH$_{3}$CH$_{2}$CN}}
\newcommand{\methylamine}{\mbox{CH$_{3}$NH$_{2}$}}

\newcommand{\methylcyanideFT}{\mbox{CH$_{3}$C$^{15}$N}}
\newcommand{\hcop}{\mbox{HCO$^{+}$}}

\shorttitle{Infall and COMs in BHR\,71}
\shortauthors{Yang et al.}

\begin{document}

\title{Constraining the Infalling Envelope Models of Embedded Protostars: BHR\,71 and its Hot Corino}

\author{Yao-Lun Yang}
\affiliation{The University of Texas at Austin, Department of Astronomy, 2515 Speedway, Stop C1400, Austin, TX 78712, USA}
\affiliation{RIKEN Cluster for Pioneering Research, Wako-shi, Saitama, 351-0106, Japan}

\author{Neal J. Evans II}
\affiliation{The University of Texas at Austin, Department of Astronomy, 2515 Speedway, Stop C1400, Austin, TX 78712, USA}
\affiliation{Korea Astronomy and Space Science Institute, 776 Daedeokdae-ro, Yuseong-gu, Daejeon 34055, Korea}
\affiliation{Humanitas College, Global Campus, Kyung Hee University, Yongin-shi 17104, Korea}

\author{Aaron Smith}
\altaffiliation{Einstein Fellow}
\affiliation{Department of Physics, Massachusetts Institute of Technology, Cambridge, MA 02139, USA}

\author{Jeong-Eun Lee}
\affiliation{Department of Astronomy \&\ Space Science, Kyung Hee University, Gyeonggi 446-701, Korea  \\
School of Space Research, Kyung Hee University, Yongin-shi, Kyungki-do 449-701, Korea}

\author{John J. Tobin}
\affiliation{National Radio Astronomy Observatory, 520 Edgemont Road, Charlottesville, VA 22903, USA}
\affiliation{Homer L. Dodge Department of Physics and Astronomy, University of Oklahoma, 440 W. Brooks Street, Norman, OK 73019, USA}

\author{Susan Terebey}
\affiliation{Department of Physics and Astronomy, California State University at Los Angeles, Los Angeles, CA 90032, USA}

\author{Hannah Calcutt}
\affiliation{Department of Space, Earth, and Environment, Chalmers University of Technology, Gothenburg, Sweden}

\author{Jes K. J{\o}rgensen}
\affiliation{Niels Bohr Institute, University of Copenhagen, {\O}ster Voldgade 5-7, DK-1350 Copenhagen K., Denmark}

\author{Joel D. Green}
\affiliation{Space Telescope Science Institute, 3700 San Martin Dr., Baltimore, MD 02138, USA}
\affiliation{The University of Texas at Austin, Department of Astronomy, 2515 Speedway, Stop C1400, Austin, TX 78712, USA}

\author{Tyler L. Bourke}
\affiliation{SKA Organization, Jodrell Bank, Lower Withington, Macclesfield SK11 9FT, UK}

\correspondingauthor{Yao-Lun Yang}
\email{yaolunyang.astro@gmail.com}

\begin{abstract}
	The collapse of the protostellar envelope results in the growth of the protostar and the development of a protoplanetary disk, playing a critical role during the early stages of star formation.  Characterizing the gas infall in the envelope constrains the dynamical models of star formation.  We present unambiguous signatures of infall, probed by optically thick molecular lines, toward an isolated embedded protostar, BHR\,71 IRS1.  The three dimensional radiative transfer calculations indicate that a slowly rotating infalling envelope model following the ``inside-out'' collapse reproduces the observations of both \hcop\,\jj{4}{3} and CS\,\jj{7}{6} lines, and the low velocity emission of the HCN\,\jj{4}{3} line.  The envelope has a model-derived age of 12\,000$\pm$3000 years after the initial collapse.  The envelope model underestimates the high velocity emission at the HCN \jj{4}{3} and \htcn\ \jj{4}{3} lines, where outflows or a Keplerian disk may contribute.  The ALMA observations serendipitously discover the emission of complex organic molecules (COMs) concentrated within a radius of 100\,au, indicating that BHR\,71 IRS1 harbors a hot corino.  Eight species of COMs are identified, including \methanol\ and \methylformate, along with H$_{2}$CS, SO$_{2}$ and HCN\,$v_{2}=1$.  The emission of methyl formate and $^{13}$C-methanol shows a clear velocity gradient within a radius of 50\,au, hinting at an unresolved Keplerian rotating disk.
\end{abstract}

\section{Introduction}
\label{sec:intro}
% 1. What we have know about the collapse and what we are still confused about?
% 2. Observation and simulation results
% 3. Use optically thick lines to probe the infalling envelope
% 4. Complex organic molecules
% 5. kinematics and chemistry probed by the COMs
% 6. section breakdown

The infall of gas and dust transforms dense cores into protostars.  The collapse of protostellar envelopes begins when the gravitational force exceeds the thermal and non-thermal pressure support due to turbulence and magnetic fields.  Several theoretical models have proposed solutions for the evolution of the collapsing envelope with different assumptions and initial conditions \citep[e.g., ][]{1969MNRAS.144..425P,1969MNRAS.145..271L,1977ApJ...214..488S}; however, their predictions need to pass the observational tests.  For example, an ``inside-out'' collapse from a singular isothermal sphere \citep{1977ApJ...214..488S} matches the data \citep{1992ApJ...394..204Z}, whereas a Larson-Penston similarity solution \citep{1969MNRAS.145..271L,1969MNRAS.144..425P} predicts a much larger linewidth compared to observations of low-mass cores \citep{1990ApJ...363..168Z}.  Based on the paradigm of \citet{1977ApJ...214..488S}, the models considering rotation (\citealt{1984ApJ...286..529T}, hereafter TSC; \citealt{1998ApJ...493..342S}) and magnetic fields \citep{1993ApJ...417..220G,1993ApJ...417..243G} provide increasingly realistic predictions.  In the last few decades, many observational studies tested collapse models with various aspects \citep[e.g., ][]{1997ApJ...475..211O,2001ApJ...562..770D,2013ApJ...772...22Y,2015ApJ...814...22E}.  Numerical simulations also provide insights on the dynamical evolution of collapsing protostellar envelope \citep[e.g., ][]{2009ApJ...699..230G,2013ApJ...763....6T,2015ApJ...801..117T,2015ApJ...806...31G,2017A&A...598A.116V}, but those models are not usually well adapted to detailed observational tests against particular sources (e.g., different mass and luminosity, insufficient resolution, etc).  Only a few studies comprehensively test the interplay between infall, kinematics, and magnetic fields with observational constraints \citep[e.g., ][]{2019ApJ...871..243Y}.

Observations that probe the kinematics of infalling envelopes place important constraints on theoretical models.  Molecular transitions that have high critical density are easily excited only in the densest part of the protostellar envelopes, tracing the kinematic structure of the inner envelopes \citep{1999ARA&A..37..311E}.  However, rotation and outflows produce comparable kinematic signatures on the line profiles of molecular emission, complicating the interpretation.  \citet{1977ApJ...214L..73L} first proposed using optically-thick molecular emission to probe the infalling gas in the envelope (see also \citealt{1977ApJ...211..122S}).  The opaque infalling gas in the foreground leads to red-shifted absorption in the line profile \citep{1994ASPC...65..183Z,1999ApJS..122..519C}.  With single dish observations, outflows and foreground large-scale clouds may also contribute to the line profile due to the large beam, confusing the interpretation of the red-shifted absorption \citep{2004ApJ...617.1157C}.  \citet{2001ApJ...562..770D} demonstrated that a smaller beam would observe the red-shifted absorption below the continuum, placing the infalling gas indisputably at the foreground of the central protostar.  Moreover, a smaller beam greatly reduces the contamination from unrelated kinematic signatures, such as the broad emission from outflows.  Thus, we would begin to observe the absorption of the compact continuum by the infalling opaque gas; if this absorption is red-shifted, it provides an unambiguous signature of infall.

The Atacama Large Millimeter/submillimeter Array (ALMA) provides the best instrument for measuring the infall.  \citet{2012A&A...544L...7P} reported the first detection with ALMA of the infall signature from the emission of methyl formate towards IRAS\,16293$-$2422\,B, fitted with a two-layer infall model \citep{1996ApJ...465L.133M}.  The \hcop\,\jj{4}{3} line toward an edge-on embedded protostar, HH\,212, also shows clear red-shifted absorption against the continuum \citep{2014ApJ...786..114L}.  \citet{2015ApJ...814...22E} demonstrated a 1D comprehensive modeling of the infall signatures detected toward B335 using the optically-thick molecular transitions, including \hcop\,\jj{4}{3}, HCN\,\jj{4}{3}, and CS\,\jj{7}{6}.  Constraining the underlying infall kinematics requires radiative transfer calculations using models of the envelope structure along with the chemical abundance profile of the selected tracers.  Analytic approximations of the chemical abundance can successfully reproduce observations of simple molecules, such as CO \citep{2005A&A...435..177J}, and agree with the results of self-consistent chemo-dynamical modeling, where the chemistry is solved along with the dynamics \citep{2004ApJ...617..360L}.  But for other molecules, such as CS, H$_{2}$CO, and CN, the chemo-dynamical model suggests substantially different abundance profiles compared to the analytic approximations.  However, a chemo-dynamical model is not always available and has a greater uncertainty for more complex molecules, where only a few observations exist to constrain the chemistry at the central region of the protostellar cores \citep{2013ChRv..113.8961A}.

% COMs
Heavier or more complex molecules, such as cyclic-C$_{3}$H$_{2}$, SO, and complex organic molecules (COMs), are in the gas phase at the inner protostellar envelope ($T\gtrsim100$\,K), exclusively tracing the kinematics at the inner region where a disk may be forming \citep{2013ChRv..113.8961A,2014Natur.507...78S}.  In the review by \citet{2009ARA&A..47..427H}, COMs are defined as carbon-bearing molecules that contain six atoms or more.  The kinematics of a rotating infalling envelope has been analyzed with the observations of heavier or more complex molecules, such as \methanol\ and \dmethanol\ for HH\,212 \citep{2017ApJ...843...27L}, CS for IRAS\,04365$+$2535 \citep{2016ApJ...820L..34S} and L483 \citep{2017ApJ...837..174O}, cyclic-C$_{3}$H$_{2}$ for L1527 \citep{2014Natur.507...78S}, OCS for IRAS\,16293$-$2422\,A \citep{2016ApJ...824...88O}, and methanol and HCOOH for B335 \citep{2019ApJ...873L..21I}.  However, a uniform picture of the kinematics traced by COMs has not been established yet.  For example, \citet{2017ApJ...837..174O} and \citet{2019A&A...629A..29J} derive different kinematic structures for the rotation signatures observed from the emission of COMs toward an embedded protostar, L483.

% BHR71
BHR\,71 is a Bok globule near the Southern Coalsack at 200\,pc \citep{1989A&A...225..192S,1994BaltA...3..199S}, hosting two protostars, IRS1 and IRS2, separated by 16\arcsec\ \citep{2001ApJ...554L..91B,2019ApJ...870...81T}.  \citet{2019ApJ...870...81T} detect opposite velocity gradients toward IRS1 and IRS2, suggesting that the binary system is likely formed via turbulent fragmentation.  BHR\,71 IRS1 dominates the luminosity of BHR\,71 with $L=13.5$\,\lsun\ \citep{2018ApJ...860..174Y}, whereas IRS2 only has 1.7\,\lsun\ \citep{2019ApJ...870...81T}.  Because of the wide separation and low luminosity of IRS2, we focus on IRS1 in this study.  BHR\,71 IRS1 (hereafter BHR\,71 if not mentioned specifically) is a Class 0 protostar based on its bolometric temperature and the fraction of its emission in submillimeter wavelengths \citep{2013ApJ...770..123G,2018ApJ...860..174Y}.  The envelope of BHR\,71 shows blue-shifted emission in the east and red-shifted emission in the west, indicating the rotation of the inner envelope \citep{2008ApJ...683..862C,2019ApJ...870...81T}.  BHR\,71 drives outflows in the north-south direction \citep{1997ApJ...476..781B,2006A&A...454L..79P}.  \citet[][hereafter Y17]{2017ApJ...835..259Y} performed 3D continuum radiative transfer calculations of a TSC envelope model modified to include outflow cavities and a disk to constrain the structure of BHR\,71, using primarily the \textit{Herschel} spectra along with archival \textit{Spitzer} spectra and photometry.  The best fitting TSC envelope suggested an age of 36\,000 years with an inclination angle of 130$^{\circ}$.  The age is defined as the time since the onset of the collapse in the TSC model; the source inclination angle is measured from the observer’s line of sight (LOS) to the rotation axis of the envelope, using the right hand rule. An inclination angle of 90 degrees indicates an edge-on system, and inclination greater than 90 degrees indicates the rotation vector is behind the plane of sky.  High resolution ALMA $^{13}$CO observations also indicate an inclination angle between 115$^{\circ}$ to 153$^{\circ}$ under the same definition \citep{2019ApJ...870...81T}.  BHR\,71 also shows prominent outflows in the north-south direction \citep{1997ApJ...476..781B}, resulting in several active shocked regions \citep{1998ApJ...509..768G,2011A&A...532A..53G,2015A&A...575A..98G}.

% section breakdown
In this study, ALMA data are used to constrain the infall kinematics toward BHR\,71, and survey the emission of COMs.
Section\,\ref{sec:observations} describes the observation and data reduction; Section\,\ref{sec:results} shows the observed continuum and molecular lines; Section\,\ref{sec:infall} discusses the infall signatures and the modeling that constrains the infall kinematics; Section\,\ref{sec:coms} presents the spectra of COMs and the line identification; Section\,\ref{sec:blobs} shows the compact HCN features along the outflow direction, and finally Section\,\ref{sec:conclusion} summarizes the conclusions.

\section{Observations}
\label{sec:observations}
The observations of BHR\,71 were obtained in Project 2016.0.00391S (PI: Y.-L. Yang) on November 18 2016 by ALMA with 45 12-m antennas and the Band 7 receiver in the C40-4 configuration.  The minimum and maximum projected baselines were 12.1\,m and 631.3\,m, respectively; the minimum and maximum measured baselines were 15.1\,m and 918.9\,m, respectively.  The water vapor during the observation was stable, varying between 0.64\,mm and 0.72\,mm.  The flux calibration source was J1107$-$4449, which has a flux uncertainty of 10\%.

The ALMA Correlator was configured to have four spectral windows, each with 1920 channels.  The local oscillator was tuned to observe HCN \jj{4}{3} (354.505473\,GHz), \hcop \jj{4}{3} (356.734242\,GHz), CS \jj{7}{6} (342.882857\,GHz), and \htcn \jj{4}{3} (345.339756\,GHz).  Table\,\ref{tbl:line_data} lists the basic properties of these lines.  The first three windows have 234.38 MHz bandwidth ($\sim$200\,\kms) and 0.122\,MHz (0.1\,\kms) spectral resolution, while the \htcn \jj{4}{3} window has 468.75\,MHz ($\sim$400\,\kms) and 0.244\,MHz (0.2\,\kms) resolution.

\begin{deluxetable*}{lcccc}
	\tabletypesize{\scriptsize}
	\tablecaption{Basic data of the modeled transitions \label{tbl:line_data}}
	\tablewidth{\textwidth}
	\tablehead{ \colhead{Parameters} & \colhead{HCN\,\jj{4}{3}} & \colhead{\hcop\,\jj{4}{3}} & \colhead{CS\,\jj{7}{6}} & \colhead{\htcn\,\jj{4}{3}}}
	\startdata
	Frequency [MHz] & 354505.478 & 356734.288 & 342882.850 & 354505.478\\
	Einstein-A$_{21}$ [s$^{-1}$] & 2.05\ee{-3} & 3.63\ee{-3} & 8.40\ee{-4} & 2.05\ee{-3} \\
	Dust opacity [cm$^{2}$\,g$^{-1}$] & 1.84 & 1.86 & 1.74 & 1.76 \\
	\hline
	\enddata
\end{deluxetable*}

\subsection{Data Reduction}
\label{sec:reduction}
The data were reduced by the ALMA Pipeline, version 38366 with the Common Astronomy Software Applications (\textsc{CASA}) version 4.7.0-1 \citep{2007ASPC..376..127M}.  We further performed self-calibration with the CASA version 4.7.2, and the imaging was performed with \textsc{CASA} version 5.1.1.  The strong continuum source, 0.56\,\jyb, with a signal-to-noise ratio (S/N) of 222, in the pipeline-reduced data, allows further self-calibration on both the phase and the amplitude.  Prior to the self-calibration, we flagged the channels that contain spectral lines, assuming that the lowest flux in the spectra represents the continuum.  We update this flag again after extracting the 1D spectra for each spectral window by using the same criterion to select the spectral lines from the 1D spectra, and re-run the entire calibration process.  The continuum comes from 11\%\ of channels (876 of 7680) after the flagging.  For the phase calibration, we gradually reduce the solution interval from \texttt{`inf'} to \texttt{`int'} to ensure the quality of the calibration solution.  For amplitude calibration, we set the solution interval to \texttt{`inf'} (e.g., one solution per scan).  The S/N of the continuum source increases from 222 to 819 after the self-calibration.  The solutions of the self calibration apply to both continuum and line data.

We perform the imaging with CASA version 5.1.1 to use the \texttt{tclean} task, which has better efficiency and flexibility than the \texttt{clean} task.  Using the ``H\"{o}gbom'' method for deconvolution and the ``briggs'' weighting with a robust parameter of 0.5, the root-mean-square (RMS) noise reaches 0.904\,m\jyb\ for the continuum.  The calibration uncertainty is 10\%.  For the line emission, we use the ``auto-thresh'' method for selecting the source emission for deconvolutions in the \texttt{tclean} process with a flux threshold of 45\,mJy, which is three times the RMS noise of the image reduced by the standard pipeline. The mask resolution is restricted to 0.5\arcsec\ to avoid selecting structures smaller than the synthesized beam.  The continuum emission remains in the spectral cubes until the line emission is imaged to prevent imaging artifacts, similar to the procedures in \citet{2015ApJ...814...22E}.  The deconvolution during the \texttt{clean} process is only performed at the pixels beyond noise threshold.  Subtracting the continuum in the $uv$-space would move a significant fraction of the absorption signal, mostly the wings of the absorption, into the threshold.  As a result, those absorption signals are excluded from the deconvolution, leading to artifacts in the images.  Then the primary beam correction applies to the image cubes for both the continuum and lines.  Finally, we flag the emission for the second time by assuming no absorption existing in the spectra except for the regions around the four targeted lines (HCN, \hcop, CS, and \htcn), and run the entire calibration and imaging processes again.  

The resulting synthesized beam is 0\farcs39$\times$0\farcs27, corresponding to 78$\times$54\,au at the distance of BHR\,71 (200\,pc).  The lowest flux in the spectra should approximate the continuum emission, which is removed in the image plane using the CASA \texttt{imcontsub} task.  The RMS noise reaches 13\,m\jyb, 14\,m\jyb, 12\,m\jyb, and 8\,m\jyb\ for the spectral windows centered on HCN\,\jj{4}{3}, \hcop\,\jj{4}{3}, CS\,\jj{7}{6}, and \htcn\,\jj{4}{3}, respectively.

\section{Results}
\label{sec:results}
\subsection{Continuum}
\label{sec:continuum}
\begin{figure}[htbp!]
	\centering
	\includegraphics[width=0.48\textwidth]{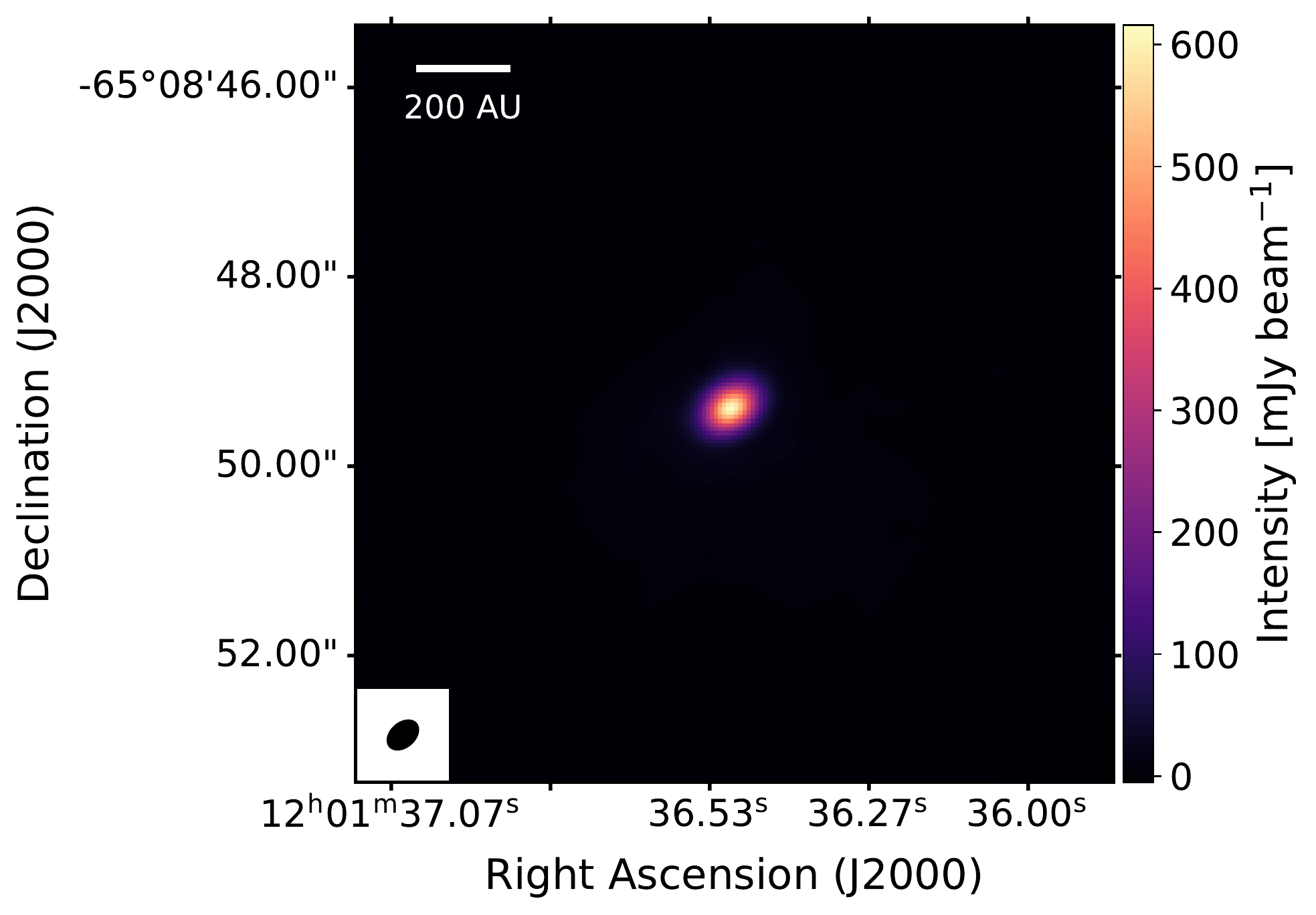}
	\includegraphics[width=0.48\textwidth]{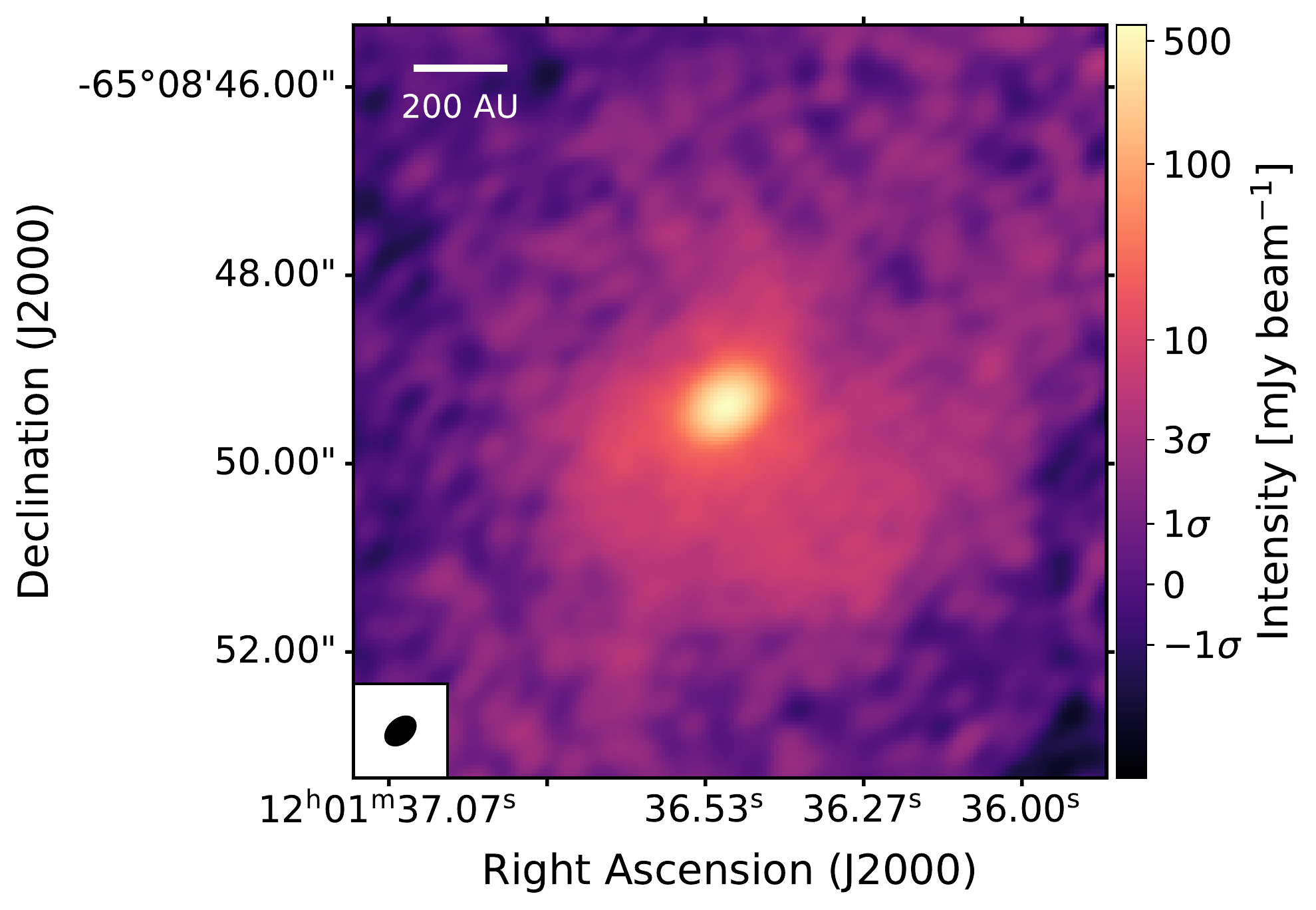}
	\caption{The continuum emission of BHR\,71 at 356\,GHz shown in linear scale (top) and symmetric logarithmic scale (bottom).  For the bottom figure, the brightness is shown in logarithmic scale for $|I_{\nu}|>1\sigma = 0.904$\,m\jyb, and in linear scale for $|I_{\nu}|\leq1\sigma$.  The synthetic beam is shown in the lower left corner.  Table\,\ref{tbl:continuum} lists the properties of the continuum source.}
	\label{fig:continuum}
\end{figure}

The ALMA observations reveal a marginally-resolved continuum source with a deconvolved size of 70\,au$\times$56\,au fitted with a 2D Gaussian profile using the CASA \texttt{imfit} task (Figure\,\ref{fig:continuum} and Table\,\ref{tbl:continuum}).  Beside the compact emission, the continuum source has faint emission (10--20\,m\jyb) extending toward a radius of 2\arcsec.  This extended emission is likely tracing the inner envelope; however, with a maximum recoverable scale of $\sim$3\arcsec, most of the extended emission is resolved out.

Dust emission can provide the dust mass if the dust emission is optically thin.  Any optically thick matter, such as a compact disk, will make the derived mass an underestimation.  The mass of optically thin dust can be evaluated with the following equation,
\begin{equation}
	M_\text{dust} = \frac{S_{\nu}d^{2}}{\kappa_{\text{356\,GHz}}B_{\nu}(\text{356\,GHz}, T_\text{dust})},
	\label{eq:dust_mass}
\end{equation}
where $S_{\nu}$ is the continuum flux density, $d$ is the distance to BHR\,71, $\kappa_{\text{356\,GHz}}$ is the dust opacity at 356\,GHz, 1.86\,cm$^{2}$\,g$^{-1}$, and $B_{\nu}(\text{356\,GHz}, T_\text{dust})$ is the Planck function at 356\,GHz given a temperature of $T_\text{dust}$.  The assumption of a single dust temperature results in $\sim$10--50\%\ uncertainty on the derived mass \citep[Fig. A.1, ][]{2008A&A...487..993K}.  Thus, we use a mass-weighted temperature of 148\,K (Equation\,8, \citealt{2008A&A...487..993K}) as the dust temperature for Equation\,\ref{eq:dust_mass}.  The derivation of the mass-weighted temperature assumes the density profile of a purely infalling envelope \citep[$\rho(r)\propto r^{-1.5}$, ][]{1976ApJ...210..377U}, the opacity of the dust covered with thin ice mantles at a gas density of 10$^{6}$\,\cc\ \citep[hereafter the OH5 dust, $\kappa_{\nu}\propto\nu^{1.8}$, ][]{1994A&A...291..943O} and $T(r)=$114\,K evaluated at 70\,au (0\farcs{35}) from the Y17 model.  The optically thin dust mass is 2.3\ee{-2}$\pm$1.2\ee{-4}\,M$_{\odot}$, suggesting a gas mass of 2.3$\pm$0.01\,M$_{\odot}$ with a gas-to-dust ratio of 100.  The error only represents the uncertainty of the flux measurement, whereas the systematic uncertainties from other origins, such as the dust opacity and calibration, are not included.

\begin{deluxetable}{lr}
	\tabletypesize{\scriptsize}
	\tablecaption{The best-fitted continuum emission \label{tbl:continuum}}
	\tablewidth{\textwidth}
	\tablehead{\colhead{Parameter} & \colhead{Value}}
	\startdata
	R.A. & 12$^{\rm h}$01$^{\rm m}$36\fs{4988} $\pm$0\fs{0001} \\
	Dec. & $-$65$^{\rm d}$08$^{\rm m}$49\fs{3819} $\pm$ 0\fs{0007} \\
	\hline
	\multicolumn{2}{c}{Convolved size} \\
	\hline
	Semi-major axis FWHM$^{a}$ & 521.3$\pm$2.1\,mas \\
	Semi-minor axis FWHM       & 394.0$\pm$1.3\,mas \\
	Position angle             & 124\fdg{91}$\pm$0\fdg{48} \\
	\hline
	\multicolumn{2}{c}{Deconvolved size} \\
	\hline
	Semi-major axis FWHM       & 349.5$\pm$3.6\,mas \\
	Semi-minor axis FWHM       & 278.9$\pm$2.7\,mas \\
	Position angle             & 113\fdg{7}$\pm$2\fdg{0} \\
	\hline
	Integrated flux & 1.129 $\pm$0.006\,Jy \\
	Peak flux       & 586.9 $\pm$2.2\,mJy\,beam$^{-1}$ \\
	Beam size       & 0\farcs39$\times$0\farcs27 (PA=131\fdg{25}) \\
	\enddata
	\tablecomments{$^{a}$Full width at half maximum}
\end{deluxetable}

\subsection{Molecular Emission}
\subsubsection{Intensity Maps}
\label{sec:intensity_maps}
\begin{figure*}[htbp!]
	\centering
	\includegraphics[width=0.48\textwidth]{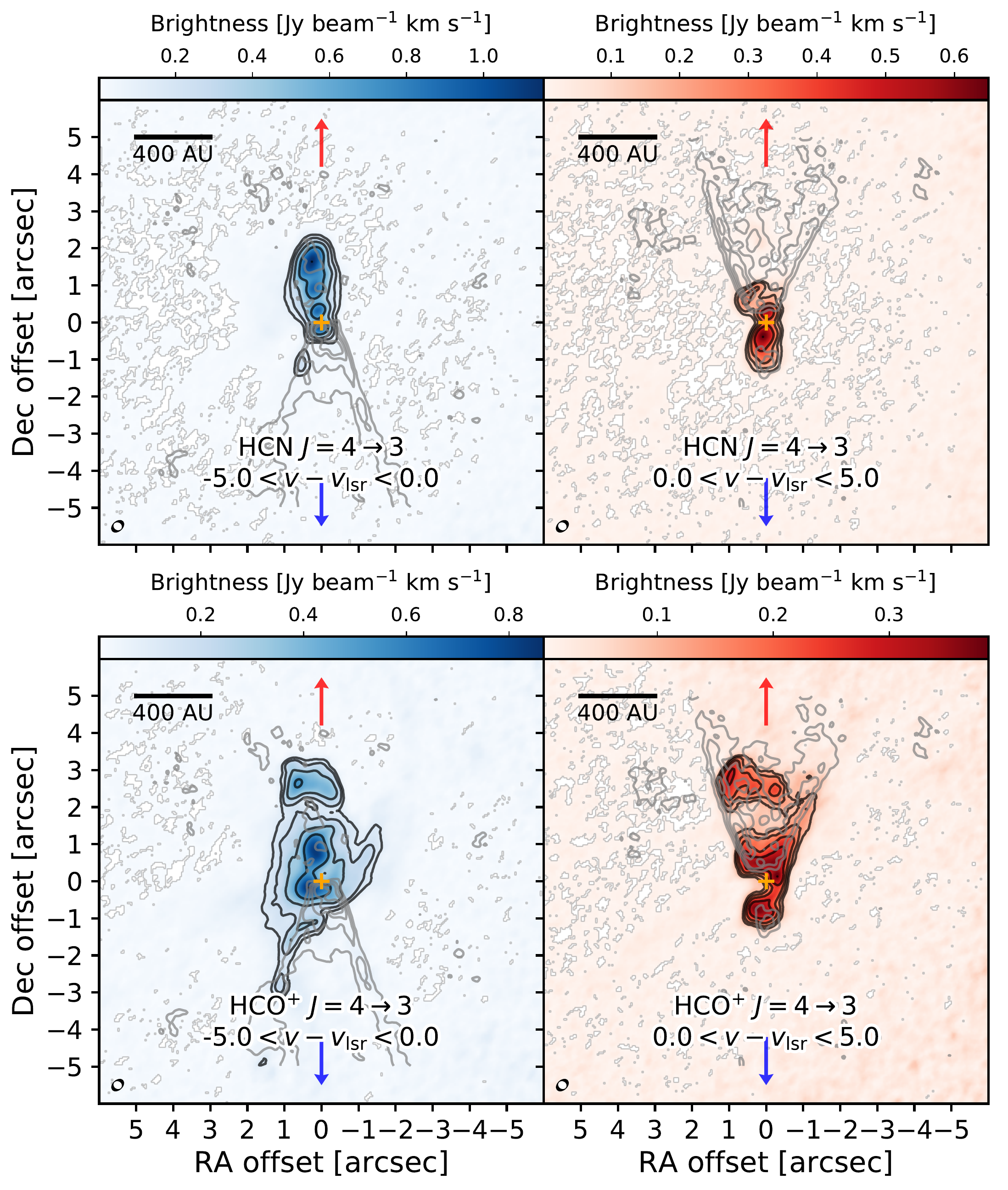}
	\includegraphics[width=0.48\textwidth]{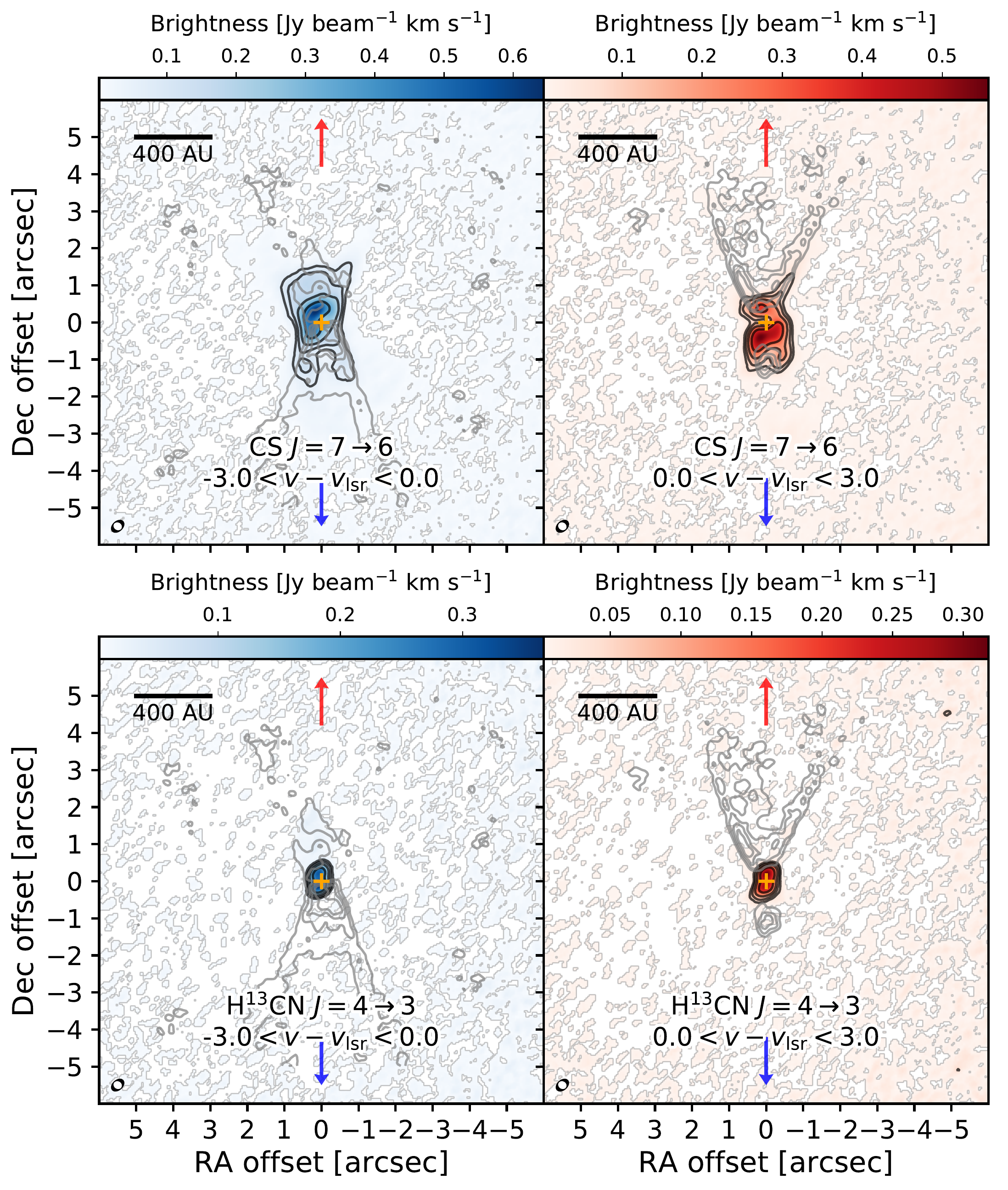}
	\caption{The moment 0 maps of the HCN\,\jj{4}{3}, \hcop\,\jj{4}{3}, CS\,\jj{7}{6}, and \htcn\,\jj{4}{3} lines shown separately for blue-shifted and red-shifted velocities with respect to the source velocity.  The orange plus signs indicate the position of the continuum source identified in Section\,\ref{sec:continuum}, while the black ellipses indicate the beam sizes.  For comparing with the CO outflows, the gray contours show the ALMA CO\,\jj{2}{1} moment maps with a beam of 0\farcs{23}$\times$0\farcs{20}, indicated by the white ellipse on top of the beam of the corresponding molecular lines.  The CO moments are calculated with the same velocity range as that of the corresponding molecular lines.  The blue CO contours are separated linearly from 0.08 to 0.56 \jyb\,\kms, while the red CO contours are separated linearly from 0.11 to 0.42 \jyb\,\kms.  The maps only show the central 12\arcsec\ of the entire observation as regions outside of this area contain no significant emission.  The contours show five equally-separated levels in logarithmic scale from 2$\sigma$ to their maximum values.  There is very little emission at $\lvert v-v_\text{source} \rvert > 3$\,\kms\ for the CS\,\jj{7}{6} and \htcn\,\jj{4}{3} lines; therefore, we reduce the velocity range to better visualize the weak emission.  The red and blue arrows indicate the large scale red- and blue-shifted outflows \citep{1997ApJ...476..781B,2019ApJ...870...81T}.}
	\label{fig:moment0}
\end{figure*}

The intensity maps of HCN, \hcop, CS, and \htcn\ emission (Figure\,\ref{fig:moment0}) illustrate the structure of the outflows and the inner envelope within the central 12\arcsec, separated into the ranges of $-$5\,\kms\ $<v-v_\text{lsr}<$ 0\,\kms\ (blue-shifted) and 0\,\kms\ $<v-v_\text{lsr}<$ 5\,\kms\ (red-shifted).  The source velocity is $-$4.45\,\kms, measured from the NH$_{3}$ spectra in \citet{1997ApJ...476..781B}.
While the emission peaks at the center of BHR\,71 across a range of velocities ($|v-v_\text{source}|\lesssim6$\,\kms), the low velocity ($|v-v_\text{source}|\lesssim3$\,\kms) emission also traces the morphology of outflows along north-south direction.  Here we describe the red- and blue-shifted velocity morphologies in detail.

The brightness of the HCN\,\jj{4}{3} line extends toward the north at the blue-shifted velocities and toward the south at the red-shifted velocities; this is the opposite to the direction of the large scale outflows observed in CO \citep{1997ApJ...476..781B,2006A&A...453..949P}.
The HCN emission is concentrated into a few compact features with a size of $\sim0\farcs5$ or less, which will be further discussed in Section\,\ref{sec:blobs}.

At red-shifted velocities, the \hcop\,\jj{4}{3} emission resembles the shape of a triangular outflow cavity, similar to the outflow traced by CO \citep{2019ApJ...870...81T}.  In contrast, at blue-shifted velocities, the emission becomes more extended than the morphology at the red-shifted velocities, containing two filamentary structures at the west and south of BHR\,71.  A boxy feature with a size of 2\arcsec$\times$3\arcsec\ appears at 3\arcsec\ north of BHR\,71 at both blue-shifted and red-shifted velocities.

The CS\,\jj{7}{6} line also traces the morphology of the outflow cavities.  At blue-shifted velocities, the emission becomes more extended to the north and only traces the outflow cavity walls at the south; at red-shifted velocities, the emission shows a similar morphology but reflects through the east-west plane, tracing the outflow cavity walls at the north and extending toward the south.  The emission of \hcop\ and CS trace a shape that is similar to that of the outflow cavities.

The \htcn\,\jj{4}{3} emission has two peaks less than 0\farcs{5} away from the center of BHR\,71, one at the north and one at the south.
The morphology of \htcn\ emission is similar at blue-shifted and red-shifted velocities.  As discussed in later sections (Section\,\ref{sec:coms_id} and Appendix\,\ref{sec:coms_identified}), the emission of SO$_{2}$ contaminates the \htcn\,\jj{4}{3} line.  The estimated brightness of this SO$_{2}$ line is about one third of the \htcn\ line, which dominates the intensity map shown in Figure\,\ref{fig:moment0}.
However, the morphology of another SO$_{2}$ line at 356755\,MHz has a similar double-peaked morphology, suggesting the emission of both SO$_{2}$ and \htcn\ may have the same origin.
Overall, all four emission lines have morphologies more complex than a single peak, tracing structures along the outflow direction.  Both CS and \hcop\ trace the outflow cavities, while HCN and \htcn\ appear as collections of compact features.

\subsubsection{Velocity Maps}
\label{sec:velocity_map}
\begin{figure}[htbp!]
	\centering
	\includegraphics[width=0.48\textwidth]{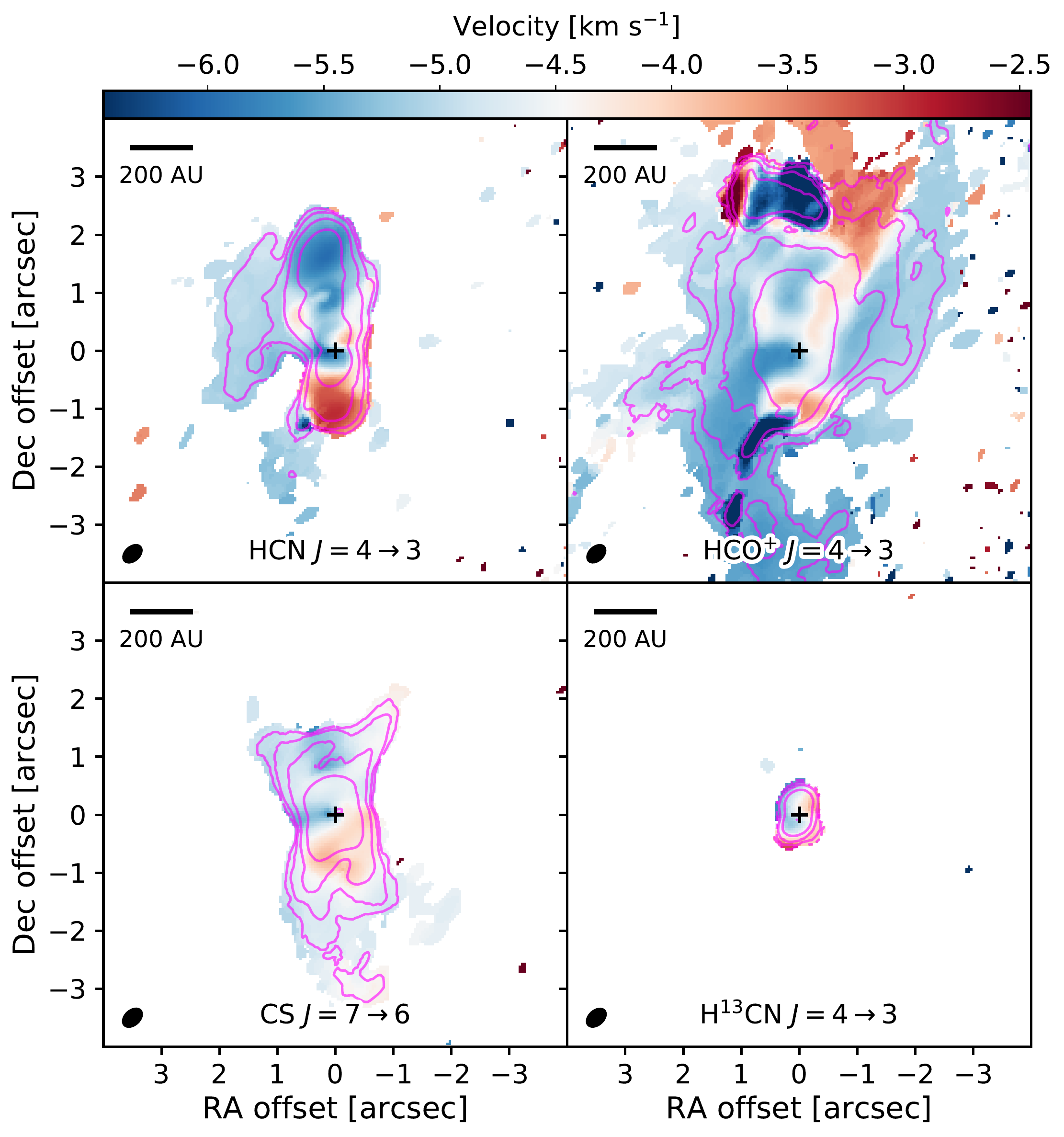}
	\caption{The moment 1 maps of the HCN \jj{4}{3}, \hcop\ \jj{4}{3}, CS \jj{7}{6}, and \htcn\ \jj{4}{3} lines shown together with the moment 0 maps in magenta contours, which show five equally-separated levels in logarithmic scale from 2$\sigma$ to their maximum values.  From left to right, and top to bottom, the 2$\sigma$ values are 0.02, 0.02, 0.02, and 0.01\,\jyb\,\kms, respectively; the maximum values are 1.47, 1.05, 0.93, and 0.62\,\jyb\,\kms, respectively.  Both moment 0 and 1 maps are calculated from $|v-v_\text{source}|\leq5$\,\kms.  The plus signs indicate the position of the continuum of BHR\,71.}
	\label{fig:moment1}
\end{figure}

Figure\,\ref{fig:moment1} shows the first moment maps, intensity-weighted velocity, of all four lines.  The HCN, CS, and \htcn\ lines show red-shifted emission at the south and blue-shifted emission at the north, opposite to the kinematics of the large scale of CO outflows \citep{1997ApJ...476..781B,2006A&A...453..949P}.  However, the high velocity emission appears in the north and south at red-shifted and blue-shifted velocities, respectively, consistent with the kinematics of the large-scale outflows (Figure\,\ref{fig:infall_channelmap_full}).

The moment 1 map of the \hcop\ emission shows a complex structure.  The red-shifted emission resides at the western side of BHR\,71, while the blue-shifted emission surrounds the source.  Toward the south, a red-to-blue velocity gradient appears to align with the filamentary structure found in the intensity map.  At the northern feature identified in the intensity map (Figure\,\ref{fig:moment0}), a strong velocity gradient from blue-shifted in the west to red-shifted in the east coincides with this feature, along with the weak red-shifted emission right next to the blue-shifted component of this feature.

\subsubsection{Channel Maps}
\begin{figure*}[htbp!]
	\centering
	\includegraphics[width=\textwidth]{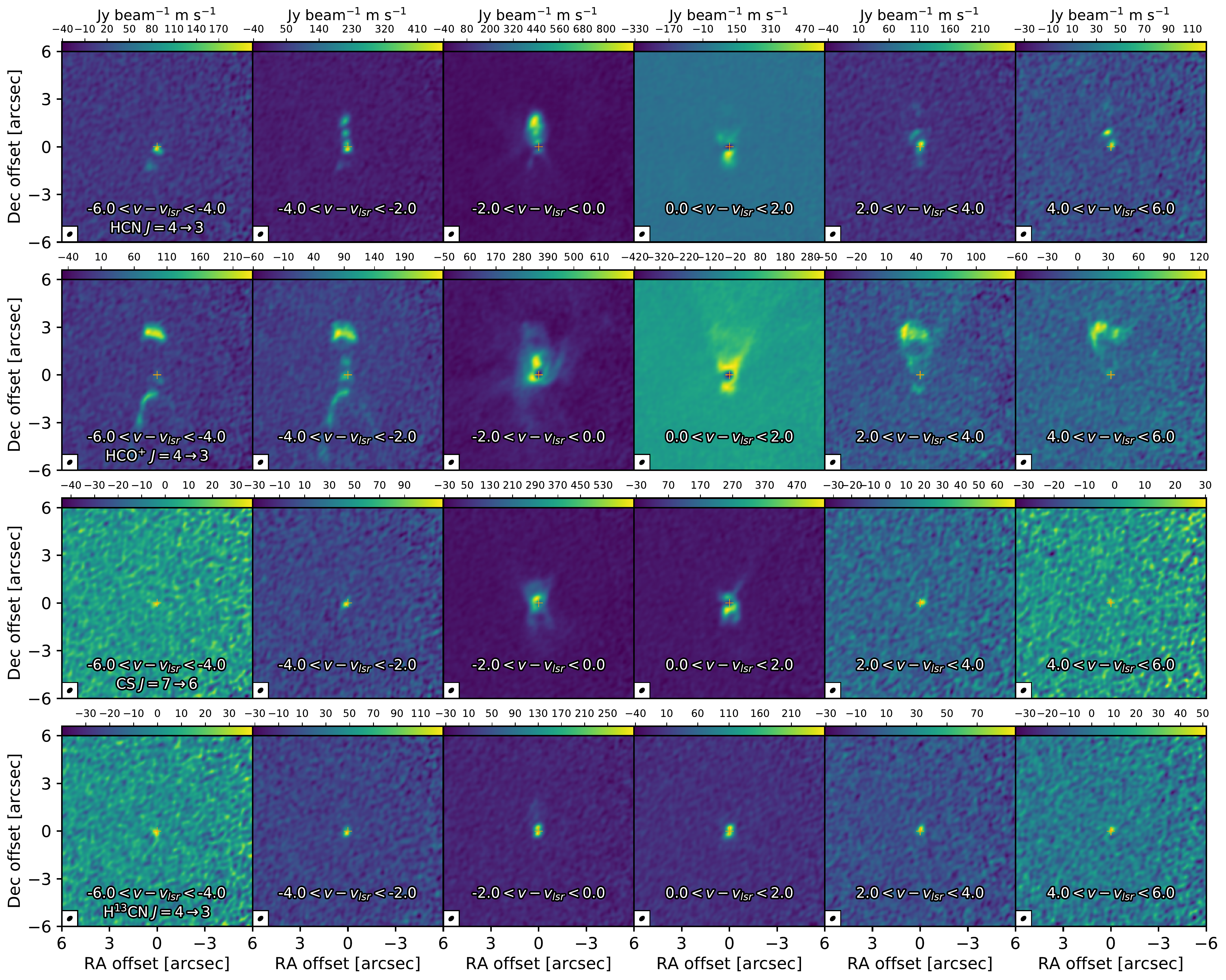}
	\caption{The channel maps of HCN\,\jj{4}{3}, \hcop\,\jj{4}{3}, CS\,\jj{7}{6}, and \htcn\,\jj{4}{3} from the top to bottom rows.  The subset images show the moment 0 map calculated with the velocity ranges shown in the images, and each image has its own scaling to show a complete structure at each channel.  The synthesized beam is shown at the lower left corner of each image.}
	\label{fig:infall_channelmap_full}
\end{figure*}

\begin{figure*}[htbp!]
	\centering
	\includegraphics[width=\textwidth]{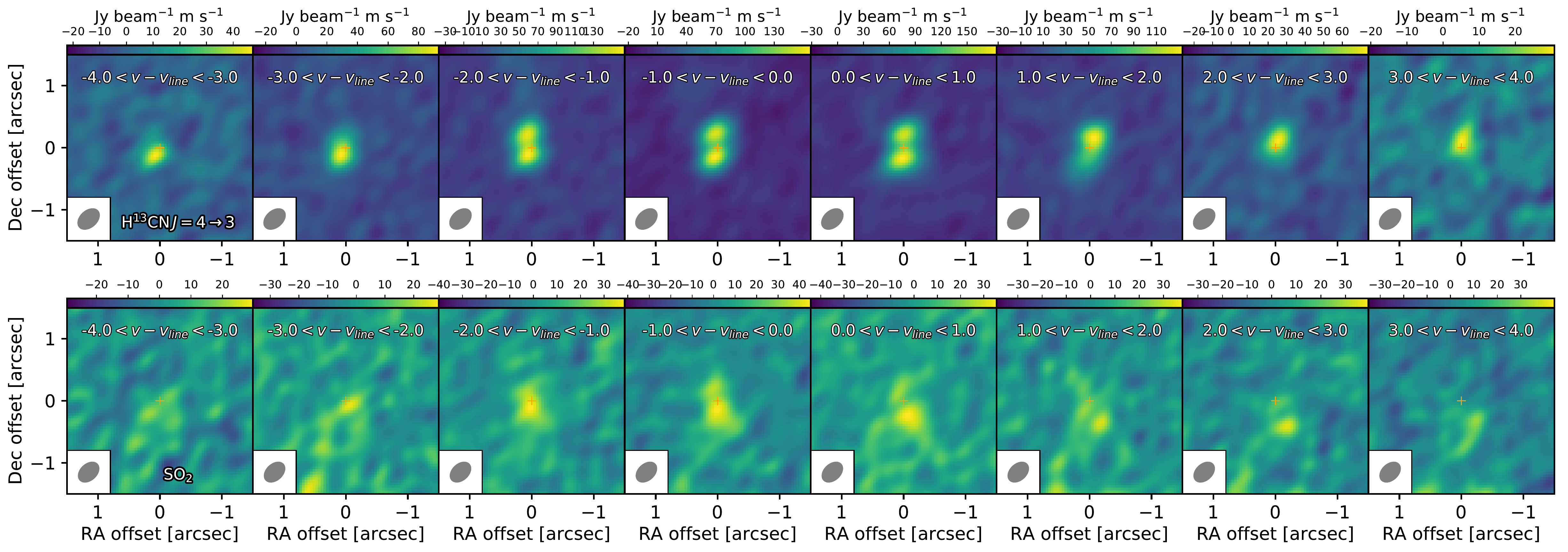}
	\caption{The zoom-in channel maps of the SO$_{2}$ line at 356755.2\,MHz and the \htcn\,\jj{4}{3} line.  Each panel shows the moment 0 map within a 1\,\kms\ interval from $-$4\,\kms\ to 4\,\kms.}
	\label{fig:channelmap_so2_h13cn}
\end{figure*}

Figure\,\ref{fig:infall_channelmap_full} shows the channel maps of all four lines, providing a detailed view of the velocity structure of these lines.  For the HCN \jj{4}{3} line, we identify four compact features at low velocity, especially $-4$\,\kms$<v-v_\text{source}<2$\,\kms.  We further discuss the origin of these HCN features in Section\,\ref{sec:blobs}.  The channel maps of the \hcop\ \jj{4}{3} line show a similar morphology to the moment 0 map.  The extended emission in the north disappears at low velocities, where more filamentary structure appears around the source.  The CS \jj{7}{6} line shows a single compact emission at most of the velocities except for the low velocities, where the emission exhibits an hour glass shape, resembling the shape of the outflow cavities.  The hour glass shape also appears in the low-velocity emission of HCN and \hcop.
The converging ends of the outflow cavity extend $\sim$0\farcs6--1\arcsec\ perpendicular to the direction of outflows, corresponding to 120--200\,au.  Since these are optically thick transitions, the radius is an upper limit.  Lastly, the \htcn\ line maps show a single compact source at high velocity, consistent with the kinematics of the large-scale outflows, while two compact sources appear at low velocities with a separation of $\sim$0\farcs5 (100\,au).  As shown in the moment 1 map of the \htcn\,\jj{4}{3} line (Figure\,\ref{fig:moment1}), the northern and southern sources have brighter emission at blue-shifted and red-shifted velocities, respectively, which shows the opposite kinematics as the high-velocity emission.  As we pointed out in Section\,\ref{sec:intensity_maps}, a SO$_{2}$ line contaminates the \htcn\ \jj{4}{3} line, which also has a double-peaked morphology in the intensity map (Appendix\,\ref{sec:coms_identified}).  However, zoom-in channel maps of these two lines (Figure\,\ref{fig:channelmap_so2_h13cn}) show different morphology variations as a function of velocity, suggesting that the observed morphology at the \htcn\,\jj{4}{3} line is mostly due to \htcn.  The red-shfited part of the SO$_{2}$ line peaks at the south, whereas the red-shifted part of the \htcn\ \jj{4}{3} line peaks at the north.

\section{Infall Kinematics}
\label{sec:infall}
% \subsection{The Infall Signature}
We use four molecular lines, HCN, \hcop, CS, and \htcn, to trace the infall kinematics in the envelope of BHR\,71.  The infall signature is best illustrated by the 1D line profiles.  Here we first characterize the observed line profiles, then present a radiative transfer model to constrain the underlying kinematics traced by these lines.  The radiative transfer uses a 2D axisymmetric envelope model calculated in full 3D to include the effect of inclination.

\subsection{The Infall Signature}
\label{sec:1d_spec}
We extract the spectra of HCN, \hcop, CS, and \htcn\ lines from the region of continuum emission (0\farcs{52}$\times$0\farcs{39}) to search for the red-shifted absorption against the continuum, indicative of the infalling gas in the envelope in front of the source \citep{1977ApJ...214L..73L,1992ApJ...394..204Z,1999ApJS..122..519C,2001ApJ...562..770D,2005ApJ...626..919E,2015ApJ...814...22E}.  The line of sight toward the continuum source would have the deepest absorption as well as the least contamination from the outflows, making it the best position to search for the infall signature.  We use the \textsc{CASA} \texttt{specflux} task to calculate the mean intensity, and fit the baseline for the second time.  For each spectral window, a linear polynomial is fitted to the line-free channels visually selected from each spectral window (see the discussion in Section\,\ref{sec:coms_result}) to achieve a better baseline calibration.

\begin{figure*}[htbp!]
	\centering
	\includegraphics[width=0.7\textwidth]{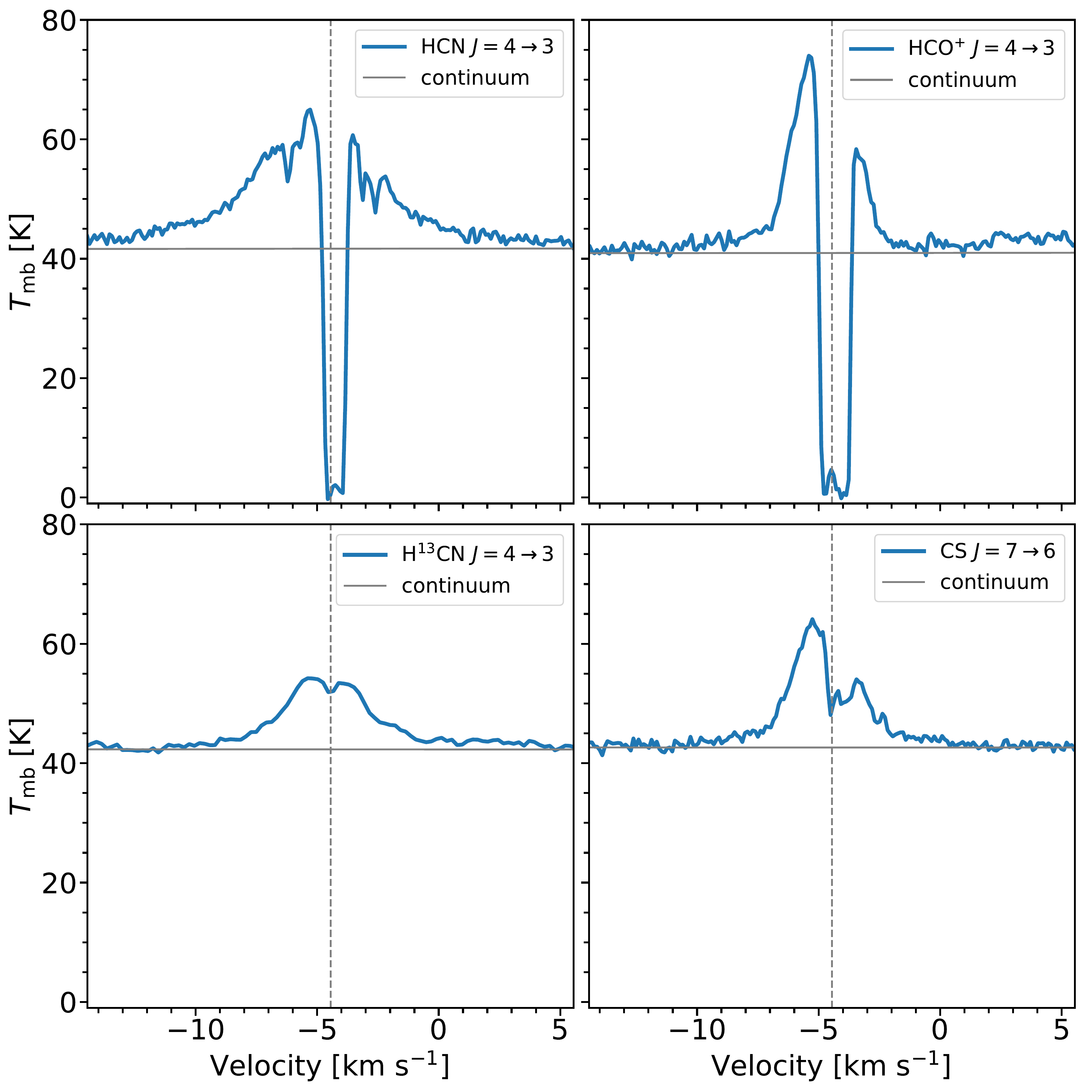}
	\caption{The spectra of HCN\,\jj{4}{3}, \hcop\,\jj{4}{3}, CS\,\jj{7}{6}, and \htcn\,\jj{4}{3} tracing the infalling envelope.  The source velocity is $-$4.45\,\kms\ \citep{1997ApJ...476..781B}.  The horizontal lines indicate the fitted continuum, while the vertical lines indicate the source velocity.}
	\label{fig:infall_lines}
\end{figure*}

Figure\,\ref{fig:infall_lines} shows the line profiles of HCN, \hcop, CS, and \htcn\ targeted for measuring the infall kinematics.  All four lines show blue asymmetric double-peaked profiles with the \hcop\,\jj{4}{3} line having the greatest asymmetry among these four, followed by CS\,\jj{7}{6}, HCN\,\jj{4}{3}, and \htcn\,\jj{4}{3}, whose two peaks have almost an equal intensity.  Both the HCN\,\jj{4}{3} and \hcop\,\jj{4}{3} lines show red-shifted absorptions below the continuum, which is an unambiguous signature of infall.
The absorption of the HCN\,\jj{4}{3} line is narrower and has a greater red-shift than that of the \hcop\,\jj{4}{3} line.  Within the absorption feature, if we take the midpoint of the velocities where the flux density is lower than the continuum as an estimate of the red-shift of the absorption feature, the absorptions in the HCN\,\jj{4}{3} and \hcop\,\jj{4}{3} lines \textit{center} at 0.17\,\kms\ and 0.14\,\kms, respectively.  The CS and \htcn\ lines also show absorptions at their line centers but not below the continuum flux.  The HCN line profile has several narrow absorption features with decreases of $\sim6$\,K, inconsistent with the hyperfine splitting of the HCN\,\jj{4}{3} line.  Therefore, the nature of these absorption features remains unclear.  The full widths at the half maximum (FWHM) of the line profiles excluding the absorption are 6.0\,\kms, 3.1\,\kms, 3.2\,\kms, and 4.1\,\kms, for the HCN\,\jj{4}{3}, \hcop\,\jj{4}{3}, CS\,\jj{7}{6}, and \htcn\,\jj{4}{3} lines, respectively, showing that the HCN line is significantly broader than the other lines.  The profile of the \hcop\ line best represents the infall signature as first proposed by \citet{1977ApJ...214L..73L}, \citet{1993ApJ...404..232Z}, and \citet{2001ApJ...562..770D}.

The absorption features seen in the \hcop\ and HCN lines have a small peak at the bottom, where the envelope should absorb most of the emission.  Because this small peak appears on top of a deep absorption, the origin of this peak must be in front of the dense envelope.  Such emission could arise from the surface layer of the envelope heated by the interstellar radiation field.

\subsection{Modeling the Infall}
\label{sec:infall_profile}
We successfully detect the infall signatures from our ALMA observations (Section\,\ref{sec:1d_spec}).  The red-shifted absorption against the continuum provides strong evidence for the presence of infalling gas along the LOS.  The absorption allows us to probe the velocity field and thus the age and rotation of the infalling envelope.  To constrain the underlying kinematics, we model the line profiles using non-LTE radiative transfer in 3D to properly consider the absorption.  By introducing the chemical abundance, a series of non-LTE radiative transfer calculations optimizes the 2D axisymmetric envelope model based on Y17 to reproduce the observations.  The modeling focus on the spectra toward the central continuum source to minimize the contamination of outflows.  However, given the moderate inclination of BHR\,71 and the broad line profiles of the HCN \jj{4}{3} and \htcn\ \jj{4}{3} lines, outflows may still contribute to the spectra at the beam toward the continuum source, especially the emission at velocities greater than $\pm$2\,\kms.  Thus, the radiative transfer modeling is mostly constrained by only the low velocity emission, and aims to investigate how well an envelope model can explain the observed infall signature.

\begin{figure}[htbp!]
	\centering
	\includegraphics[width=0.48\textwidth]{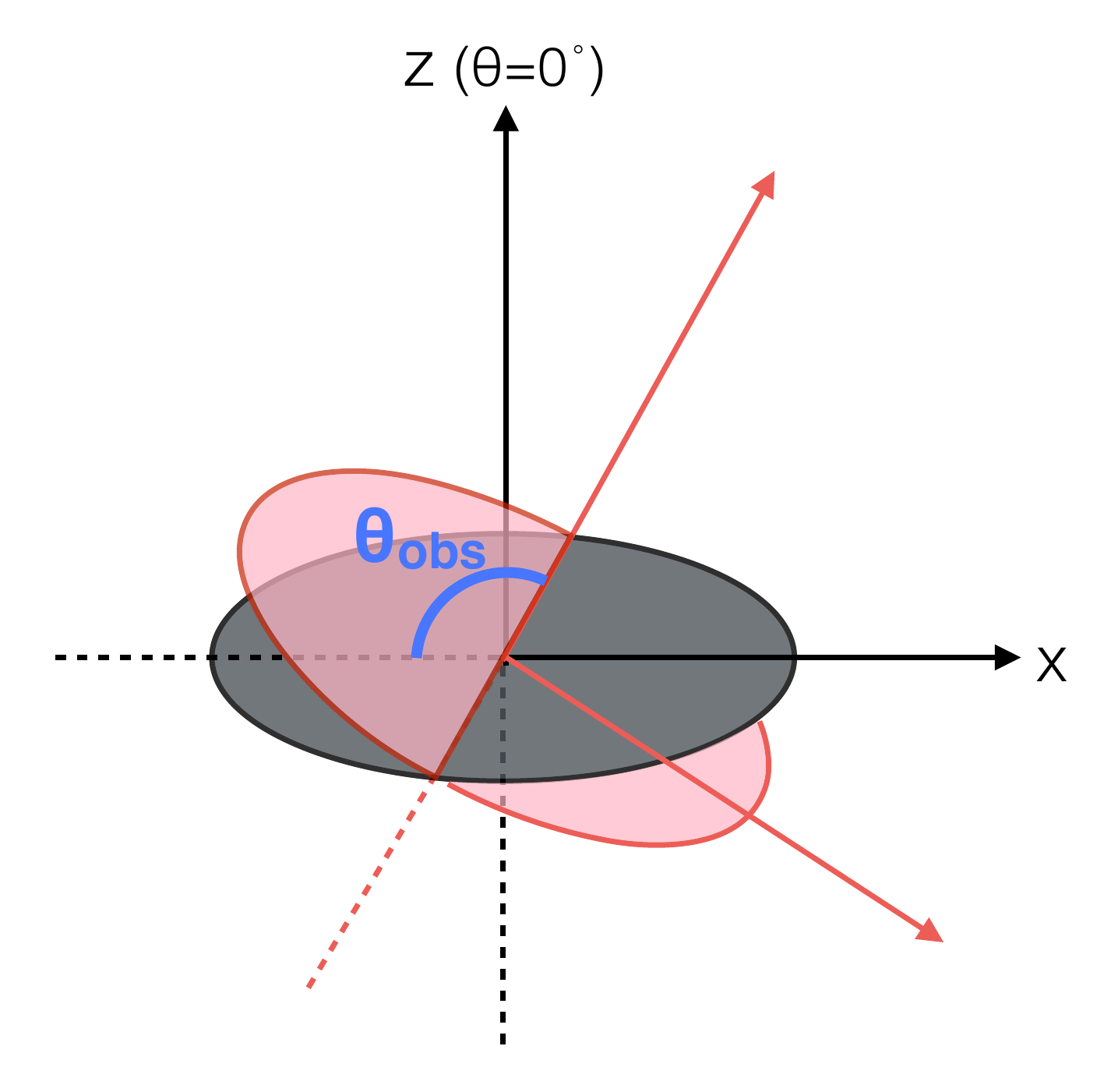}
	\caption{An illustration of a rotated system similar to the orientation of BHR\,71.  The y-axis is perpendicular to the figure.  The black/grey system represent the spherical coordinates of the BHR\,71 model with $\theta$ as the polar coordinates and the $+$z-axis as the rotation axis of the system.  As viewed by observers from $-$x-axis, BHR\,71 is rotated away from the line of sight ($-$x) by 130$^{\circ}$, illustrated by the red/pink system.  The inclination angle viewed by observers is denoted as $\theta_\text{obs}$.}
	\label{fig:inclination_cartoon}
\end{figure}

There are two types of inclination angles adopted in this study, one in the observer coordinates and another one in the model coordinates.  Figure~\ref{fig:inclination_cartoon} shows the rotated BHR\,71 system as viewed by observers.  For characterizing models, we typical show quantities in the model coordinates and from $\theta=0^{\circ}$ to $90^{\circ}$; for comparisons with the observations, the model is viewed from $\theta_\text{obs}$ ($130^{\circ}$).  For the model, viewing from $\theta = 50^{\circ}$ is equivalent to the view from $\theta_\text{obs}$ ($130^{\circ}$) because of the symmetry of the envelope model with the outflows at the north-south direction.

\subsubsection{Updating the Continuum Model}
\label{sec:cont_model}
\begin{figure}[htbp!]
	\centering
	\includegraphics[width=0.48\textwidth]{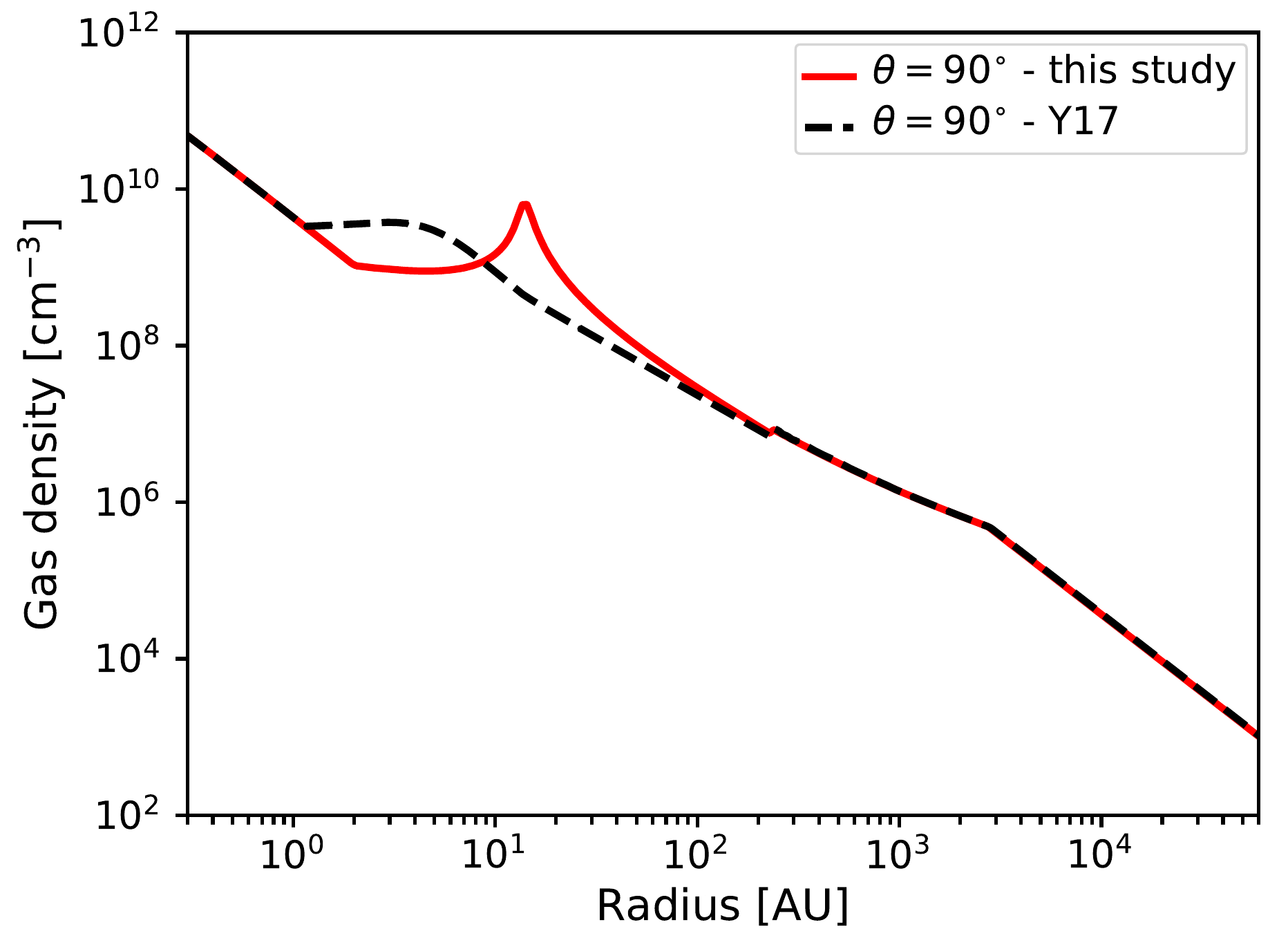}
	\caption{The radial density profile of the updated continuum model of BHR\,71 (red) compared with the profile shown in Y17 (dashed black).  Only the density profiles along the midplane of the envelope are shown here.  The envelope age is 36\,000 years.}
	\label{fig:cont_model}
\end{figure}

Y17 presented a TSC envelope with the geometry of outflow cavities.  The prescription of the model is 2D axisymmetric but the calculations are carried out in 3D to model the inclination of the rotation axis.  We discovered a numerical error in calculating the density around the centrifugal radius (13\,au in the Y17 model), causing an underestimation of the density around the centrifugal radius.  The updated radial density profile (Figure\,\ref{fig:cont_model}) has a clear density peak at the centrifugal radius similar to the analytical solution of the disk in \citet{1976ApJ...210..377U} and \citet{1981Icar...48..353C}.
Dominated by the envelope, the radial brightness profile extracted from the synthetic image at 160\,\micron\ still agrees with the \textit{Herschel} observation after correcting this numerical error, which determines the age of the TSC model \citep{2017ApJ...835..259Y}.

\subsubsection{The Radiative Transfer Model}
\begin{figure}[htbp!]
	\centering
	\includegraphics[width=0.48\textwidth]{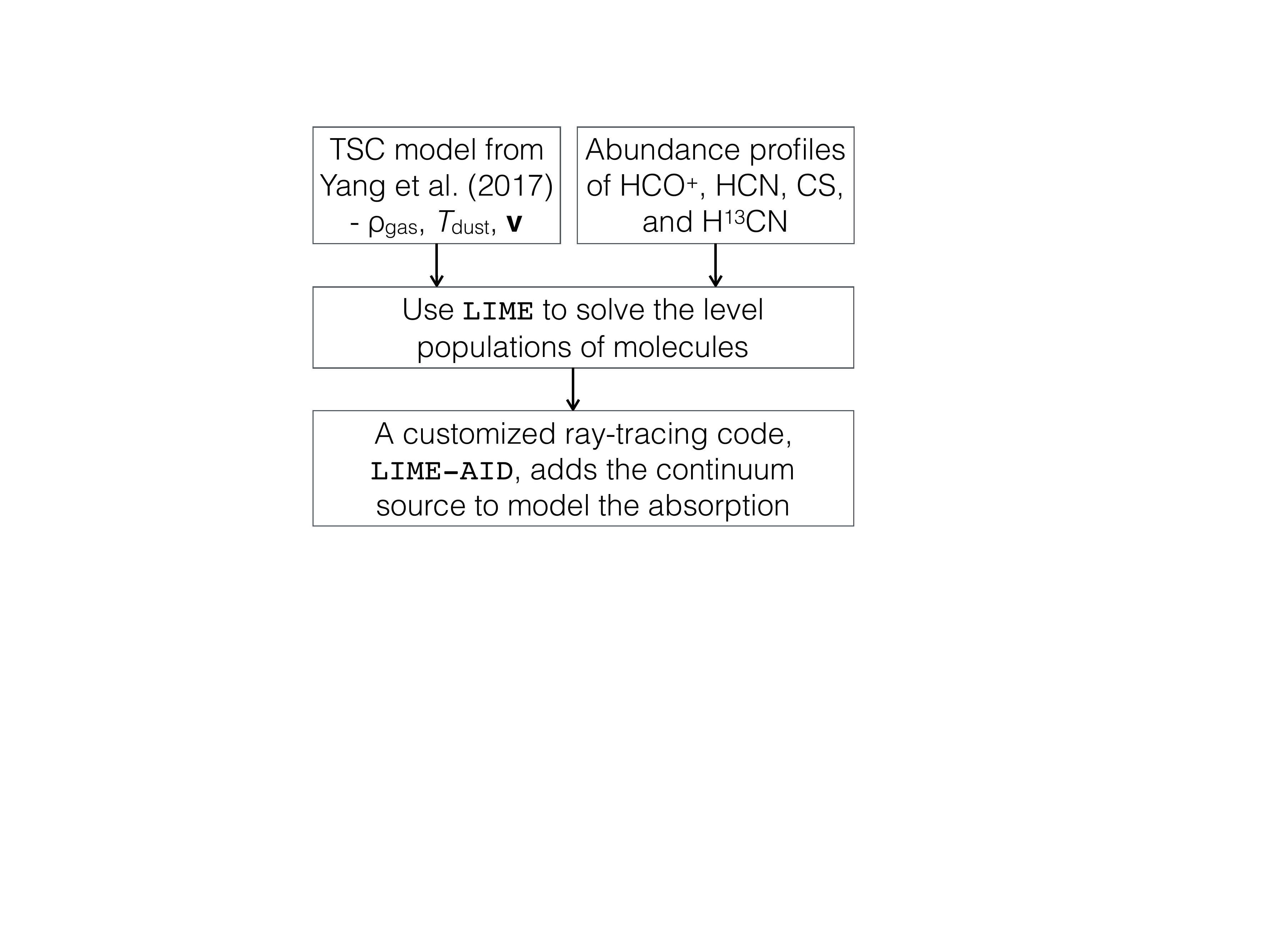}
	\caption{The workflow of modeling the infall profile.  The $\rho_\text{gas}$, $T_\text{dust}$, and $\bm{v}$ represent the gas density, dust temperature, and velocity vector, respectively.}
	\label{fig:workflow}
\end{figure}

We perform radiative transfer calculations to constrain the kinematics of the infalling envelope.  Figure\,\ref{fig:workflow} illustrates the work flow for modeling the infall signature.  Using \textsc{hyperion} \citep{2011A&A...536A..79R}, Y17 constrains the parameters of the 2D axisymmetric TSC envelope model with the far-infrared \textit{Herschel} spectra and the archival data of BHR\,71, which provides the gas density, velocity vector fields, and dust temperature ($T_\text{dust}$) as functions of radius $r$ and polar angle $\theta$.  The temperatures of gas and dust are in equilibrium when the gas density is greater than 10$^{4}$--10$^{5}$\,\cc \citep{2004ApJ...614..252Y}, where the Y17 model suggests $\sim$10$^{4}$\,au for both the radius of $n_\text{gas}=10^{4}$\,\cc\ and the radius of $T_\text{dust}=10$\,K.  Thus, we set the gas temperature to the dust temperature when $T_\text{dust}>10$\,K, and then apply an external heating correction when $T_\text{dust}<10$\,K, following the model in \citet{2004ApJ...614..252Y}.

\begin{figure}[htbp!]
	\centering
	\includegraphics[width=0.48\textwidth]{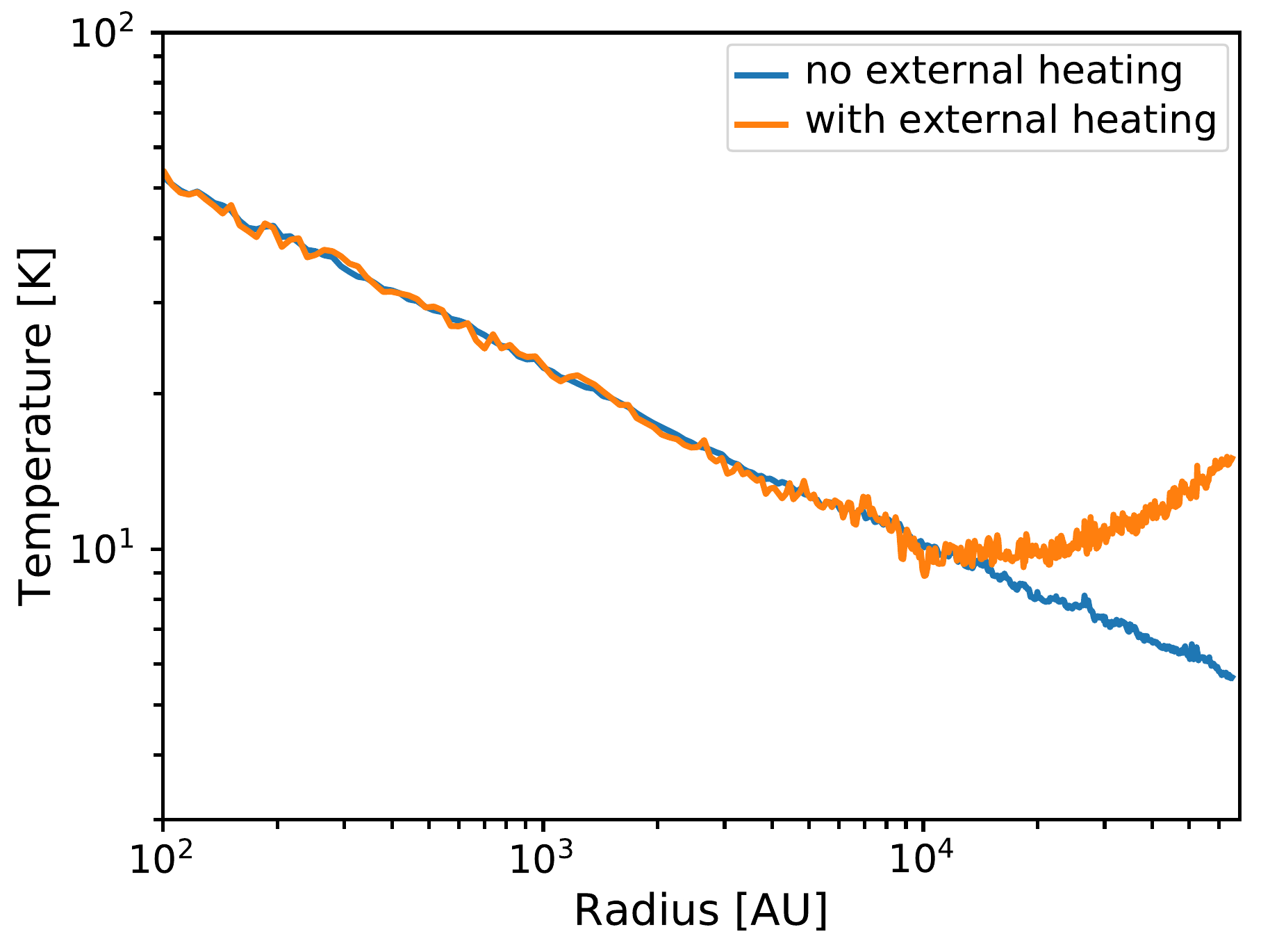}
	\caption{The radial temperature profiles along the midplanes of the envelopes with cavities.  The same external heating correction also prescribes to the entire envelope.  The orange line shows the profile with the external heating, while the blue line shows the profile without the external heating.}
	\label{fig:external_heating}
\end{figure}

An external heating correction applies for the cells with temperatures lower than 10\,K.  The corection is a linear function that increases with the radius so that the temperature at the edge of the envelope becomes 15\,K, described as
\begin{align}
	T(r)                     & = T(r) + \Delta T_{\text{ext}}(r), \text{if }T(r) < \text{10 K, and} \nonumber    \\
	\Delta T_{\text{ext}}(r) & = \frac{(r-r_\text{10\,K})}{(r_\text{max}-r_\text{10\,K})}(15 - T(r_\text{max})), 
	\label{eq:external_heating}
\end{align}
where $r_\text{max}$ is the envelope outer radius of 0.315\,pc, and $r_{\text{10\,K}}$ is the radius where the temperature becomes less than 10\,K.  A typical $r_\text{10\,K}$ is $\sim$10\,000\,au, similar to the values derived from self-consistent radiative transfer calculations of dense cores \citep{2004ApJ...614..252Y}.  Figure\,\ref{fig:external_heating} shows the radial temperature profiles before and after applying the external heating.  While the external heating correction has no significant impact on the synthetic line profile, we apply the external heating as described in Equation\,\ref{eq:external_heating} to the model for consistency.

With all other parameters taken from the best-fitting dust model, the abundances of the molecules are the main free parameters in this study.  The main parameters for the Y17 model of the TSC envelope are sound speed, rotational speed, and age, which are 0.37\,\kms, 5\ee{-13} rad\,s$^{-1}$, and 34\,000 years, respectively.  In Section\,\ref{sec:abundance_profile}, we investigate the impact of different abundance profiles on the observed infall signature.  The following paragraphs describe the procedures of calculating the synthetic line profiles.  We employ the Line Modeling Engine \citep[\textsc{lime};][]{2010A&A...523A..25B} to obtain the molecular level populations and then post-process %ray-trace
the model with a customized ray-tracing code to include a central continuum source.  \textsc{lime} iteratively solves the equation of radiation transport assuming non-local thermodynamical equilibrium (non-LTE), and produces image cubes for the specified molecular transitions.

Gridding is critical to ensure a realistic representation of the input model.  \textsc{lime} automatically determines the gridding based on the temperature and density profiles.  However, discontinuities, such as the cavities, in models often lead to unrealistic cell structure in the gridding.  Thus, we run \textsc{lime} with the density profile \textit{without} the cavities to ensure a robust gridding.  We assume a zero abundance in the outflow cavities, as these molecular lines show little emission along the outflows at high velocities.  Molecules around the cavity walls may contribute to the observed spectrum.  Modeling the emission from outflow cavity walls requires an accurate model for the kinematics and abundance, which has been elusive and is beyond the scope of this study.  The synthetic continuum emission may be slightly overestimated, but negligibly due to the low density in the cavities.

We set up the \textsc{lime} model with 50\,000 grid points distributed in a spherical envelope of 30\,000\,au according to the density profile.  The minimum spatial scale is 1\,au.  While using more grid points will reduce the potential clustering, where the grid points show an unusual overdensity at a localized region, we find no difference in the synthetic line profiles calculated with 50\,000 grid points and 500\,000 grid points.  We also set the maximum radius in \textsc{lime} smaller than the maximum radius in the model, 64\,973\,au, to better sample the high density regions.  Extending the model to the full size of the envelope has negligible effect on the synthetic line profiles.  \textsc{lime} uses sink points at the surface of the model to collect the emitting rays for imaging.  Because we post-process the derived level populations output to add the central continuum source, the number of the sink points is irrelevant.

Simulating the absorption against the continuum requires a central continuum source, which is beyond the capability of \textsc{lime}.  Therefore, we developed a ray-tracing package, \textsc{lime-aid} (\textsc{LIME}-Additional Intensity Decoder), to include continuum sources in our model, taking the level populations from \textsc{lime} along with the temperature, density, velocity, and abundance to calculate the LOS image cubes. The code itself is a derivative of the Cosmic Ly$\alpha$ Transfer code \citep[\textsc{colt};][]{2015MNRAS.449.4336S}, with the Monte Carlo scattering procedures replaced by ray-tracing.  The code also runs natively on unstructured data \citep{2017MNRAS.472..205S}, i.e., the Voronoi tessellation of points, which is efficiently and robustly constructed with the Computational Geometry Algorithms Library \citep{cgal_2018}.  Appendix\,\ref{sec:limeaid} provides a detailed description of \textsc{lime-aid}.

\subsubsection{The Kinematics of Infalling Envelopes}
\begin{figure*}[htbp!]
	\centering
	\includegraphics[width=\textwidth]{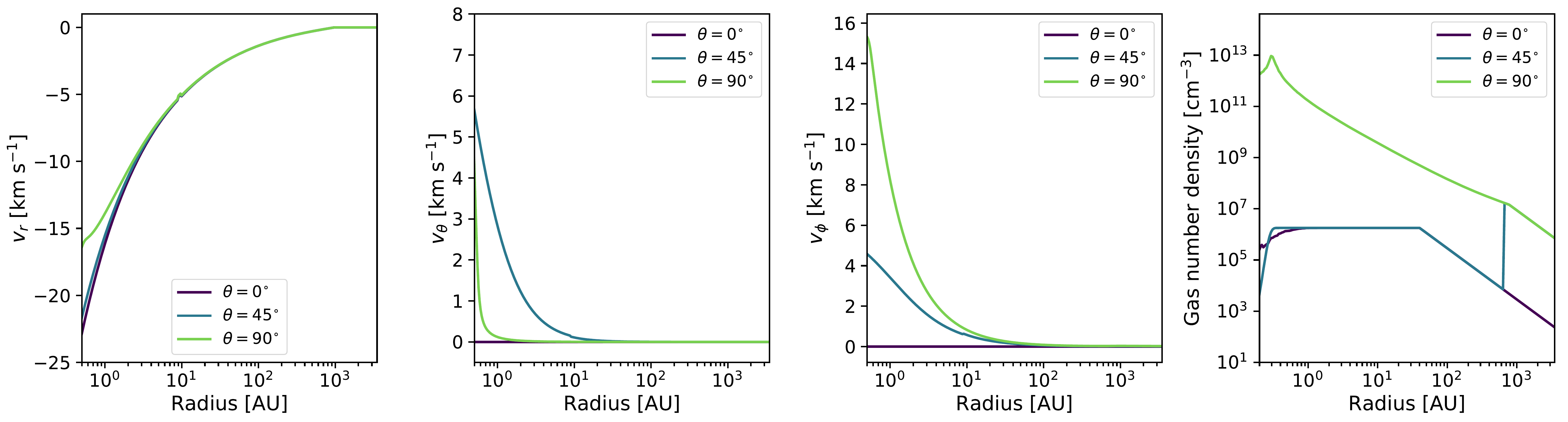}
	\caption{The radial, polar, and azimuthal velocity profiles as well as the gas density as a function of radius from the best-fitted TSC envelope (from left to right), which is the same model in Y17 with an age of 12\,000\,years instead.  The dust sublimation at high temperature leads to the decrease of density at the innermost region along the $\theta=0^{\circ}$ and 45$^{\circ}$. The velocity profiles along three different polar angles are shown, where $\theta=0^{\circ}$ is face-on and $\theta=90^{\circ}$ is edge-on.  The negative velocity for $v_\text{r}$ indicates that the gas is moving toward the center.}
	\label{fig:tsc_breakdown}
\end{figure*}

In a TSC envelope, matter rotating infalls onto the midplane of the envelope.  Figure\,\ref{fig:tsc_breakdown} illustrates the velocity in spherical coordinates along three different polar angles.  The velocity vector is defined with respect to the origin of the model, the center of the envelope (i.e., negative  in $v_\text{r}$ indicates infall).  Along the mid-plane of the envelope ($\theta=90^{\circ}$), the centrifugal force becomes significant at small radii compared to other $\theta$.  Thus, the radial velocity decreases most slowly along the mid-plane of the envelope, where the azimuthal velocity increases most rapidly as the radius decreases.  In the polar direction ($\theta=0^{\circ}$), gas infalls without rotation; therefore, the radial velocity increases the fastest as the radius decreases, while the polar and azimuthal velocities remain zero.  The TSC envelope does not follow the dynamical evolution within the centrifugal radius, where a disk may form.  Thus, a Keplerian disk does not evolve from the TSC model.

\subsubsection{Constraining the Abundance Profile}
\label{sec:abundance_profile}
The molecular abundance is a strong function of temperature and density in the protostellar envelope.  At the outer envelope, the molecules become photodissociated by the external radiation field.  As the radius decreases, molecules become abundant due to the shielding from dust.  However, molecules may freeze onto the dust grains at low temperature and high density.  The freeze-out timescale for molecules is inversely proportional to the gas density and the squared root of the temperature \citep{2004ApJ...617..360L}.  In the envelope, the density increases much more rapidly than the temperature as the radius decreases.  Thus, molecules become increasingly frozen out toward smaller radii.  When the temperature is higher than the evaporation temperature of the molecules, the abundance increases again as the molecules are released from the ices.  \citet{2004A&A...416..603J} use a ``drop function,'' a constant abundance along with a region of lower abundance, to simplify the freeze-out process for modeling the CO line profiles.  Here we start with the simple drop function for the abundance profile to try to fit the observed infall profiles.

\begin{figure}[bthp!]
	\centering
	\includegraphics[width=0.48\textwidth]{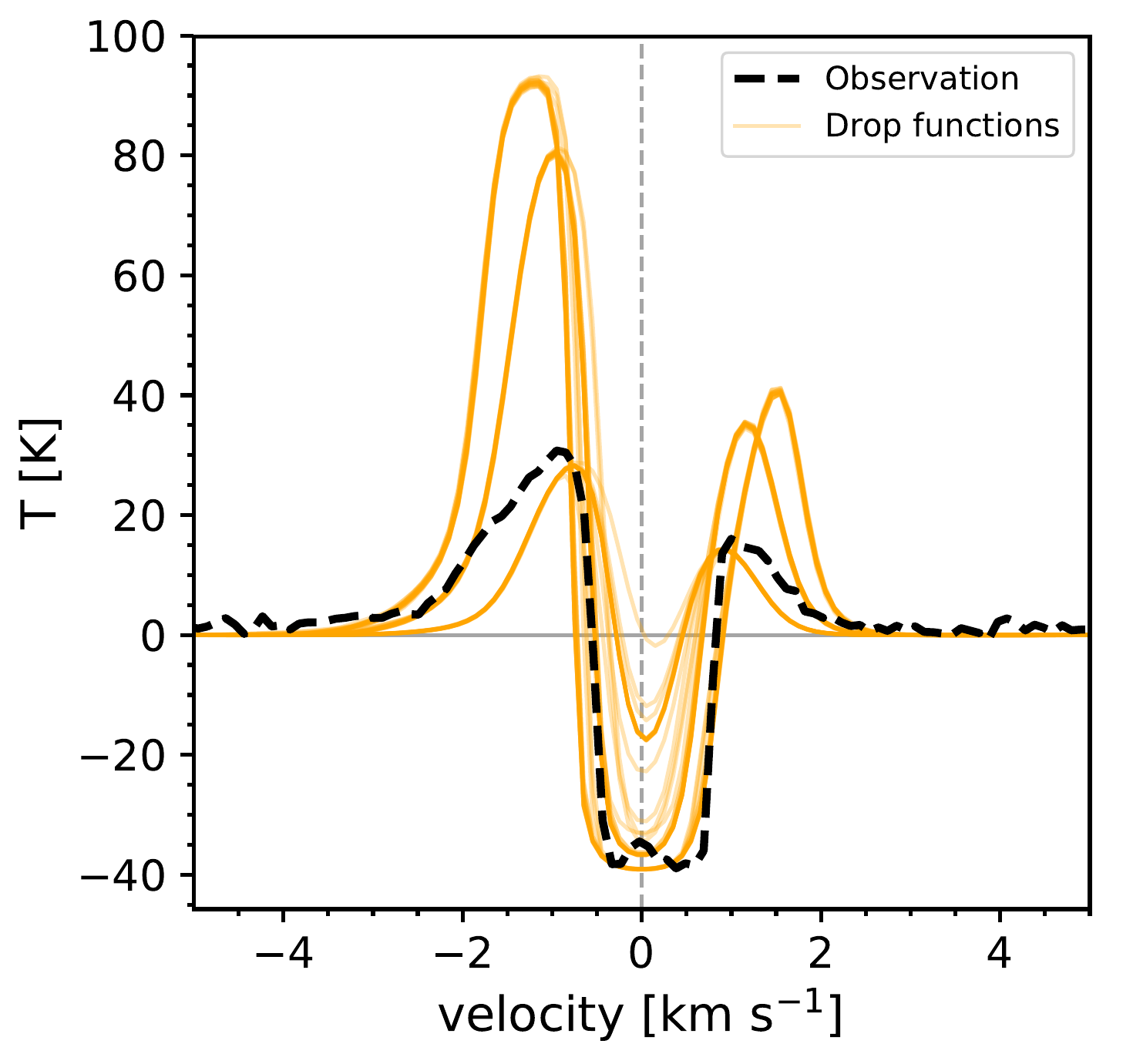}
	\caption{The observed \hcop\,\jj{4}{3} line profile along with the spectra modeled with the drop function abundance.  A total of 54 models are presented here.  The lack of agreements with the observations suggests the need of a more complex abundance profile than the drop function.}
	\label{fig:drop_grid_spectra}
\end{figure}

The drop function depends on the four parameters, the evaporation temperature ($T_{\rm evap}$), the depletion density ($n_{\rm depl}$), the density at which the molecule depletes ($X_{0}$), and the depleted abundance ($X_{\rm depl}$).  We test the feasibility of the drop function on the \hcop\,\jj{4}{3}, whose profile resembles a typical infall signature, a red-shifted absorption against the continuum along with a clear blue-asymmetric double-peaked profile.  The desorption temperature of CO determines the production of \hcop; therefore, we set the evaporation temperatures of \hcop\ to either a typical temperature of 20\,K \citep{2004ApJ...617..360L} or to a slightly higher temperature of 30\,K \citep{2002A&A...389..908J}.
Additionally, we add a destruction temperature of 100\,K, where water becomes evaporated and destroys \hcop\ \citep{2013ApJ...779L..22J}, acting as an inner cutoff radius of the abundance profile.
We explore the parameter space using 54 models with $10^{-10}\leq X_{0}\leq 10^{-7}$, $10^{-11}\leq X_{\rm depl}\leq 10^{-8}$, $10^{5.5}\leq n_{\rm depl}$ (\cc) $\leq 10^{8}$, and $T_\text{evap}=20, 30$\,K, distributed linearly in logarithmic scale.
Figure\,\ref{fig:drop_grid_spectra} shows the spectra of all 54 synthetic line profiles compared with the observation.  The models with the drop function result in too much emission at high velocities, when the synthetic spectra agree with the absorption feature.  The synthetic spectra underestimate the high velocity wings, when matching the peaks of the observations.  Our experiments suggest that the drop function fails to provide the required flexibility to model the high resolution ALMA spectra.  \citet{2005ApJ...626..919E} also found that a simple step function requires unlikely combinations of parameters to fit the observations of B335.

\begin{figure}[htbp!]
	\centering
	\includegraphics[width=0.48\textwidth]{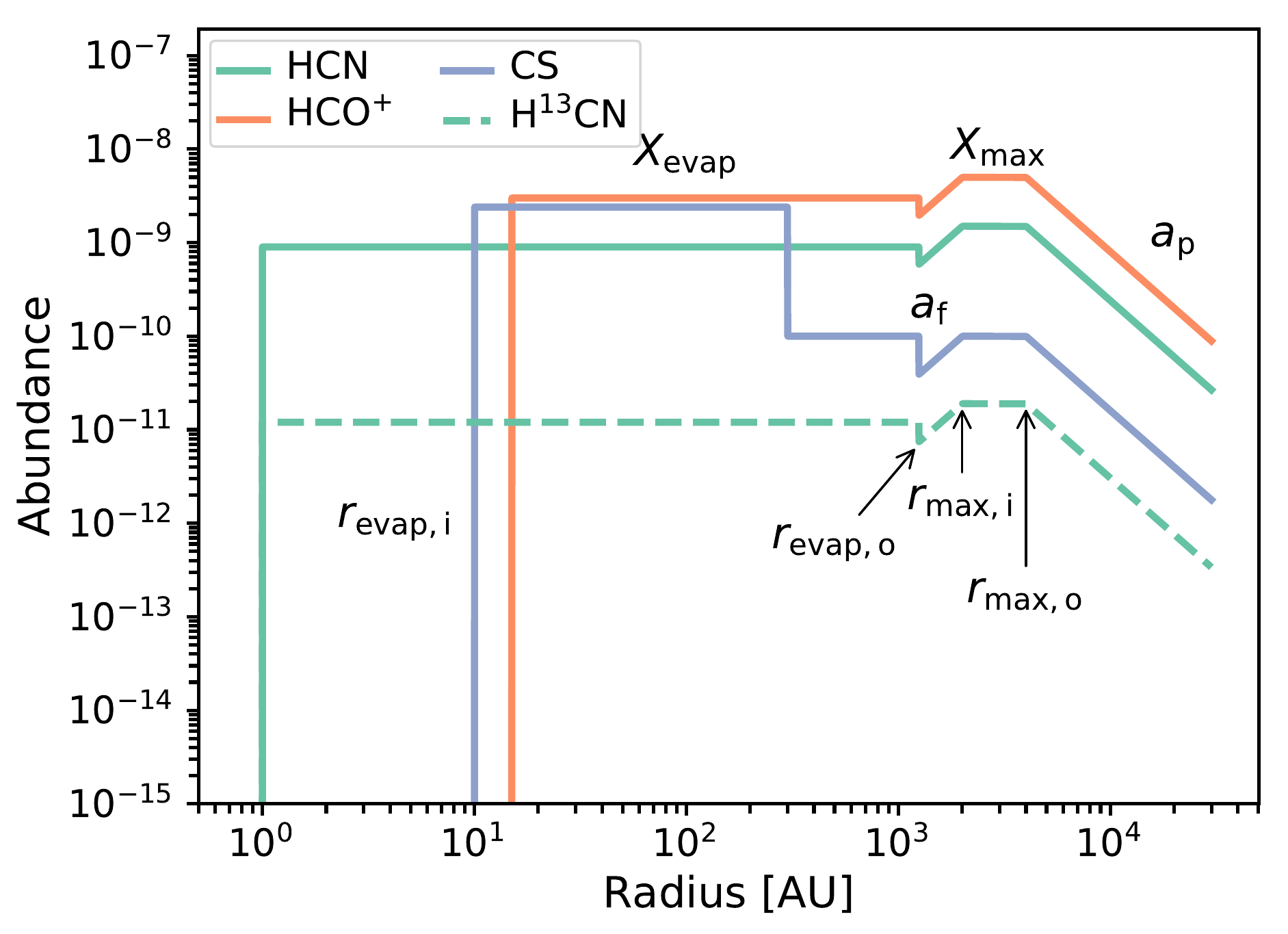}
	\caption{The best-fitted parametrized abundance profiles for HCN, \hcop, CS, and \htcn.  The \htcn\ abundance is one third of the abundance of HCN.}
	\label{fig:abundance_profile}
\end{figure}

To overcome the limitation of the drop function abundance, we construct a parameterized abundance profile that resembles the features in the chemo-physical modeling \citep{2004ApJ...617..360L}.  Such a profile has the flexibility to reproduce not only the abundance of \hcop\ but also HCN, CS, and \htcn.  Here we use \hcop\ as an example to describe the effect of each parameter (Table\,\ref{tbl:abundance}).  \hcop\ is a daughter molecule of CO and H$_{3}^{+}$ formed by the reaction, described as
\begin{equation}
	{\rm CO}+{\rm H}_{3}^+ \rightarrow \hcop+{\rm H}_{2}.
	\label{eq:hcop_formation}
\end{equation}
Thus, the freeze-out of CO dominates the abundance of \hcop\ when the temperature is lower than the evaporation temperature of CO, 20\,K \citep{2004ApJ...617..360L}.  When the temperature becomes higher than the evaporation temperature of CO, the abundance of \hcop\ increases.  The low abundance of H$_{3}^{+}$ due to the low ionization in the high density inner regions makes the abundance in the evaporation zone lower than the abundance in the outer envelope.  Instead of a sudden drop in the abundance in the ``drop function'', we set the abundance proportional to $r^{2}$ in the freeze-out zone until CO becomes evaporated at the inner high temperature region.  The high CO abundance enhances the production of \hcop, and other carbon-bearing molecules.  When the temperature becomes higher than 100\,K, water, a major destroyer of \hcop, sublimates, reducing the abundance of \hcop.  For other molecules, such as HCN and CS, the gas-phase chemical equilibrium may also reduce their abundance at high temperature.  Toward the edge of the envelope, the external radiation field photodissociates the \hcop.  Thus, we introduce a decrease of the abundance at the outer radius as $r^{-2}$.  Lastly, we set the abundance at the inner evaporation zone ($X_\text{evap}$) lower than the outer region, which has the maximum abundance ($X_\text{out}$).  $X_\text{out}$ is not necessarily greater than $X_\text{evap}$ for other molecules.  Figure\,\ref{fig:abundance_profile} illustrates the constructed abundance profiles and the corresponding parameters.

\begin{table*}
	\centering
	\caption{The parameters of the abundance profiles (Figure\,\ref{fig:abundance_profile})}
	\begin{tabular}{r|p{2.5in}|cccc}
		\toprule
		Parameter         & Description                                  & \hcop      & HCN         & CS          & \htcn      \\
		\midrule
		$X_\text{out}$    & Maximum abundance                            & 5.0\ee{-9} & 8.0\ee{-10} & 1.0\ee{-10} & 2.7\ee{-10} \\
		$X_\text{evap}$   & Abundance of the evaporation zone            & 2.0\ee{-9} & 1.0\ee{-8}  & 2.5\ee{-9}  & 3.3\ee{-9}  \\
		$r_\text{max,i}$  & Inner radius of the maximum abundance region & 1000\,au   & 1000\,au    & 1000\,au    & 1000\,au    \\
		$r_\text{max,o}$  & Outer radius of the maximum abundance region & 1500\,au   & 1200\,au    & 1500\,au    & 1200\,au    \\
		$r_\text{evap,i}$ & Inner radius of the evaporation zone         & 50\,au     & 1\,au       & 10\,au      & 1\,au       \\
		$r_\text{evap,o}$ &  Outer radius of the evaporation zone & \multicolumn{4}{c}{set to 1250 au} \\
		$a_\text{f}$      &  Power-law index of the freeze-out zone & \multicolumn{4}{c}{set to 2.0} \\
		$a_\text{p}$      &  Power-law index of the outer decreasing region dominated by photodissociation & \multicolumn{4}{c}{set to $-$2.0} \\
		\bottomrule
	\end{tabular}
	\label{tbl:abundance}
\end{table*}

The chemical evolution of CO also controls the abundance profiles of HCN and CS.  When CO freezes onto dust grains, HCN and CS abundances are also reduced.  Once CO becomes evaporated, the gas-phase production of HCN and CS increases \citep{2004ApJ...617..360L}.  For HCN, the gas-phase abundance eventually becomes unsustainable for chemical equilibrium so that its abundance decreases in the higher temperature region (i.e., inner radius).  \citet{2004ApJ...617..360L} find three peaks in the abundance profile of HCN in the protostellar stage, corresponding to the evaporation of CO, HCN, and CN; however, our experiment suggests that the abundance with only one evaporation zone can reproduce the observations.  The abundance of \htcn\ follows the abundance of HCN divided by the isotope ratio of 77 \citep{1994ARA&A..32..191W}.  CS has two peaks in its abundance profile.  The first one occurs where CO is evaporated, while the second one occurs at an inner radius where CS, which has a higher evaporation temperature, is evaporated.  The abundance of CS remains high at the inner radius, where most of the oxygen still freezes onto dust grains as atomic oxygen and water.  Thus, sulfur atoms tend to stay in CS rather than SO, which would be preferred if oxygen abundance increases.  At large radii, CS has a low abundance due to the destruction by UV photons.

\begin{figure}[htbp!]
	\centering
	\includegraphics[width=0.48\textwidth]{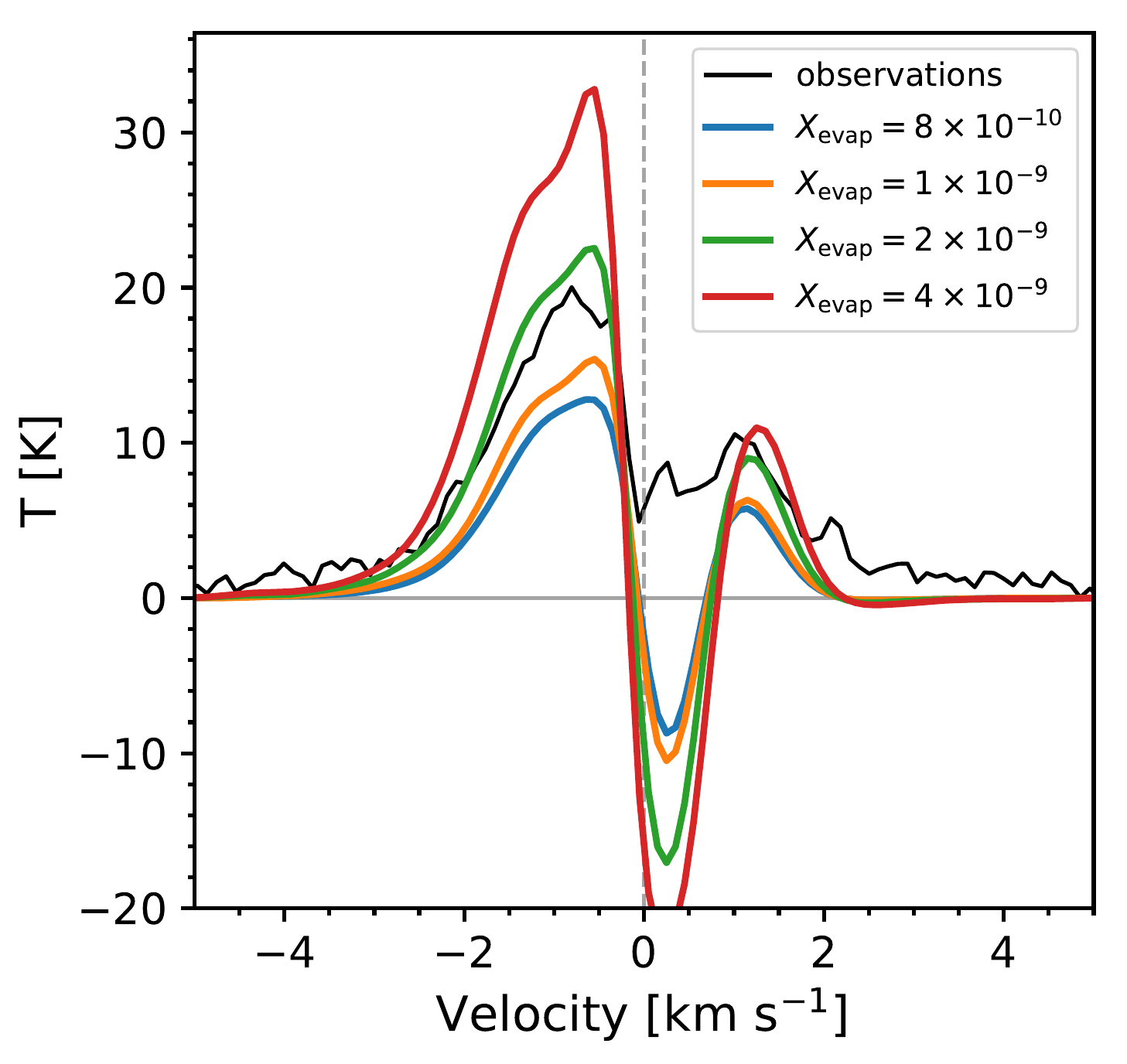}
	\caption{The synthetic CS line profiles as a function of the abundance in the evaporation zone.}
	\label{fig:cs_aevap}
\end{figure}

Our experiments of the CS abundance profile suggest a need for two evaporation zones to describe the observations.  With only one evaporation zone, the synthetic spectrum typically overestimates the absorption.  Our experiments show that only decreasing the abundance in the evaporation zone can effectively reduce the absorption, but it also reduces emission at the same time (Figure\,\ref{fig:cs_aevap}).  As \citet{2004ApJ...617..360L} demonstrate, the abundance of CS has two peaks inside the region where the temperature is greater than the CO evaporation temperature.  Thus, we set up two evaporation zones corresponding to 20\,K for CO and 35\,K for CS for the abundance of CS.

In general, the properties of the evaporation zone (abundance, inner and outer radii) dominate the resulting synthetic spectra, while the combination with the outer abundance profile have secondary effects to tune the line profile.
To constrain the abundance profile, we set the initial parameters with the insights of the chemical modeling in \citet{2004ApJ...617..360L}.  Then we manually search for the best-fitting abundance profile while exploring the effect of each parameter.  During the search, the evaporation radius is fixed to 1000\,au, where the temperature is $\sim$20\,K in the midplane of the envelope.  The slopes of freeze-out and photodissociation are fixed to 2 and $-$2 to reduce the number of degrees of freedom.

\subsection{The Best-fitting Model}
\label{sec:bestfit}

\begin{figure*}[htbp!]
	\centering
	\includegraphics[width=0.7\textwidth]{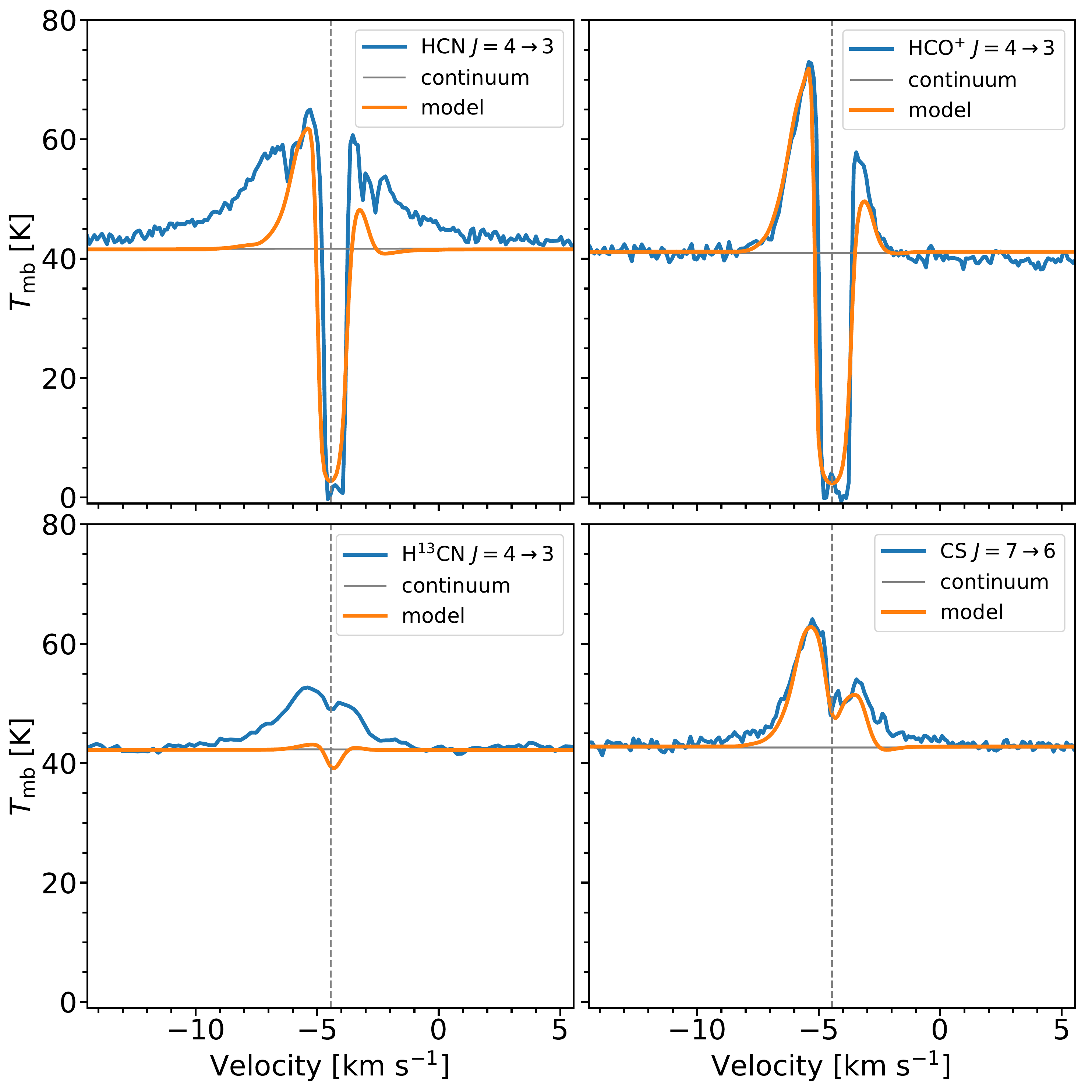}
	\caption{The synthetic spectra of HCN\,\jj{4}{3}, \hcop\,\jj{4}{3}, CS\,\jj{7}{6}, and HCN\,\jj{4}{3} in orange along with the observed line profile in blue and the continuum fitted from the observations in gray.  The emission of identified COMs is subtracted for the spectra of \hcop~\jj{4}{3} and \htcn~\jj{4}{3} lines, which makes the spectra slightly different than the spectra shown in Figure\,\ref{fig:infall_lines}.}
	\label{fig:infall_colt}
\end{figure*}

Figure\,\ref{fig:infall_colt} shows the best-fitting infall profiles sampled with the same $uv$-coverage of our ALMA observation using the CASA tasks \textsc{simobserve} and \textsc{simanalyze}.  We take the best-fitted parameters from Y17 as the initial guess of the model, and iterate the abundance profiles and ages to find the best-fitting models.  The \textsc{lime-aid} program includes the central continuum source according to the 2D continuum fitted from the line-free channel in each spectral window instead of an averaged continuum over all spectral windows.  Table\,\ref{tbl:cont_spws} lists the properties of the continuum emission fitted at each molecular line.

Sampling the synthetic image cube with the ALMA $uv$-coverage is the last step of modeling.  Due to the imperfect $uv$-coverage of the ALMA observations, the resulting spectra after simulating the ALMA visibility show a systematic decrease of 2\,K on the continuum, and the decrease becomes as high as 4\,K at the peak of the line, which is consistent with some contribution from the extended line emission and the compact continuum source.  Thus, we tune the input continuum flux densities to match the observed continuum.  Table\,\ref{tbl:cont_spws} lists the best-fitted continuum fluxes.

\begin{table*}[htbp!]
	\caption{The observed and fitted continuum emission at each molecular line}
	\centering
	\begin{tabular}{ccccc}
		\toprule
		Line & Size & PA & Observed $F_\text{cont}$ & Fitted $F_\text{cont}$ \\
			 & [\arcsec] & [$^{\circ}$] & [Jy] & [Jy] \\
		\midrule
		HCN\,\jj{4}{3}   & 0.31$\pm$0.03$\times$0.27$\pm$0.03 & 97$\pm$39 & 1.11 & 1.12 \\
		\hcop\,\jj{4}{3} & 0.37$\pm$0.03$\times$0.28$\pm$0.03 & 106$\pm$17 & 1.16 & 1.20 \\
		CS\,\jj{7}{6}    & 0.32$\pm$0.03$\times$0.27$\pm$0.03 & 99$\pm$22 & 1.03 & 1.06 \\
		\htcn\,\jj{4}{3} & 0.33$\pm$0.02$\times$0.27$\pm$0.02 & 118$\pm$15 & 1.04 & 1.05 \\
		\bottomrule
	\end{tabular}
	\label{tbl:cont_spws}
\end{table*}

As we will further discuss in Section\,\ref{sec:coms_id}, the emission of acetone (\acetone) and deuterated methanol (\dmethanol) contributes to the line profile of the \hcop\ \jj{4}{3} line, and the emission of SO$_{2}$ contaminates the \htcn\ \jj{4}{3} line profile; thus, we subtract the modeled emission of those molecules from the observed line profiles for an accurate comparison against the models.  We also summarize the effect of cavities and inclination in Appendix\,\ref{sec:infall_effect}.  Figure\,\ref{fig:abundance_profile} shows the best-fitting abundance profiles of \hcop, HCN, CS, and \htcn\ with the best-fitting parameters listed in Table\,\ref{tbl:abundance}.  Moreover, Table\,\ref{tbl:model_params} lists other important parameters of the model.

\begin{table*}[htbp!]
	\caption{Important parameters of the model of BHR\,71}
	\centering
	\begin{tabular}{rll}
		\toprule
		$t_\text{age}$ & Age of the protostellar system after the start of collapse & 12\,000 years$^{a}$ \\
		$c_\text{s, eff}$ & Effective sound speed of the envelope including the turbulent velocity & 0.37\,\kms \\
		$c_\text{turb}$ & Turbulence velocity & 0.34\,\kms \\
		$\Omega_{\circ}$ & Initial angular speed of the cloud & $2.5 \times 10^{-13}\,\text{rad}\,{\rm s}^{-1}$ \\
		$R_\text{env}$ & Envelope outer radius & 0.315~pc \\
		$R_\text{cen}$ & Centrifugal radius & 0.6\,au$^{a}$ \\
		$\theta_\text{cav}$ & Cavity opening angle & 20$^{\circ}$ \\
		$\theta_\text{obs}$ & Inclination angle of the protostar & 130$^{\circ}$ \\
		\bottomrule
		\multicolumn{3}{p{6in}}{Note: Please see \citet{2017ApJ...835..259Y} for detailed prescription of the composite model of BHR\,71, especially the treatment of the disk.} \\
		\multicolumn{3}{p{6in}}{$^{a}$The age is reduced to fit the molecular lines.  The centrifugal radius is also changed correspondingly.}
	\end{tabular}
	\label{tbl:model_params}
\end{table*}

\subsubsection{The Implied Protostellar Age}
\label{sec:age}
\begin{figure}[htbp!]
	\centering
	\includegraphics[width=0.47\textwidth]{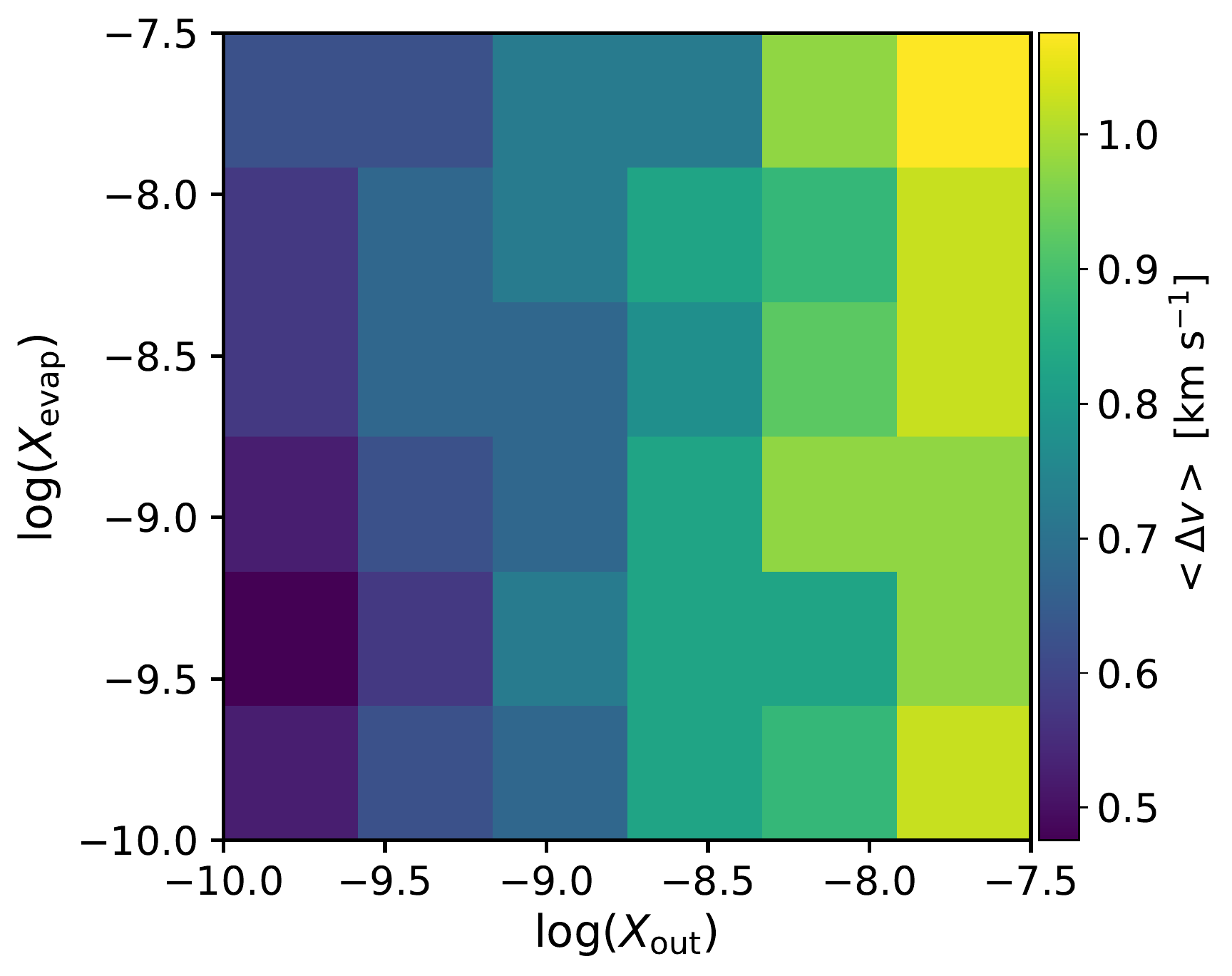}
	\caption{The squared difference of the peak velocities, $(\Delta v)^{2}$, from a grid of models of the \hcop\,\jj{4}{3} line.  The definition of $\Delta v$ is described in Section\,\ref{sec:age}.  All models have an age of 36\,000 years but with different evaporation abundances ($X_\text{evap}$) and maximum abundances ($X_\text{out}$) as indicated in the figure.  In comparison, the $\Delta v$ becomes 0.4\,\kms\ for the best-fitting model of the \hcop\,\jj{4}{3} line with an age of 12\,000 years.}
	\label{fig:peak_diff}
\end{figure}

The parametrized abundance profile provides a great flexibility to model the infall signature.  However, the velocities where the line profile peaks are less variable to the changes in the abundance profile.  We take the averaged difference of the peak velocities, $<\Delta v> = (| v_\text{peak, blue, obs}-v_\text{peak, blue, model}| + | v_\text{peak, red, obs}-v_\text{peak, red, model}|) / 2$, as an indicator of the goodness of the fitting for the peak positions.  Figure\,\ref{fig:peak_diff} shows the $<\Delta v>$ map of a grid of models for the \hcop\,\jj{4}{3} line where only the chemical abundance profile varies.  This grid of models has an age of 36\,000 years, the age derived in Y17.  While the abundance parameters vary more than two orders of magnitude, $<\Delta v>$ only decreases from 1.1\,\kms\ to 0.5\,\kms.  Moreover, the model with lower abundances produces a better match to the peak positions because the line profile becomes much flatter and only the low velocity emission appears; however, this model fails to reproduce the intensity of the observed infall signatures.  But, with an younger age of 12\,000 years, the $<\Delta v>$ becomes 0.2\,\kms\ for the \hcop\,\jj{4}{3} line without compromising of the intensity.  Thus, the age parameter is more effective on changing the velocities where the line profile peaks, making the peak positions a strong diagnostics of age.

A TSC model with an age of 36\,000 years produces infall signatures that peak at higher velocities than that of the observations.
The \hcop\ and HCN lines require an age less than 20\,000 years, while the CS line requires an age of 12\,000 years.  
The envelope with a younger age still agrees with the radial brightness profile of BHR\,71, because the $\chi^{2}$ value essentially levels off below 36\,000 years \citep{2017ApJ...835..259Y}.  To further characterize the uncertainty of the model-derived age, we run a grid of models whose ages ranging from 10\,000 to 15\,000 years with 1000 years increment, and compare the velocities of the peak positions against the CS line profile, which demands a younger age.  The minimum averaged velocity difference occurs at the age of 12\,000 years with $<\Delta v> = 0.05$\,\kms.  The difference becomes 0.10\,\kms\ and 0.11\,\kms\ for the ages of 13\,000 years and 11\,000 years.  If we take the spectral resolution, 0.1\,\kms\ as a threshold for the uncertainty, the derived age is 12\,000$\pm$2000 years.

With the updated age, the total infalling gas mass becomes 0.15$\pm$0.04\,\msun, instead of 0.43\,\msun\ from the age best-fitted with the continuum SED, which is only 6.5\%\ of the gas mass derived from the continuum.  If we consider the optically thick dust emission, the discrepancy becomes even greater.  Such discrepancy hints that the infall rate may be higher in the past.

\subsubsection{Turbulence Velocity}
\begin{figure}[htbp!]
	\includegraphics[width=0.48\textwidth]{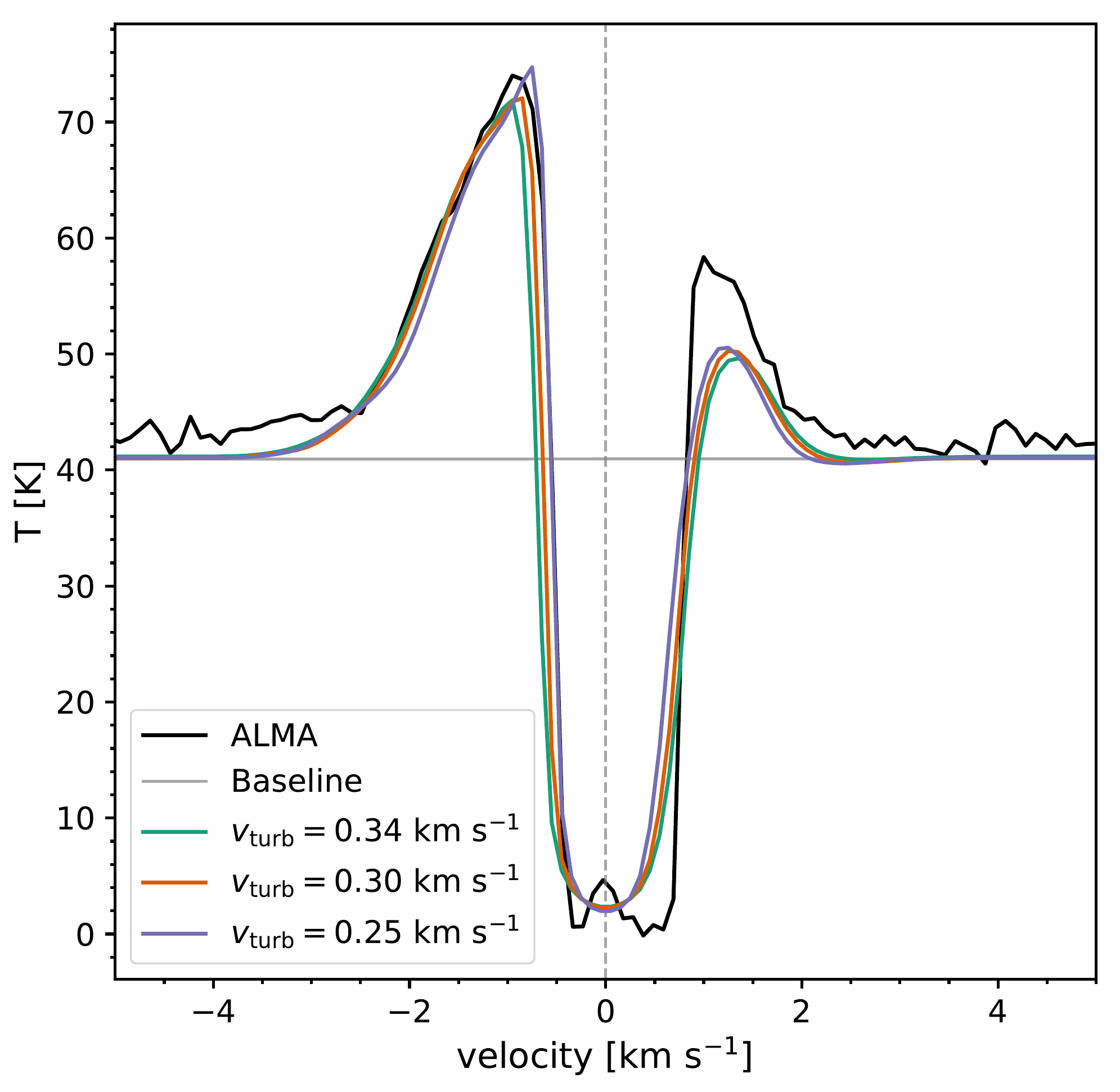}
	\caption{The synthetic line profiles of the \hcop\ \jj{4}{3} line with three different turbulence velocities.  A $v_\text{turb}$ of 0.34\,\kms\ is the same as the one used by Y17, but a $v_\text{turb}$ of 0.25\,\kms\ reproduces the line profiles reasonably well.}
	\label{fig:vturb}
\end{figure}

We iteratively test different abundance profiles for each molecule to find the best-fitting model.  While the synthetic profiles qualitatively reproduce the observations, the modeled absorption feature has always been slightly broader than the observations, where changing the abundance profile has little effect.  Thus, we decrease the turbulence velocity from 0.34\,\kms, derived in Y17 based on the ammonia observations by \citet{1995MNRAS.276.1067B}, to 0.25\,\kms\ (Figure\,\ref{fig:vturb}).  In comparison, Y17 derive an effective sound speed of 0.37\,\kms, contributed by a thermal velocity dispersion with $T=13$\,K derived by \citet{1995MNRAS.276.1067B} and a turbulence velocity of 0.34\,\kms.  The discrepancy in the turbulence velocity is similar to the uncertainty in the line width measurements of the ammonia observations, 0.1\,\kms.  

\subsubsection{The Compact Continuum Emission}
\label{sec:cont_discussion}
\begin{figure}[htbp!]
	\centering
	\includegraphics[width=0.47\textwidth]{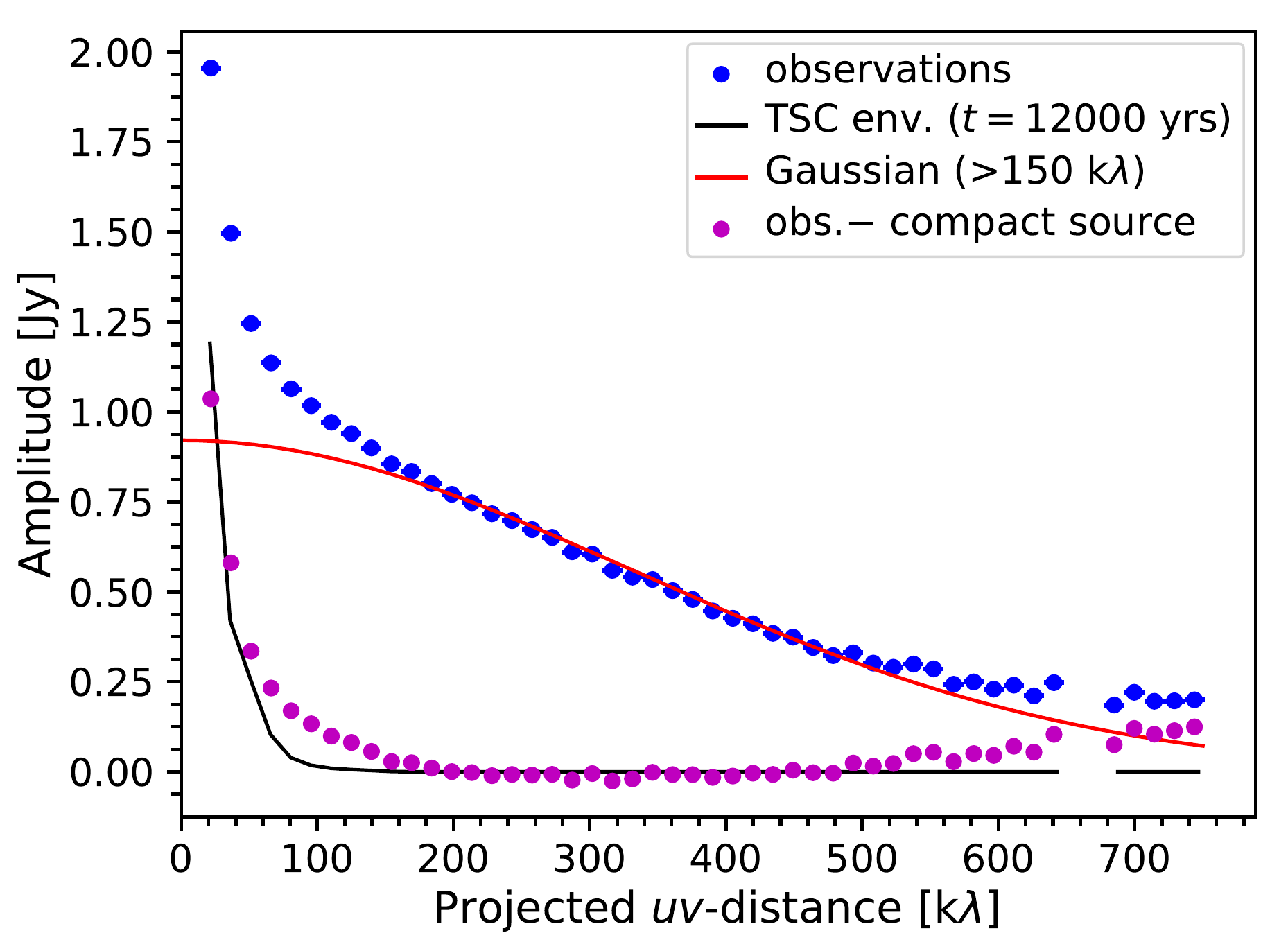}
	\caption{The $uv$-visibility of the observed continuum compared with the TSC envelope model of an age of 12\,000 years.  The weighted visibility is binned into 50 bins, where the mean and uncertainty are calculated.  The observations and envelope model are shown in blue and black, respectively.  The red line indicates a Gaussian fit to the visibility with $uv$-distance $>$150\,k$\lambda$.  The Gaussian is centered at zero with an amplitude of 0.92\,J7 and a Gaussian width of 332\,k$\lambda$.  The magneta dots show the visibility after subtracting the fitted Gaussian component.}
	\label{fig:uv_cont}
\end{figure}

The shortcomings of the infalling envelope model at the inner $\sim$100\,au and the COMs emission suggest a compact rotational structure with a radius of $\sim$50\,au, just below the spatial resolution of our observations.  To test the possibility of a compact source, we compare the observed $uv$-visibility with our envelope model.  The continuum visibility shows a peak at small $uv$-distance and a slowly declining tail at the long $uv$-distance (Figure~\ref{fig:uv_cont}).  The compact continuum source fitted from the image (Figure~\ref{fig:continuum}) has FWHMs of 0\farcs{35}$\times$0\farcs{28}, which correspond to Gaussian widths of 312\,k$\lambda$ and 391\,k$\lambda$ in $uv$-visibility, respectively.  A Gaussian fit to the observed $uv$-visibility at $uv$-distance greater than 150\,k$\lambda$ finds a Gaussian with a width of 332\,k$\lambda$, which is consistent with the compact continuum source (Figure\,\ref{fig:uv_cont}).  Our envelope model qualitatively agrees with the visibility after subtracting the compact source.  Both models of the envelope and the compact continuum source underestimate the visibility at large $uv$-distance, suggesting a more compact unresolved source within the continuum shown in Figure\,\ref{fig:continuum}.

\subsubsection{The Shortcomings of the Model}
% the high-v wings in HCN and the failure of H13CN
Our models reproduce the line profiles of HCN \jj{4}{3} (except for the high velocity wings), \hcop\ \jj{4}{3}, and CS \jj{7}{6} lines, but fail to reproduce the \htcn\ \jj{4}{3} line.  The TSC model has its limitataions.  For instance, the model ignores the angular momentum transportation within the centrifugal radius, where a disk should be forming.  Thus, the rotation velocity within the centrifugal radius \replaced{becomes flar}{remains constant} instead of approaching Keplerian under the TSC model (see Figure\,\ref{fig:tsc_breakdown}).  The Larson-Penston type of collapse is an alternative to the TSC model.  \citet{1977ApJ...218..834H} describes the similarity solution after the core formation following a Larson-Penston collapse (see also \citealt{1998ApJ...493..342S}).  With a higher infall velocity, a Larson-Penston collapse results in a greater peak separation than that of the observations \citep{1992ApJ...394..204Z}.  Adopting a Larson-Penston collapse to our 3D radiative transfer modeling is challenging as most of the models are in 1D \citep[e.g., ][]{1998ApJ...493..342S}.  Numerical simulations are another alternative from the TSC model \citep[e.g., ][]{2015ApJ...806...31G,2015ApJ...801..117T}.  However, most of the numerical models are not optimized to a certain source.  Fully characterizing the numerical models require exploration of parameter space and realizations along the dynamical evolution, which is beyond the scope of this study.

\paragraph{The High Velocity Wings of the HCN Line}
The best-fitting HCN model has high abundance at the region as close as 1\,au to the central protostar, driven by the broad peak at the blue-shifted velocity and the chemical modeling \citep{2004ApJ...617..360L}.  However, line width increases rather moderately, indicating that the inner evaporation radius is ill-constrained.  None of the models under our setup can reproduce the broad high velocity emission.  Outflows may contribute to the high velocity HCN emission; however, the high velocity emission is mostly concentrated at the center (Figure\,\ref{fig:infall_channelmap_full}) instead of following the morphology of the outflows.  Thus, the high velocity emission suggests an unexpected structure at the inner envelope, possibly an unresolved disk (see Section\,\ref{sec:unresolved_disk}).

\paragraph{The \htcn\ Line}
To model the profile of the \htcn\ \jj{4}{3}, we start with scaling down the HCN abundance by 77, the isotope ratio of $^{12}$C/$^{13}$C \citep{1994ARA&A..32..191W}.  However, this abundance profile produces a very weak line profile inconsistent with the observation.
Then we test different $^{12}$C/$^{13}$C ratios and find the best fit for a ratio of 3 for the model best-fitting the observation, which has a ratio of 3 for $^{12}$C/$^{13}$C.  While a $^{12}$C/$^{13}$C ratio of 3 is highly unlikely, such a high ratio suggests that the \htcn\,\jj{4}{3} line requires additional structures on top of our model.  Given the compact morphology of the \htcn\ emission (Figure\,\ref{fig:moment0}), a highly concentrated structure at the center of BHR\,71 may reproduce the observations.  The model for another protostar, B335, also shows a similar discrepancy between the models and the \htcn\,\jj{4}{3} line \citep{2015ApJ...814...22E}.  The possible underlying structure remains unknown, but future observations of the emission of complex organic molecules which are concentrated in the central region may provide insights to this inconsistency.  Besides the overall strength of the \htcn\,\jj{4}{3} line, our model predicts a significant absorption, whereas our observations show only a self-absorption of $\sim$5\,K.  Typically, increasing the exponent of the power-law for the outer photodissociation region reduces the absorption, but in the case of \htcn, where we drastically decrease the abundance at the outer envelope (Figure\,\ref{fig:abundance_profile}), the absorption still largely disagrees with the observation.  The best-fitting CS abundance has a large evaporation zone ranging from 10\,au to 400\,au.  To match the observed absorption feature, we not only need to make the abundance decrease more rapidly at the photodissociation region but also to reduce $X_\text{out}$.

\begin{figure}[htbp!]
	\centering
	\includegraphics[width=0.48\textwidth]{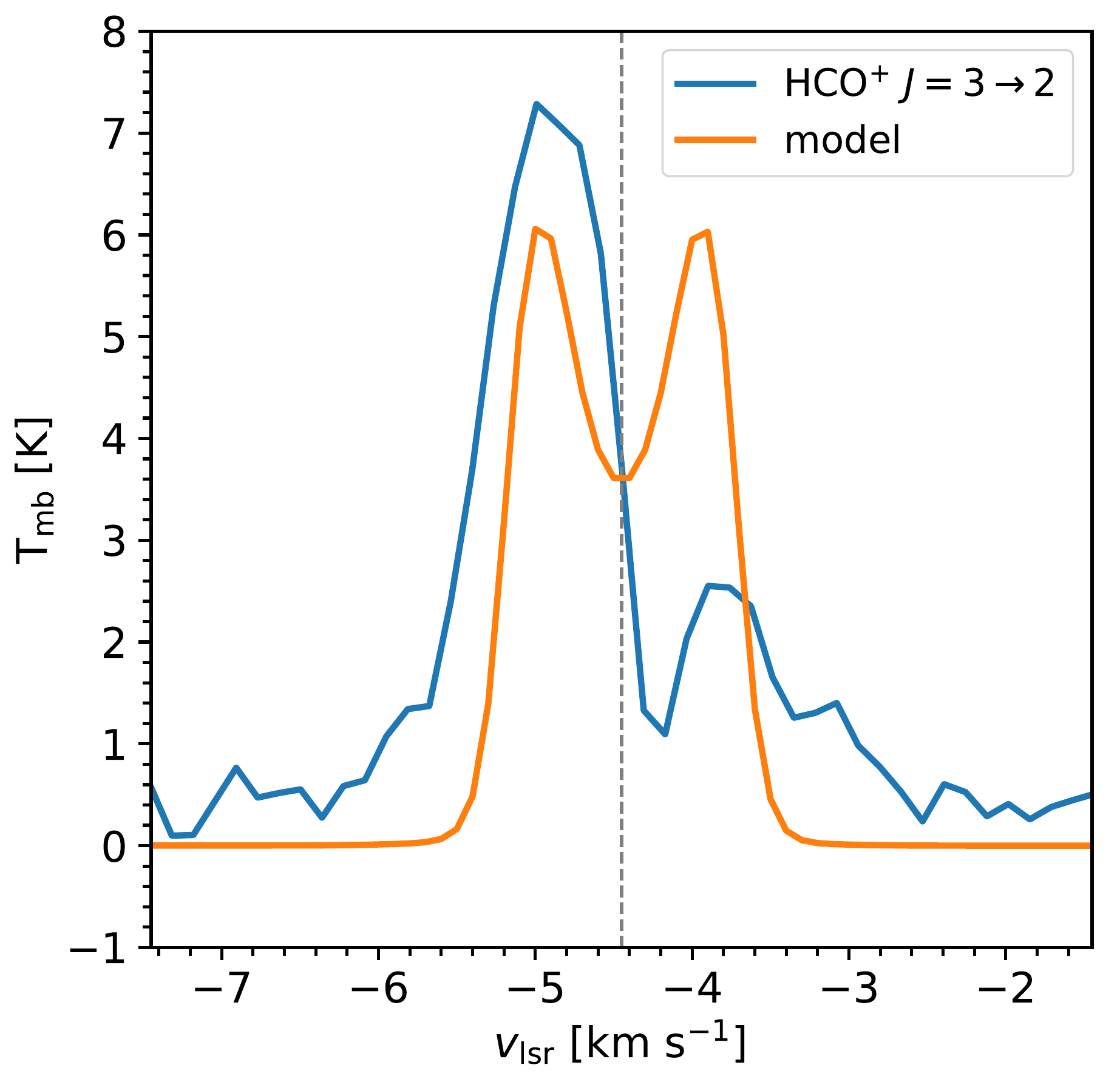}
	\caption{The synthetic \hcop\,\jj{3}{2} line profile (orange) compared with the archival APEX observation (blue).  The model is the same best-fitted model for the \hcop\,\jj{4}{3} line extracted from a 23\farcs{3} circular aperture without any beam correction.}
	\label{fig:single_hco+3-2}
\end{figure}

\paragraph{Comparison to the Single-dish Observations}
Figure\,\ref{fig:single_hco+3-2} shows the synthetic spectrum of the \hcop\ \jj{3}{2} line using the best-fitting model of \hcop\ compared with the archival APEX observation, which has a 23\farcs{3} beam.  While the synthetic emission has a strength comparable to the observation, the line profile is significantly different.  The observation shows significant blue-asymmetry between the two peaks, whereas the model suggests an equal strength double-peaked profile.  Given that our model is optimized for the ALMA observations, which is only sensitive to the structure smaller than 3\arcsec, we do not optimize our model for the single disk observation.  However, a comparable strength of the double-peaked profile suggests that the TSC envelope cannot fully describe the structure of BHR\,71.  From the \textit{Herschel}/SPIRE 500\,\micron\ image, the envelope of BHR\,71 exhibits an asymmetric structure elongated along northwest-southeast direction \citep[e.g., ][]{2017ApJ...835..259Y,2019ApJ...870...81T}.  If the foreground is more opaque than the TSC model predicts, the absorption would be shifted toward the velocities where it is observed.  Also, at larger scales, the envelope deviates significantly from spherical symmetry \citep{2010ApJ...712.1010T}.  Observations that measure the chemical abundance of \hcop, such as the observation of H$^{13}$CO$^{+}$, and probe the intermediate scales between the single dish data and our ALMA observation may revise our model to fit the data at both low and high resolutions.

\section{Complex Organic Molecules}
\label{sec:coms}
The bandwidths of our observations ($\sim$200\,\kms\ and $\sim$400\,\kms) allow us to not only measure infall indicators but also survey the emission of complex molecules at similar frequencies.  Throughout the dynamical evolution of dense cores from prestellar to protostellar, the chemical state of the molecular gas evolves as the material experiences variations in both density and temperature, which determines the available chemical reactions.  Over the last decade, observations have revealed a new class of protostars that exhibit rich spectra of complex organic molecules (COMs) from a warm inner region of their envelopes ($T\gtrsim$100\,K), so called ``hot corinos'' \citep{2004ASPC..323..195C,ceccarelli2007extreme}.  The formation of COMs depends on the environments, including temperature, density, and the sites of formation, which are constantly modified by the dynamical evolution of the star formation \citep{2008ApJ...682..283G,2013ChRv..113.8961A}.  While ALMA is revolutionizing the discovery of hot corinos \citep[e.g., The ALMA Protostellar Interferometric Line (PILS) Survey, ][]{2016A&A...595A.117J}, the link between the physics of star formation and the hot corino chemistry is relatively unknown.  In this section, we analyze the emission of molecules that do not trace the infall in the ALMA spectra extracted from the continuum size (0\farcs{52}$\times$0\farcs{39}) smoothed to 1\,\kms\ to optimize the line detectability.

\subsection{The Spectra of Complex Organic Molecules}
\label{sec:coms_result}
\begin{figure*}[htbp!]
	\centering
	\includegraphics[width=\textwidth]{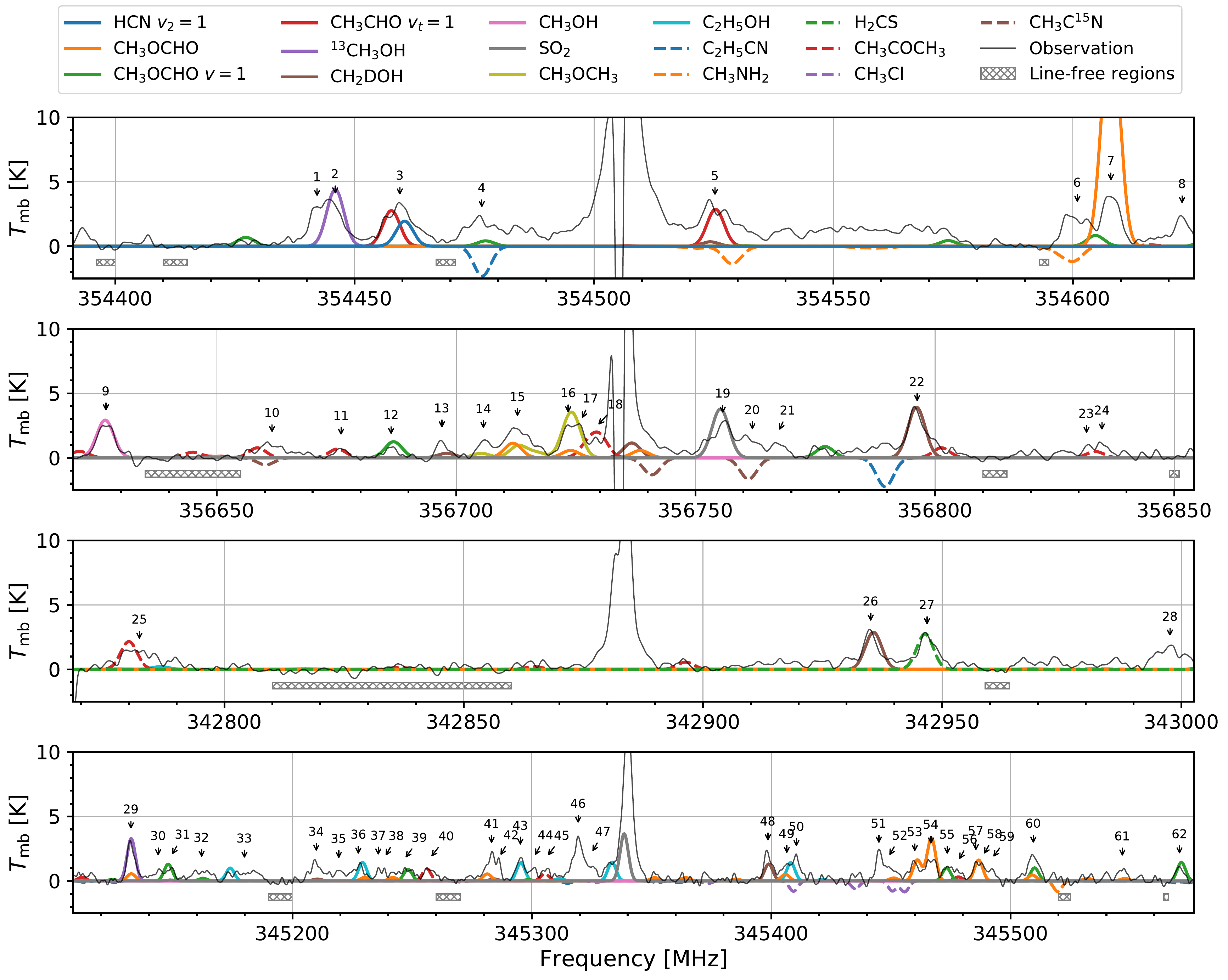}
	\caption{The 1D spectra of BHR\,71 with molecular species labelled.  Confirmed identifications are shown as positive spectra, while the tentative species are shown as negative spectra.  The observed spectra are smoothed with an 1D Gaussian kernel which has a FWHM of 1\,\kms\ to achieve a higher S/N on the weak emission.  The numbers of the identified features are consistent with the line number shown in Table\,\ref{tbl:line_fitting}.  Line 46 is identified as the emission of CH$_{3}$OD, which is not included in the database queried by \textsc{xclass}, hence not modeled.  The horizontal bars beneath the spectra indicate the line-free regions selected for baseline fitting and calibration.}
	\label{fig:1d_spectra_line_id}
\end{figure*}

The entire spectra of our ALMA observations shows many emission lines besides the emission that probes the infall (Figure\,\ref{fig:1d_spectra_line_id}).  On average, these emission mostly come from the central 0\farcs{50}$\pm$0\farcs{13}$\times$0\farcs{21}$\pm$0\farcs{09} (100\,au$\times$42 au) region, while several emission lines shows extensions in the north-south direction ($\sim$1\arcsec\ in size), consistent with higher abundances of COMs due to the increased temperature in the outflow cavity.  The mean line width of these emission is 4.5$\pm$2.5\,\kms\ with a median width of 3.7\,\kms; however, the widths of some lines may be overestimated due to the blending with other lines.  Appendix\,\ref{sec:line_fitting} describes the fitting and properties of these emission lines.

\subsection{Line Identification}
\label{sec:coms_id}
We use \textsc{splatalogue}\footnote{\href{http://www.splatalogue.net/}{http://www.splatalogue.net/}} and \textsc{xclass} \citep{2017A&A...598A...7M} to identify the lines seen in Figure\,\ref{fig:1d_spectra_line_id}.  The spectroscopic data are taken from the Cologne Database of Molecular Spectroscopy (CDMS; \citealt{2001A&A...370L..49M,2005JMoSt.742..215M,2016JMoSp.327...95E}) and the Jet Propulsion Laboratory (JPL; \citealt{1998JQSRT..60..883P}).  Here we list only the references that cover the frequencies relevant to this study.  For HCN $v_{2}=1$, the data were taken from \citet{2003ApJ...585L.163T}.  For \methanol, the data were taken from \citet{2008JMoSp.251..305X}; for \dmethanol, the data were taken from \citet{2012JMoSp.280..119P}; for \tmethanol, the data were taken from \citet{1997JPCRD..26...17X}.  For \methylformate, the data were taken from \citet{2009JMoSp.255...32I}, which includes the data from \citet{1984ApJS...55..633P,1999ApJ...521..255O,2007JMoSp.246..158C,2008JMoSp.251..293M}.  For \acetaldehyde, the data were taken from \citet{1996JPCRD..25.1113K}.  For \dimethylether, the data were taken from \citet{2009A&A...504..635E}.  For \acetone, the data were taken from \citet{2002ApJS..142..145G}.  For CH$_{3}$Cl, the data were taken from \citet{1986JMoSp.116..251W}.  For CH$_{3}$C$^{15}$N, the data were taken from \citet{2009A&A...506.1487M} together with the data from \citet{1996ApJ...471.1067P}.  For \methylamine, the data were taken from \citet{1992JMoSp.156..383K}.  For gauche-\ethanol, the data were taken from \citet{2008JMoSp.251..394P} together with the data from \citet{1996JMoSp.175..246P}.  For \ethylcyanide, the data were compiled from \citet{1994ApJS...93..589P,2009ApJS..184..133B}.  For SO$_{2}$, the data were taken from \citet{2005JMoSp.232..213M} with additional data from \citet{1985JPCRD..14..395L,1985JMoSp.111...66H,1998JMoSp.191...17B}.  Lastly, for H$_{2}$CS, the data were taken from \citet{2019A&A...621A.143M} together with the data from \citet{1971JMoSp..39..136J,2008ApJS..176..543M}.

To confirm identifications, we observe the following guidelines in \citet{2005ApJ...619..914S}
\begin{itemize}
	\item \textit{Rest frequency}: The identified transition has to have its rest frequency measured in laboratory or calculated to have an uncertainty at least on the order of 1 part in 10 million, corresponding to $\sim$0.04\,MHz for our data.
	\item \textit{Frequency agreement}: To assign a robust identification from the selected possible transitions, the assigned transition needs to agree with the observed line within the half width at half maximum of the line.
	\item \textit{Relative intensity}: After a transition is assigned, the relative strengths of all other transitions for the identified molecules need to match with the observation.
\end{itemize}

For each line, we search the transitions with a maximum upper energy of 1000\,K, and test if the corresponding species produces spectra consistent with the observations using \textsc{xclass}, assuming a typical model of an excitation temperature of 100\,K and a line width of 3.5\,\kms.  The \textsc{xclass} program considers the optical depths of both dust and molecules.  The best-fitting envelope model (Section\,\ref{sec:cont_model}) provides an estimate of the dust column density.  The emission of each molecule does not interact with those of others.  Table\,\ref{tbl:line_id} lists the results of the line idenfication.  The emission of CH$_{3}$OD is identified with the results in \citet{1988ApJS...67..135A} instead of the modeling with \textsc{xclass}.

\startlongtable
\begin{deluxetable*}{r l l l l l l}
 \tabletypesize{\scriptsize}
 \tablecaption{Line Identification \label{tbl:line_id}}
 \tablewidth{\textwidth}
 \tablehead{\colhead{Formula} & \colhead{Name} & \colhead{Frequency (MHz)} & \colhead{Transition\tablenotemark{a}} & \colhead{Einstein-A\tablenotemark{b}} & \colhead{$E_{\rm u}$ (K)} & \colhead{Line No.}}
 \startdata
 \hline
 \multicolumn{7}{c}{\textbf{Confirmed identifications}} \\ 
 \hline
 \methanol                       & Methanol           &  356626.667(0.016)    & [\J, \K]= [23, 4]$\rightarrow$[22, 5]                                   &  8.041(-5) & 728 & 9   \\
 \hline
 \tmethanol                      & Methanol           &  345132.599(0.05)     & [4, 0, 4]$\rightarrow$[3, 1, 3]                                         &  8.239(-5) & 36  & 29  \\
 \tmethanol                      & Methanol           &  354445.95(0.05)      & [4, 1, 3]$\rightarrow$[3, 0, 3]                                         &  1.274(-4) & 44  & 2   \\
 \hline
 \dmethanol                      & Methanol           &  356697.9372(0.0099)  & [9, 2, 8]$\rightarrow$[9, 0, 9]                                         &  3.536(-6) & 133 & 13  \\
 \dmethanol                      & Methanol           &  356796.1421(0.0034)  & [8, 6, 3]$\rightarrow$[7, 6, 2]                                         &  6.277(-5) & 234 & 22  \\
 \dmethanol                      & Methanol           &  356796.1421(0.0034)  & [8, 6, 2]$\rightarrow$[7, 6, 1]                                         &  6.277(-5) & 234 & 22  \\
 \dmethanol                      & Methanol           &  345398.9037(0.0112)  & [16, 2, 14]$\rightarrow$[15, 3, 13]                                     &  4.275(-5) & 310 & 48  \\
 \dmethanol                      & Methanol           &  342935.6452(0.0118)  & [17, 2, 16]$\rightarrow$[17, 1, 17]                                     &  1.227(-4) & 343 & 26  \\
 \hline
 \methanold \tablenotemark{b}    & Methanol           &  345319.587(0.359)    & [\J, \K, parity]=[8, 2, $-$]$\rightarrow$[8, 1, $+$]      &  8.1 ($\mu^{2}\Sigma S$) & 100 & 46  \\
 \hline
 \methylformate                  & Methyl formate     &  345465.3445(0.0023)  & [16, 13, 4]$\rightarrow$[16, 12, 5], Sym=E                              &  2.513(-5) & 192 & 54  \\
 \methylformate                  & Methyl formate     &  354608.0913(0.003)   & [33, 0, 33]$\rightarrow$[32, 1, 32]                                     &  1.097(-4) & 293 & 7   \\
 \methylformate                  & Methyl formate     &  354608.0919(0.003)   & [33, 1, 33]$\rightarrow$[32, 1, 32]                                     &  6.792(-4) & 293 & 7   \\
 \methylformate                  & Methyl formate     &  354608.0923(0.003)   & [33, 0, 33]$\rightarrow$[32, 0, 32]                                     &  6.792(-4) & 293 & 7   \\
 \methylformate                  & Methyl formate     &  354608.0928(0.003)   & [33, 1, 33]$\rightarrow$[32, 0, 32]                                     &  1.097(-4) & 293 & 7   \\
 \methylformate                  & Methyl formate     &  345461.011(0.1)      & [28, 13, 15]$\rightarrow$[27, 13, 14], Sym=E                            &  4.94(-4)  & 352 & 53  \\
 \methylformate                  & Methyl formate     &  345486.602(0.1)      & [28, 13, 16]$\rightarrow$[27, 13, 15], Sym=E                            &  4.941(-4) & 352 & 58  \\
 \methylformate                  & Methyl formate     &  356711.841(0.1)      & [29, 17, 12]$\rightarrow$[28, 17, 11]                                   &  4.561(-4) & 448 & 15  \\
 \methylformate                  & Methyl formate     &  356711.841(0.1)      & [29, 17, 13]$\rightarrow$[28, 17, 11]                                   &  4.561(-4) & 448 & 15  \\
 \hline
 \methylformatev                 & Methyl formate     &  345162.544(0.1)      & [11, 8, 3]$\rightarrow$[10, 7, 3], Sym=E                                &  7.091(-5) & 269 & 32  \\
 \methylformatev                 & Methyl formate     &  345148.04(0.1)       & [28, 6, 23]$\rightarrow$[27, 6, 22], Sym=A                              &  5.947(-4) & 452 & 31  \\
 \methylformatev                 & Methyl formate     &  356686.906(0.1)      & [29, 6, 24]$\rightarrow$[28, 6, 23]                                     &  6.584(-4) & 469 & 12  \\
 \methylformatev                 & Methyl formate     &  345473.217(0.1)      & [28, 9, 19]$\rightarrow$[27, 9, 18], Sym=E                              &  5.626(-4) & 481 & 55  \\
 \methylformatev                 & Methyl formate     &  345510.004(0.1)      & [28, 9, 20]$\rightarrow$[27, 9, 19], Sym=A                              &  5.635(-4) & 481 & 60  \\
 \methylformatev                 & Methyl formate     &  345571.314(0.1)      & [27, 5, 22]$\rightarrow$[26, 5, 21], Sym=E                              &  6.055(-4) & 481 & 61  \\
 \methylformatev                 & Methyl formate     &  345248.21(0.1)       & [28, 10, 19]$\rightarrow$[27, 10, 18], Sym=E                            &  5.481(-4) & 493 & 39  \\
 \hline
 \acetone                        & Acetone            &  342780.0361(0.0357)  & [17, 17, 0]$\rightarrow$[16, 16, 1]                                     &  1.85(-3)  & 147 & 25  \\
 \acetone                        & Acetone            &  342780.0345(0.0357)  & [17, 17, 1]$\rightarrow$[16, 16, 0]                                     &  1.85(-3)  & 147 & 25  \\
 \acetone                        & Acetone            &  345256.0368(0.0368)  & [18, 15, 4]$\rightarrow$[17, 14, 3]                                     &  1.287(-3) & 151 & 40  \\
 \acetone                        & Acetone            &  345305.3651(0.0367)  & [18, 15, 3]$\rightarrow$[17, 14, 4]                                     &  1.288(-3) & 151 & 45  \\
 \acetone                        & Acetone            &  356675.1969(0.0814)  & [19, 13, 6]$\rightarrow$[18, 12, 7]                                     &  6.123(-4) & 158 & 11  \\
 \acetone                        & Acetone            &  356658.4123(0.0187)  & [27, 10, 17]$\rightarrow$[26, 11, 16], Sym=EA                           &  1.215(-3) & 275 & 10  \\
 \acetone                        & Acetone            &  356658.4780(0.0187)  & [27, 11, 17]$\rightarrow$[26, 10, 16], Sym=EA                           &  1.215(-3) & 275 & 10  \\
 \acetone                        & Acetone            &  356658.5351(0.0184)  & [27, 10, 17]$\rightarrow$[26, 11, 16], Sym=AE                           &  1.215(-3) & 275 & 10  \\
 \acetone                        & Acetone            &  356658.6016(0.0184)  & [27, 11, 17]$\rightarrow$[26, 10, 16], Sym=AE                           &  1.215(-3) & 275 & 10  \\
 \acetone                        & Acetone            &  356729.9828(0.0153)  & [27, 10, 17]$\rightarrow$[26, 11, 16], Sym=EE                           &  1.215(-3) & 275 & 18  \\
 \hline
 \dimethylether                  & Dimethyl ether     &  356705.204(0.004)    & [8, 4, 4]$\rightarrow$[7, 3, 5], Sym=AE                                 &  9.249(-5) & 55  & 14  \\
 \dimethylether                  & Dimethyl ether     &  356712.951(0.002)    & [8, 4, 5]$\rightarrow$[7, 3, 5], Sym=EE                                 &  6.255(-5) & 55  & 15  \\
 \dimethylether                  & Dimethyl ether     &  356724.457(0.003)    & [8, 4, 4]$\rightarrow$[7, 3, 5], Sym=AA                                 &  2.095(-4) & 55  & 16  \\
 \dimethylether                  & Dimethyl ether     &  356724.864(0.003)    & [8, 4, 5]$\rightarrow$[7, 3, 5], Sym=AE                                 &  1.17(-4)  & 55  & 16  \\
 \hline
 \ethanol                        & gauche-Ethanol     &  345173.9493(0.0112)  & [\J, \Ka, \Kc, $v_t$]=[7, 7, 1(0), 0]$\rightarrow$[6, 6, 1(0), 1]\tablenotemark{c}   &  2.515(-4) & 140 & 33  \\
 \ethanol                        & gauche-Ethanol     &  345229.238(0.05)     & [\J, \Ka, \Kc, $v_t$]=[21, 1, 21, 0]$\rightarrow$[20, 1, 20, 0]         &  3.728(-4) & 242 & 36  \\
 \ethanol                        & gauche-Ethanol     &  345295.3553(0.0022)  & [\J, \Ka, \Kc, $v_t$]=[21, 1, 21, 1]$\rightarrow$[20, 1, 20, 1]         &  3.731(-4) & 246 & 43  \\
 \ethanol                        & gauche-Ethanol     &  345408.1651(0.0022)  & [\J, \Ka, \Kc, $v_t$]=[21, 0, 21, 1]$\rightarrow$[20, 0, 20, 1]         &  3.743(-4) & 246 & 49/50 \\
 \hline
 $\rm SO_{2}$                    & Sulfur dioxide     &  356755.1899(0.0014)  & [10, 4, 6]$\rightarrow$[10, 3, 7]                                       &  3.281(-4) & 90  & 19  \\
 $\rm H_{2}CS$                   & Thioformaldehyde   &  342946.4239(0.0500)  & [10, 0, 10]$\rightarrow$[9, 0, 9]                                       &  6.08(-4)  & 91  & 27  \\
 \hline
 HCN$~{\rm v}_{2}=1$             & Hydrogen cyanide   &  354460.4346(0.0007)  & \J=4$\rightarrow$3                                                      &  1.869(-3) & 1067 & 3  \\
 \hline
 \multicolumn{7}{c}{\textbf{Tentative identifications}} \\
 \hline
 \acetaldehyde$~{\rm v}_{t}=1$   & Acetaldehyde       &  354522.7114(0.1995)  & [6, 6, 1]$\rightarrow$[7, 5, 3]                                         &  3.184(-6) & 304 & 5   \\
 \acetaldehyde$~{\rm v}_{t}=1$   & Acetaldehyde       &  354457.6646(0.0409)  & [18, 2, 16]$\rightarrow$[17, 2, 15]                                     &  1.585(-3) & 375 & 3   \\
%  \hline
%  \methylamine                    & Methylamine        &  354528.4832(0.0057)  & [Sym, \J, \K]=[B$_2$, 8, 2]$\rightarrow$[B$_1$, 7, 2]                   &  2.156(-5) & 93   & 5  \\
%  \methylamine                    & Methylamine        &  354600.5152(29.979)  & [Sym, \J, \K]=[B$_2$, 8, 5]$\rightarrow$[B$_1$, 7, 5]                   &  1.399(-5) & 174  & 6  \\
%  \methylamine                    & Methylamine        &  354600.5152(29.979)  & [Sym, \J, \K]=[B$_1$, 8, 5]$\rightarrow$[B$_2$, 7, 5]                   &  1.399(-5) & 174  & 6  \\
%  \methylamine                    & Methylamine        &  354600.5152(29.979)  & [Sym, \J, \K]=[B$_1$, 8, 6]$\rightarrow$[B$_2$, 7, 6]                   &  1.016(-5) & 217  & 6  \\
%  \methylamine                    & Methylamine        &  354600.5152(29.979)  & [Sym, \J, \K]=[B$_2$, 8, 6]$\rightarrow$[B$_1$, 7, 6]                   &  1.016(-5) & 217  & 6  \\
 \hline
 CH$_{3}$C$^{15}$N               & Methyl cyanide     &  356761.0470(0.002)   & [\J, \K]=[20, 1]$\rightarrow$[19, 1]                                    &  3.962(-3) & 187  & 20 \\
 \hline
 \ethylcyanide                   & Ethyl cyanide      &  354476.6596(0.0015)  & [40, 3, 38]$\rightarrow$[39, 3, 37]                                     &  3.767(-3) & 361  & 4  \\
 \hline
 CH$_{3}$Cl                      & Methyl chloride    &  345408.9874(0.0023)  & [\J, \K, $F$]=[13, 3, 12.5]$\rightarrow$[12, 3, 11.5]                   &  7.764(-4) & 183 & 49/50  \\
\enddata
\tablenotetext{a}{The typical quantum numbers are listed as [\J, \Ka, \Kc] unless specified.}
\tablenotetext{b}{The number in the prentice indicates the power of ten (e.g. 8.041(-5) represents 8.041\ee{-5}).}
\tablenotetext{c}{\citet{1988ApJS...67..135A}.}
\tablenotetext{d}{Two degenerate transitions occur for \Kc=$1\rightarrow1$ and $0\rightarrow0$.}

\end{deluxetable*}

Some species have multiple lines detected in our observations, such as methyl formate (both ground state and excited state), dimethyl ether, acetone, methanol, and ethanol, allowing us to further constrain their excitation temperatures and column densities.  With \textsc{xclass}, LTE radiative transfer calculations produce synethetic spectra, whcih is optimized to the observations.  We further exclude possible optically thick emission from the optimization, based on the line profile, such as significant deviations from Gaussianity and absorption.  The properties of methyl formate (both ground state and excited state) and dimethyl ether are fitted simultaneously due to blending between their spectra.  Other species, such as acetone, deuterated methanol (\dmethanol), and ethanol, are fitted separately.  Table\,\ref{tbl:coms_abundance} lists the fitted excitation temperatures and column densities for selected species.  For methanol and its isotopologs, only \dmethanol\ has sufficient detections to determine its properties; thus, the fitting of column densities of \methanol\ and \tmethanol\ assumes the excitation temperature of \dmethanol.  Given the distribution of the fitted temperatures, we take 100\,K as the typical temperature to fit the column densities for other identified species, also shown in Table\,\ref{tbl:coms_abundance}.  

Figure\,\ref{fig:1d_spectra_line_id} shows the best-fitting spectra of the identified COMs along with the spectra of the tentatively identified COMs, which only serves as a consistency test while our observation cannot confirm the identifications of those species.  The lines excluded from the optimization due to suspected high optical depth have significant deviation from the modeled spectra, such as Line 7 and 54.  While methylamine is NOT identified, its spectra show some agreement with the observations (see the discussion in Appendix\,\ref{sec:coms_tentative}), which requires future line surveys to confirm.
Our identification of complex organic molecules and their compact morphology at $<1\arcsec$ region indicate that BHR\,71 has a hot corino.  Besides the COMs, ALMA ACA observations detect the emission of carbon-chain molecules (\L{}. Tychoniec in prep., priv. comm.), suggesting that BHR\,71 may have different degrees of complex chemistry at different spatial scales.

\begin{deluxetable*}{rlcccc}
    \tabletypesize{\scriptsize}
    \tablecaption{The abundance of identified COMs \label{tbl:coms_abundance}}
    \tablewidth{0pt}
    \tablehead{\colhead{Species} & \colhead{Formula} & \colhead{No. of identified lines} & \colhead{$T_\text{ex}$ [K]} & \colhead{Column density (cm$^{-2}$)} & \colhead{Abundance to $N_{\rm H_{2}}$} }
    \startdata
    \multicolumn{6}{c}{Species with fitted temperatures and column densities} \\
    \hline
    Methanol                    & CH$_{2}$DOH $v_{t}=0$                    & 3 & 100                   & 6.9\ee{16}  &   4.1\ee{-7}  \\
    Methanol                    & $^{13}$CH$_{3}$OH $v_{t}=0$              & 2 & 100\tablenotemark{a}  & 1.3\ee{16}  &   7.6\ee{-8}  \\
    Methanol                    & CH$_{3}$OH $v_{t}=0$                     & 1 & 100\tablenotemark{a}  & 2.4\ee{18}  &   1.4\ee{-5}  \\
    Methyl formate              & CH$_{3}$OCHO $v=0$                       & 4 & 100                   & 4.4\ee{16}  &   2.6\ee{-7}  \\
    Methyl formate ($v_{t}=1$)  & CH$_{3}$OCHO $v_{t}=1$                   & 7 & 150                   & 3.7\ee{16}  &   2.2\ee{-7}  \\
    Acetone                     & CH$_{3}$COCH$_{3}$ $v=0$                 & 5 & 130                   & 9.7\ee{15}  &   5.7\ee{-8}  \\
    Dimethyl ether              & CH$_{3}$OCH$_{3}$ $v=0$                  & 3 & 60                    & 1.0\ee{16}  &   5.9\ee{-8}  \\
    gauche-Ethanol              & C$_{2}$H$_{5}$OH $v=0$                   & 5 & 260                   & 1.3\ee{16}  &   7.6\ee{-8}  \\
    \hline
    \multicolumn{6}{c}{Species with fitted column densities assuming $T_\text{ex}=100$\,K} \\
    \hline
    Sulfur dioxide              & SO$_{2}$ $v=0$                           & 1 & 100\tablenotemark{b}  & 2.7\ee{15}  &   1.6\ee{-8}  \\
    Thioformaldehyde            & H$_{2}$CS $v=0$                          & 1 & 100\tablenotemark{b}  & 9.9\ee{14}  &   5.8\ee{-9}  \\
    Hydrogen cyanide            & HCN $v_{2}=1$                            & 1 & 100\tablenotemark{b}  & 4.0\ee{17}  &   2.4\ee{-6}  \\
   \enddata
   \tablecomments{These column density are fitted with an excitation temperature of 100~K. The gas column density within a 0.5\arcsec\ radius beam is 1.7\ee{23}~cm$^{-2}$, calculated from the best-ftting envelope model.}
   \tablenotetext{a}{The excitation temperatures of methanol and its isotopologues are set to the fitted excitation temperature of \dmethanol.}
   \tablenotetext{b}{The excitation temperature is assumed \replaced{to be}{as} 100\,K, the typical temperature from the fitted species.}

   \end{deluxetable*}

\section{Addressing the Shortcomings of the Infall Model}
\subsection{An Observational Test of the Best-fitting Infall Kinematics}

\begin{figure*}[htbp!]
	\centering
	\includegraphics[width=0.7\textwidth]{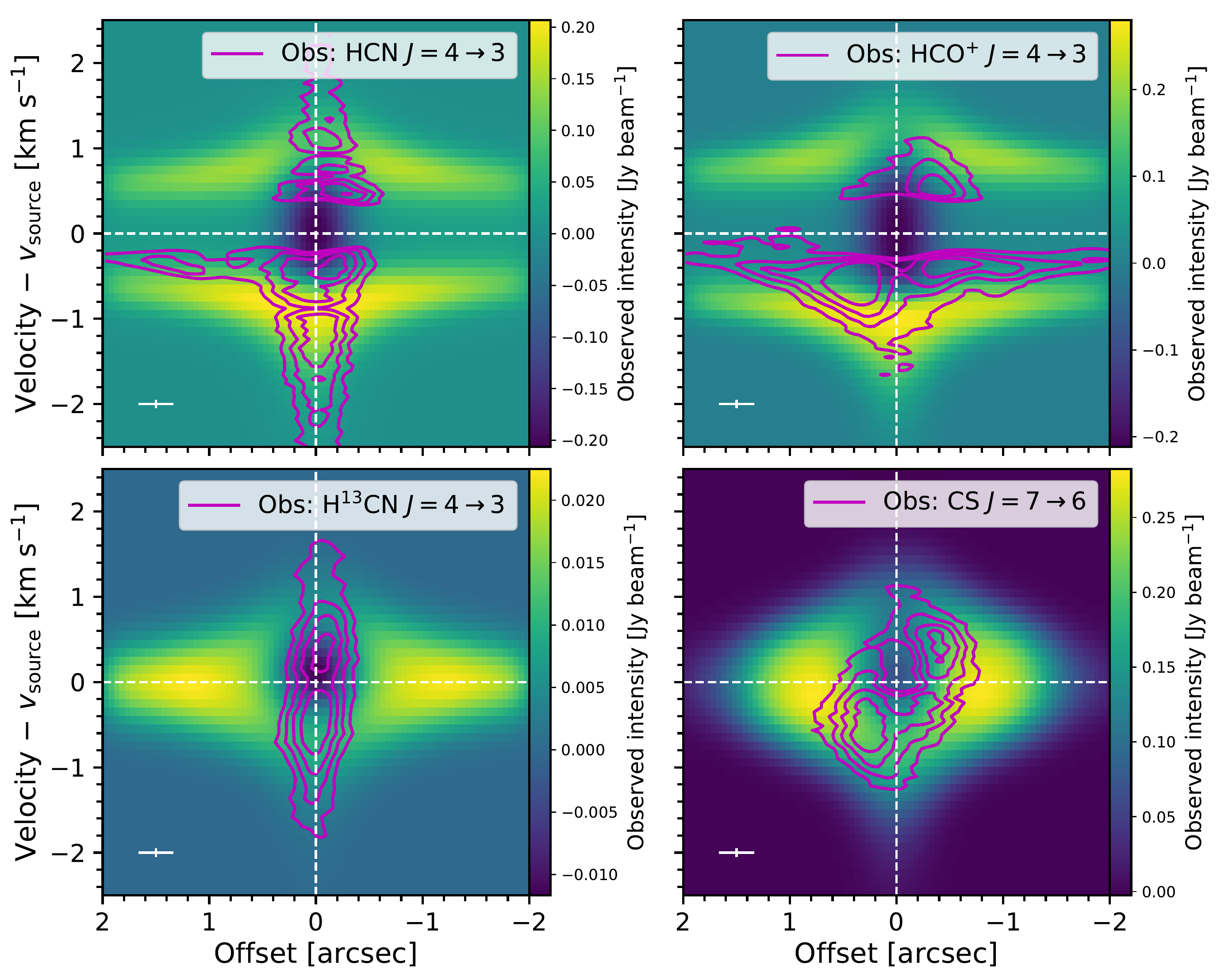}
	\caption{The PV diagrams of the best-fitting envelope models (images) compared with the observed PV diagrams (magenta contours), extracted with a slice across the midplane of the envelope (east-west).  The contour levels increase linearly from 0.05 to 0.26, 0.05 to 0.55, 0.02 to 0.17, and 0.05 to 0.28 for the HCN \jj{4}{3}, \hcop\ \jj{4}{3}, \htcn\ \jj{4}{3}, and CS \jj{7}{6} lines, respectively. The synthetic PV diagrams are convolved with the spectral and spatial resolutions of the observations, which are indicated as white errorbars.}
	\label{fig:infall_pv}
\end{figure*}

While the infall signature toward the continuum source provides the least contaminated signature to constrain the underlying infall kinematics, the variations of infall signatures across the envelope will further assess the effects of rotation and abundance profile.  Figure\,\ref{fig:infall_pv} shows the PV diagrams of the observations and the best-fitting models.  Beside the \htcn\,\jj{4}{3} line, the modeled morphology at the zero offset agrees with the observations relatively well compared with the offset positions since the models are constrained at the central position.  Overall, our models have emission more spatially extended than the observations, suggesting a denser inner region or a lower density at the outer region than the model predicts.  Since the envelope model has been constrained with the far-infrared SED, which mainly probes the outer envelope, a denser inner region is a more plausible scenario.  Under the TSC framework, a higher density at the inner region requires a younger age or a higher effective sound speed, resulting in a more compact density peak, or a higher rotation speed at the inner region, leading to a larger centrifugal radius.  However, the uncertainty in the chemical abundance should be characterized before any quantitative comparison.

\begin{figure*}[htbp!]
	\includegraphics[width=\textwidth]{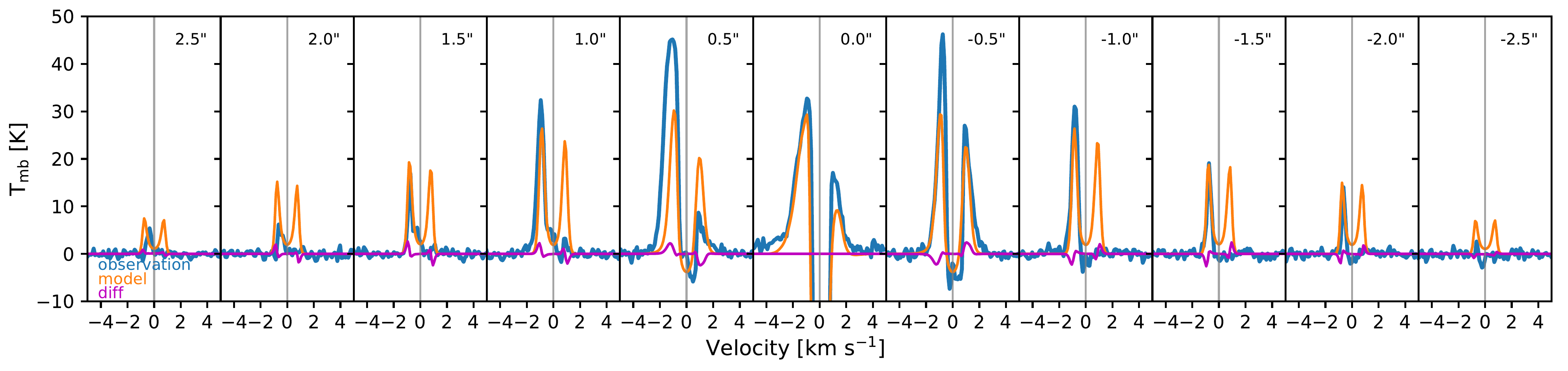}
	\includegraphics[width=\textwidth]
	{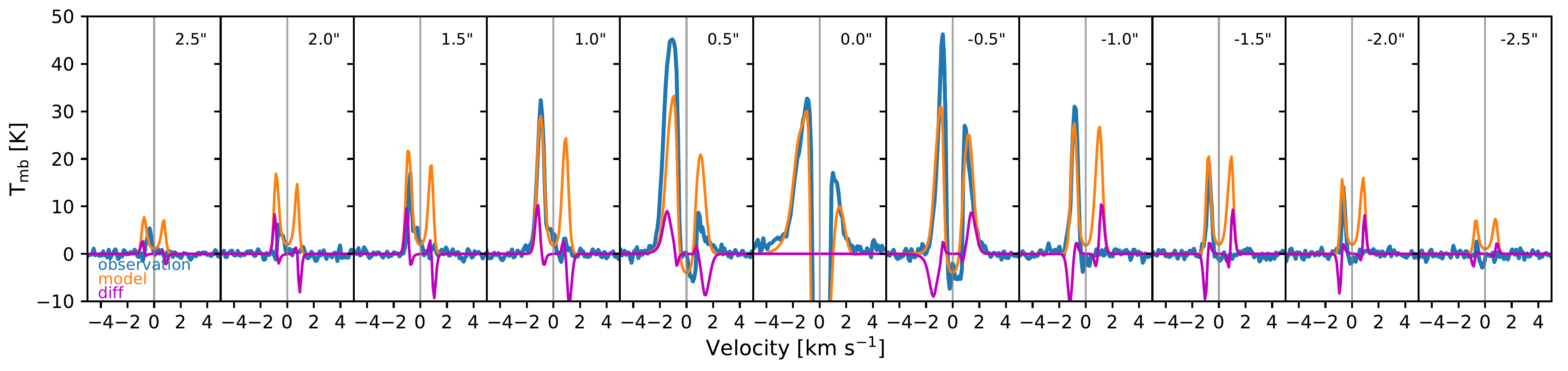}
	\caption{Two grids of the \hcop\,\jj{4}{3} line profiles extracted with 0\farcs5$\times$0\farcs5 box regions along the midplane of the envelope (east-west).  \added{The top panel show the best-fitting model, while the bottom panel show the same model with a higher $\Omega$ of 10$^{-12}$\,rad\,s$^{-1}$, whose line profiles peak at slightly higher velocity.}  Those regions offset from the center ranging from 2\farcs5 to $-$2\farcs5 in RA.  The observation is shown in blue, while the \replaced{model is}{models are} shown in orange.  The magenta lines show the difference in the synthetic spectra to the opposite side of the offset (i.e., the line for 0\farcs{5} is the difference of profiles toward 0\farcs{5} and $-$0\farcs{5}).  }
	\label{fig:hcop_grid}
\end{figure*}

The variation of the infall signature across the midplane of the envelope presents another way to investigate the infall kinematics.  Here we focus on the \hcop\,\jj{4}{3} line best fitted by our model.  The blue asymmetric double-peaked profile only appears within the inner 0\farcs{5} region, whereas only a single blue peak exists beyond the inner 0\farcs{5} region (Figure\,\ref{fig:hcop_grid}\added{, solid lines}).  In the observations, the strength of the red peak increases from east to west, which is expected due to the rotation; however, the observations show a much greater difference than our model\deleted{, suggesting a greater rotation at the inner region}.  The widths of the two peaks in the infall signature also change from east to west, especially the blue peak, which becomes broader at the 0\farcs{5} offset and narrower at the $-$0\farcs{5} offset.  
Except for pointings toward the center and $-$0\farcs{5}, the model disagrees with the observations.  The synthetic profile shows a higher intensity when the pointing is displaced by 0\farcs{5} from the center due to the reduction of the optical depth.  As we move further away from the center along the midplane, the decrease of molecular abundance starts to dominate the behavior of the synthetic profile, resulting in a gradual decrease in the overall intensity.

The comparison between our models and the observations highlights two issues.  First, our TSC envelope model underestimates the variation of the red-shifted peak across the inner 0\farcs{5} region due to the rotation.  As discused in previous paragraphs, a larger centrifugal radius may reduce the disgreement in the PV diagrams between the observations and the models (Figure\,\ref{fig:infall_pv}), suggesting a higher rotation at the inner region.  In fact, a TSC envelope with twice the best-fitting rotation speed, 5\ee{-13} rad\,s$^{-1}$, leads to a greater asymmetry for the \hcop\ line profiles at one side of the midplane to another; for the \hcop\ line profile at the 0\farcs{5} offset, the peak of the difference (magenta line in Figure\,\ref{fig:hcop_grid}) becomes $\sim$10\,K.  The observation shows a difference of $\sim$20\,K, suggesting a much faster rotation at the inner 0\farcs{5} region.  However, the linewidth variation as a function of the offset cannot be obviously explained by a higher rotation.

Secondly, the \hcop\ model fails to explain the disappearance of the red-shifted emission beyond the inner 0\farcs{5} region.  Moreover, the synthetic infall signatures overestimate the intensity at off-center positions for all four lines, suggesting more complex abundance profiles.  The temperature gradient in the envelope governs the degree of asymmetry between the blue-shifted and red-shifted peaks, and also produces the red-shifted emission as long as the warm gas in the foreground of the infalling envelope falls within the LOS.  If we increase the depletion in the freeze-out zone of the abundance profile to fit the red-shifted peaks, both the blue-shifted and red-shifted peaks would have the same sudden disappearance beyond the inner 0\farcs{5} region, which is not observed.  The lack of red-shifted emission may also explain the significant blue-asymmetry in the \hcop\,\jj{3}{2} line observed by APEX (Figure\,\ref{fig:single_hco+3-2}), while our model predicts an equal strength double-peaked profile.

\added{A model with a higher rotation speed of 10$^{-12}$\,rad\,s$^{-1}$ results in a moderately brighter red peak at the $-$0\farcs{5} position, whereas the brightness of the red peak remains the same at the 0\farcs{5} position (Figure\,\ref{fig:hcop_grid})n.  The blue peak shows a similar behavior with a moderately brightness  at 0\farcs{5} position but the same brightness at $-$0\farcs{5} position.  Thus, the behavior of the infall signatures within the inner 0\farcs{5} region is unlikely due to a higher rotation speed of the TSC model.  A single Gaussian line shifted toward blue and red at the 0\farcs{5} and $-$0\farcs{5} positions, respectively, with the self-absorption at the source velocity could be consistent with the observations.  Such model would have a greater velocity shift than that of in the TSC model, suggesting a higher rotation.  Perhaps the higher rotation only occurs at the inner part of the envelope.}

\deleted{A model with a higher rotation speed would result in a greater difference in the strengths of the two peaks in the line profiles across the midplane of the envelope, matching the observations (Figure\,\ref{fig:hcop_grid}).  However, a high rotation model would result in a weaker red-shifted peak toward the central position, whereas our best-fitting model (with a relatively low rotation) has already underestimated the red-shifted peak.  A high rotation may only be needed at the inner part of the envelope, which is beyond the framework of the TSC model.}

\added{Figure\,\ref{fig:tsc_breakdown} shows that the TSC model underestimates the rotation speed within the centrifugal radius.  However, the observations show a higher rotation out to $\sim$100\,au, which cannot be reproduced by the envelope model.  A model with an older age would have a higher rotation as well as a larger centrifugal radius, indicating a larger disk as our analyses hint.  In contrast, the molecular line profiles require an age as young as 12\,000 years (Section\,\ref{sec:age}).  Resolving this issue requires a model that consistently describes the kinematics of both the envelope and the disk as well as the observations targeting the embedded disk, such as the observations of C18O.}

\subsection{A Compact Rotational Structure Traced by COMs}
\label{sec:coms_kinematics}
\begin{figure*}[htbp!]
	\centering
	\includegraphics[width=\textwidth]{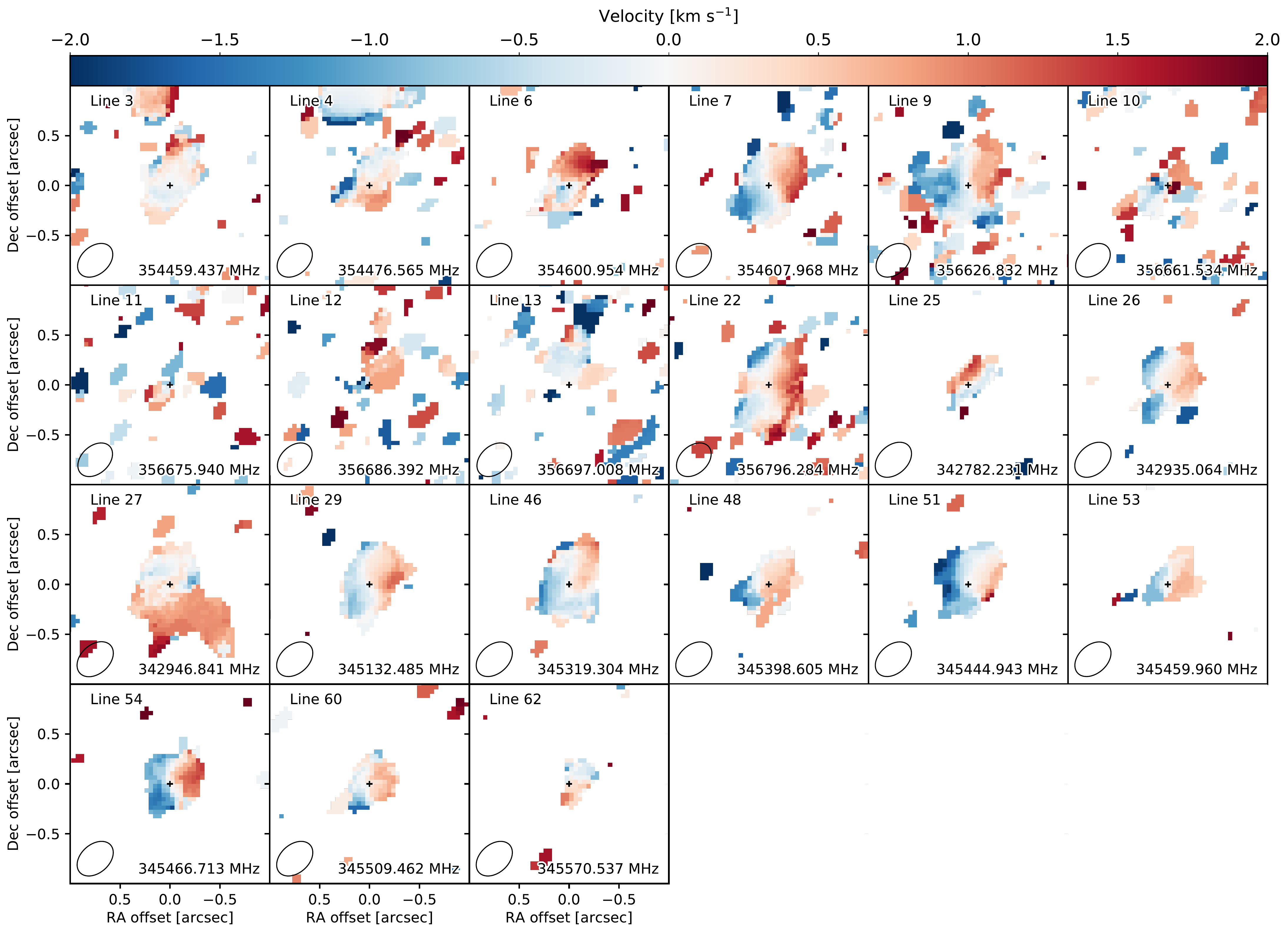}
	\caption{The moment 1 maps of the unblended COM emission, calculated with a 4\,\kms\ window centered at the source velocity.}
	\label{fig:coms_moment1}
\end{figure*}

\begin{figure}[htbp!]
	\centering
	\includegraphics[width=0.47\textwidth]{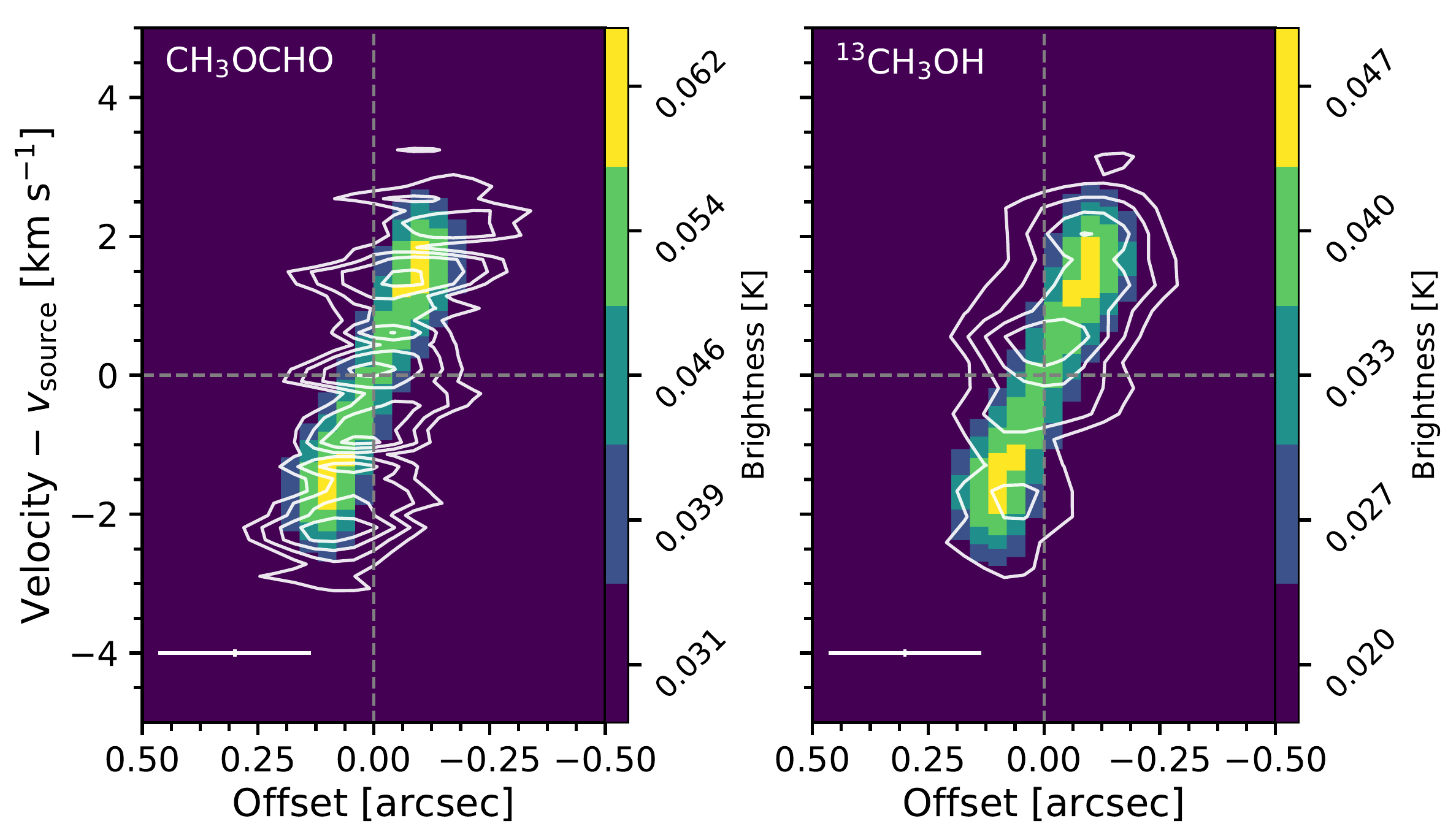}
	\includegraphics[width=0.47\textwidth]{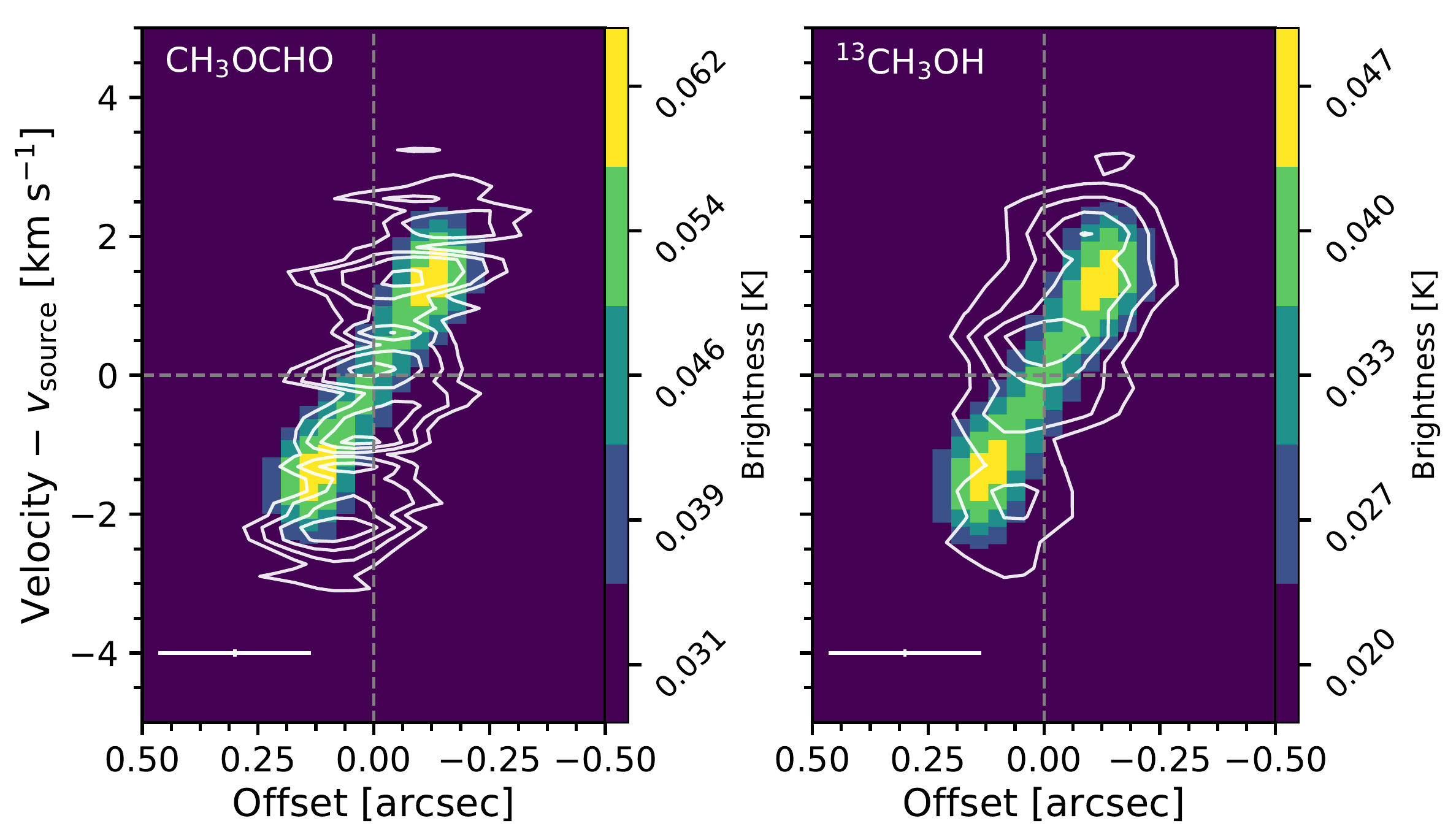}
	\caption{The PV diagrams of the emission of $^{13}$C-methanol and methyl formate (white contours) along with the synthetic PV diagrams of Keplerian disks (colored images) with a power-law radial brightness profile (top) and a uniform brightness (bottom).  The PV slice is perpendicular to the outflows.  \deleted{The orange dahsed lines indicate the maximum Keplerian rotation velocity as a function of the offset.} \added{The colorbar shows the brightness of the modeled images, which are normalized to the maximum brightness of the observed PV diagrams.  The ticks of the colorbar indicate the contour values of the observed PV diagrams.}}
	\label{fig:coms_pv}
\end{figure}

Figure\,\ref{fig:coms_moment1} shows the moment 1 maps of a selection of unblended COM emission, showing a velocity gradient in the west-east direction from blue-shifted to red-shifted, consistent with the velocity gradient of seen in the emission of C$^{18}$O \citep{2019ApJ...870...81T}.  Two of the COM emission lines, \tmethanol\ at 354608.0\,MHz and \methylformate\ at 345132.5\,MHz, show the velocity gradient with high S/N, allowing further analyses with PV diagrams (Figure\,\ref{fig:coms_pv}).

The emission of COMs better traces the kinematics at the inner region as COMs only become abundant in the gas phase at $T>100$\,K; therefore, the emission only comes from the inner envelope, where a disk may be forming.  To investigate the kinematics probed by COMs, we construct the PV diagrams along a slice perpendicular to the outflow direction, north to south \citep{1997ApJ...476..781B,2006A&A...453..949P}, minimizing the contamination from the outflows.  Assuming that the emission of COMs is optically thin due to their low column density, the PV diagrams of \tmethanol\ and \methylformate\ allow us to qualitatively constrain the rotational kinematics.

The PV diagrmas are first tested against a simple model of a Keplerian rotating disk.  In an ideal case, a Keplerian disk would result in blue-shifted velocities only at one side of the disk and red-shifted velocities only at the other side; however, with a finite spatial resolution, the emission becomes smeared out at the low velocities; therefore, low velocities emission may appear at either side of the source.  The emission of a Keplerian disk at any east-west offset from the center, $b$, can be written as
\begin{align}
	T_\text{B}(b,v)                      & = \int^{\theta_\text{max}}_{\theta_\text{min}} T_\text{B}\left(\frac{b}{\text{cos}\,\theta}\right)\,e^{-\left(\frac{v-v_\text{los}}{\Delta v}\right)^{2}}\,b\,\text{sec}^{2}\theta\,\text{d}\theta; \nonumber \\
	v_\text{los}                         & = \sqrt{GM\text{cos}^{3}\theta/b}\,\text{cos}\,\theta_\text{incl}; \nonumber \\
	\theta_\text{max}, \theta_\text{min} & = \pm \text{cos}^{-1}(b/R),         
	\label{eq:disk}
\end{align}
where $\theta$ is the angle between the LOS toward the source and the LOS according to $b$ with respect to the source itself, $R$ is the outer radius of the disk, and $\theta_\text{incl}$ is the inclination angle of the disk, 130$^{\circ}$.  The central mass, $M$, is set to the total infall mass inferred from our envelope model, 0.15\,\msun, and the line width, $\Delta v$, is assumed to be 1\,\kms.  To match the spatial and spectral resolutions of the observations, a 2D Gaussian kernel is applied to the emission of the Keplerian disk.  

The maximum velocity in the PV diagram determines the inner radius of the disk, 15\,au, while the maximum spatial extent constrains the outer radius of the disk, 50\,au.  Then, we test two models of the radial brightness temperature profile, a constant brightness temperature throughout the disk, and a power-law profile, which represents the decrease of density toward outer radius.  The disk is likely to be optically thick; however, to qualitatively test the Keplerian disk against our observations we assume the power-law profile has included the effect of optical depth.  A more realistic model would have two power-law profiles jointed at a certain radius, representing the transition from optically thick to optically thin.  But such model has more parameters, requiring a better spatial resolution than that of our observations.  For a qualitative comparison, the power-law profile is set to $r^{-2}$, while \citet{2011ApJ...743..112S} found a power-law slope of $-1$--$-3$ using the two power-law model.  Figure\,\ref{fig:coms_pv} shows the synthetic PV diagrams along with the observations, calculated with Equation\,\ref{eq:disk} independent from the modeling of the infalling envelope.  The two representative models show similar PV diagrams as the observations, suggesting that a Keplerian disk may contribute to the rotation signatures seen in the emission of COMs.  Between the two brightness profiles, the power-law profile leads to the emission peaked at a lower velocity, better reproducing the observations.  Although this comparison suggests a Keplerian would reproduce the PV diagrams of COMs, the minor difference between the PV diagrams of the two profiles highlights the need of high resolution observations to characterize the nature of this rotation signature.

\subsection{The Origin of the High Velocity Emission in the HCN and H$^\textit{13}$CN lines: Disks or Outflows?}
\label{sec:unresolved_disk}
While the infall model reproduces most of the ALMA observations, it fails to match the high velocity wings in the HCN \jj{4}{3} line and greatly underestimates the \htcn\ \jj{4}{3} line.  The compact morphology of the \htcn\ \jj{4}{3} line and the high apparent \htcn/HCN ratio suggest a compact optically thick structure at the center of BHR\,71, which could be an unresolved disk motivated by the rotation observed in the emission of COMs (Section\,\ref{sec:coms_kinematics}).  A compact disk may produce the high velocity emission observed in the HCN \jj{4}{3} and \htcn\ \jj{4}{3} lines; however, outflows may contribute to the high velocity emission as well.  The morphology of outflows (Figure\,\ref{fig:moment0}) and the moderate inclination angle indicate that the line-of-sight toward the central continuum source includes the outflows.  To investigate the possible origin for the high velocity emission, we first test the possibility of an unresolved disk; then we discuss the cases for a disk and outflows.

\begin{figure*}[htbp!]
	\includegraphics[width=\textwidth]{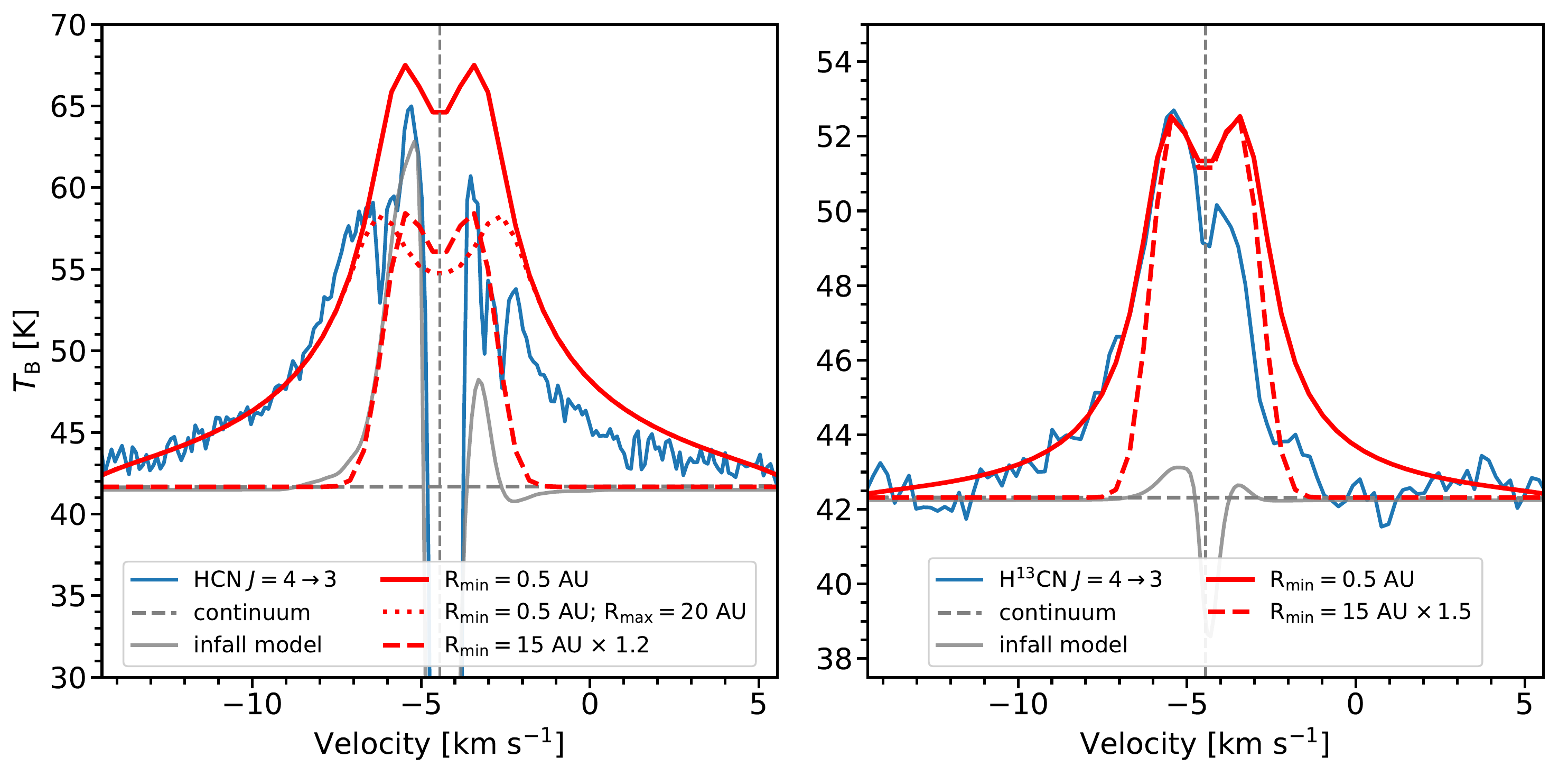}
	\caption{The spectra of the HCN\,\jj{4}{3} and \htcn\,\jj{4}{3} lines (blue) along with the synthetic spectra of the infalling envelope (gray) and the Keplerian disk models, including a disk from 15 to 50\,au (red dashed), a disk from 0.5 to 50\,au (red solid), and a disk from 0.5 to 20\,au (red dotted).  The models with a disk from 15 to 50\,au are scaled from the profiles listed in Equation\,\ref{eq:brightness_profiles} by 1.2 and 1.5 for the HCN and \htcn\ lines, respectively.}
	\label{fig:disk}
\end{figure*}

According to Equation\,\ref{eq:disk}, the emission of a Keplerian disk shows a symmetric double-peaked profile (Figure\,\ref{fig:disk}).  The disk emission is calculated separated from the infalling envelope, resulting in the lack of absorption at the source velocity.  We start from a power-law radial brightness profile with the disk suggested from the COMs emission ($r_\text{min}=15$\,au and $r_\text{max}=$50\,au; Section\,\ref{sec:coms_kinematics}), and vary the slope of the power-law profile to reproduce the observations.  The optimization is done by eyes.  With a disk from 15\,au to 50\,au, the model only produces emission up to $\pm$3\,\kms\ regardless of the power-law profile, whereas the observed profiles extend to velocities up to $\pm$10\,\kms.  Matching the wings of the HCN line requires an inner radius of 0.5\,au.  While a disk extending from 0.5 to 50\,au reproduces the high velocity wings for the HCN and \htcn\ lines reasonably well, the model overestimates the low velocity emission of the HCN line.  A smaller outer radius of 20\,au would reduce the low velocity emission, producing a better fit to the HCN line.

While Keplerian disks produce the line profiles consistent with the observations, the underlying best-matched radial brightness temperature profiles are rather unphysical.  The best-matched brightness profiles can be written as
\begin{align}
	T_\text{B}(r) & = 3.6 \times 10^{4} \left(\frac{r}{\text{au}}\right)^{-2} \text{ [K] for HCN, and } \nonumber \\
	T_\text{B}(r) & = 6.3 \times 10^{3} \left(\frac{r}{\text{au}}\right)^{-1.7} \text{ [K] for H}^{13}\text{CN.}  
	\label{eq:brightness_profiles}
\end{align}
Both profiles have unphyiscally high temperatures at 0.5\,au, 10$^{3}$--10$^{4}$\,K, and low tempeartures at 50\,au, 8--14\,K, suggesting that a Keplerian disk may not be the origin of the high velocity emission.

Outflows also produce high velocity emission.  At an inclination angle of 130$^{\circ}$, the line-of-sight toward the center is likely to includes some outflowing gas.  However, confirming the contribution from outflows requires future modelings that combine the kinematics of both the infalling envelope and outflows.

\section{Peculiar Kinematics of the HCN compact features}
\label{sec:blobs}
\begin{figure}[htbp!]
	\centering
	\includegraphics[width=0.48\textwidth]{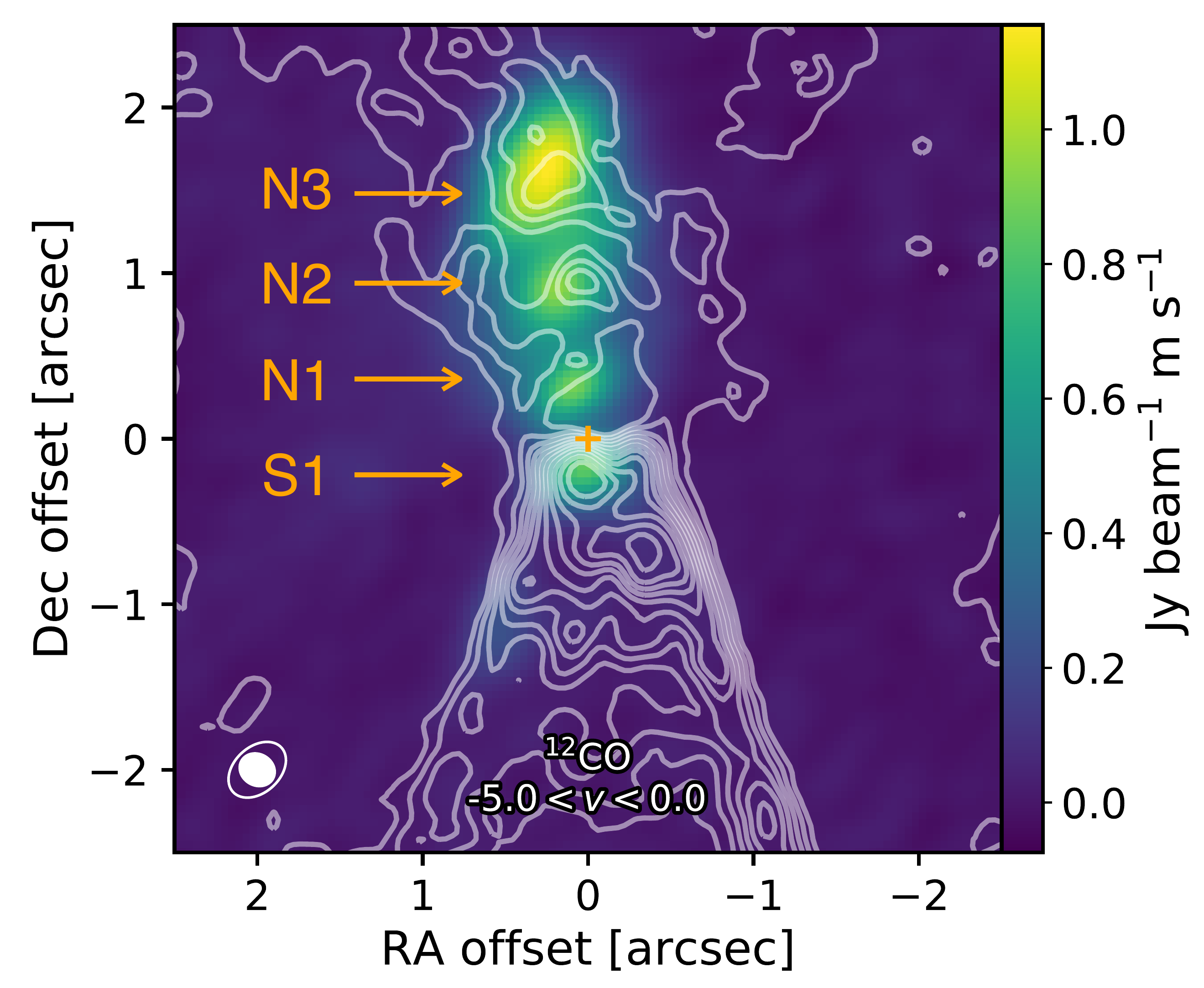}
	\caption{The moment 0 map of the HCN emission shown as an image with the moment 0 of the CO\,\jj{2}{1} emission plotted as contours (M. Dunham, priv. comm.).  Both moments are calculated from -5 to 0\,\kms.  Shown at the bottom left corner, the solid ellipse indicates the beam size of the CO observation, while the the greater ellipse outline indicates the beam size of the HCN observation.}
	\label{fig:hcn_co_hires}
\end{figure}

The channel maps of the HCN \jj{4}{3} emission reveal four compact features, labelled as N3, N2, N1, and S1 from north to south, peaking at 0.5--1.5\,\kms\ relative to the source velocity, located along the outflow direction.  Table\,\ref{tbl:hcn_blobs} lists the sizes of these features fitted with a 2D Gaussian profile.  Interestingly, if we focus on $\pm$2\,\kms\ around the source velocity, the three northern features peak at the blue-shifted velocity, while the southern feature peaks at the red-shifted velocity, an opposite kinematics to that of the outflow probed by CO \citep{1997ApJ...476..781B,2006A&A...453..949P}.  At the velocities greater than 4\,\kms, the HCN emission only appears in the south at the blue-shifted velocity and in the north at the red-shifted velocity, following the kinematics of the outflows.
N2 and N3 also appear in the CO observation with a similar resolution as that of HCN (priv. comm. with M. Dunham) at the same velocity range (Figure\,\ref{fig:hcn_co_hires}).  At low velocity, the PV diagram along the outflow direction shows triangular shape at these features, most significantly at N2 and N3.

\begin{table*}
	\centering
	\caption{Source fitting of the HCN features identified in Figure\,\ref{fig:hcn_co_hires}}
	\begin{tabular}{ccccc}
		\toprule
		Name & R.A.              & Dec.                  & FWHMs [mas]              & PA                      \\
		\midrule
		N3   & 12\h01\m36\fs{54} & $-$65\dd08\m47\fs{81} & 82$\pm$3$\times$59$\pm$2 & 154\fdg{8}$\pm$4\fdg{2} \\
		N2   & 12\h01\m36\fs{52} & $-$65\dd08\m48\fs{58} & 66$\pm$1$\times$50$\pm$1 & 163\fdg{2}$\pm$2\fdg{2} \\
		N1   & 12\h01\m36\fs{52} & $-$65\dd08\m49\fs{05} & 57$\pm$2$\times$49$\pm$2 & 145\fdg{5}$\pm$9\fdg{1} \\
		S1   & 12\h01\m36\fs{50} & $-$65\dd08\m49\fs{59} & 52$\pm$1$\times$39$\pm$1 & 97\fdg{8}$\pm$1\fdg{9}  \\
		\bottomrule
	\end{tabular}
	\label{tbl:hcn_blobs}
\end{table*}

\begin{figure*}[htbp!]
	\centering
	\includegraphics[width=\textwidth]{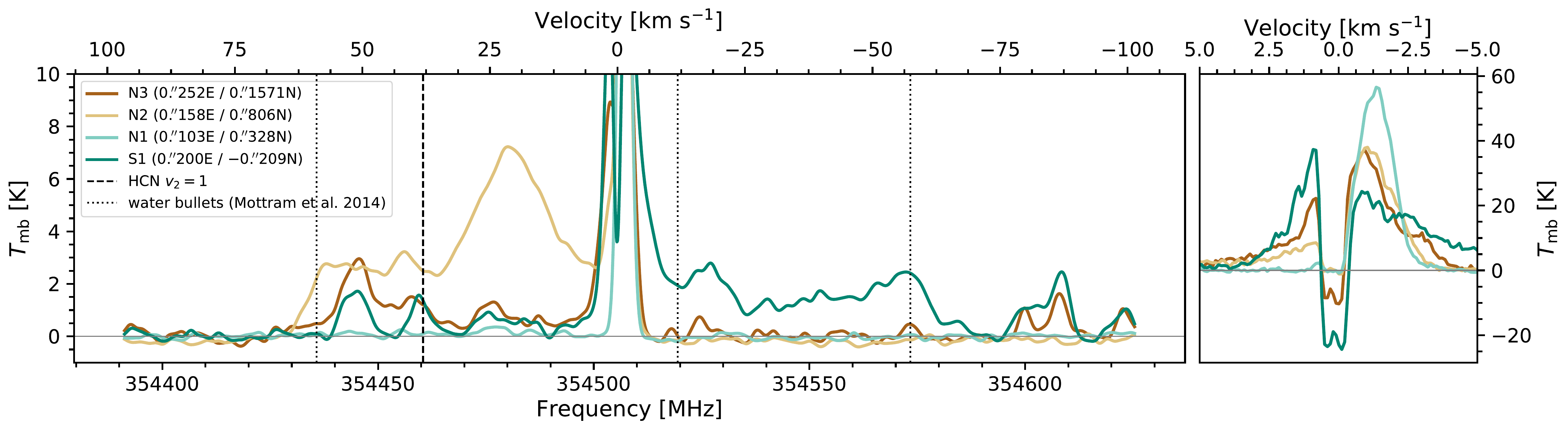}
	\caption{\textbf{Left:} The full spectra of the four HCN features extracted from the fitted 2D Gaussian sources (Table\,\ref{tbl:hcn_blobs}).  The velocities of the water bullets measured by \citet{2014A&A...572A..21M} are shown in dotted lines, while the rest frequency of the HCN $v=2$ \jj{4}{3} line is shown in dashed line.  \textbf{Right:} The spectra of the same four HCN features shown only around the line centroid of the HCN\,\jj{4}{3} line.}
	\label{fig:hcn_blobs}
\end{figure*}

\begin{figure}[htbp!]
	\centering
	\includegraphics[width=0.48\textwidth]{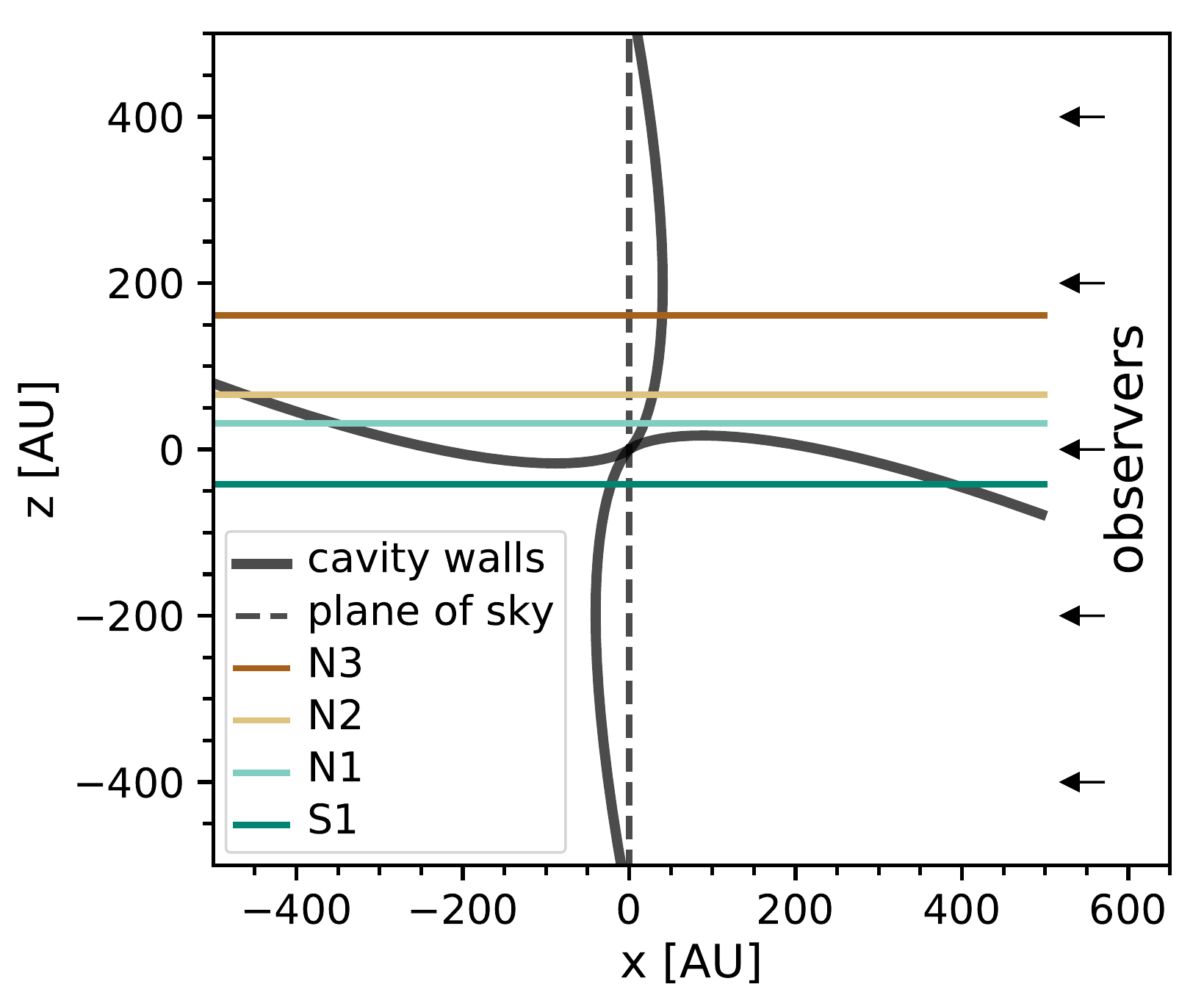}
	\caption{An illustration of outflow cavity (black thick solid line) and the plane of sky (black dashed line) along with the line of sights of the HCN features (colored lines), assuming an outflow opening angle of 22.5$^{\circ}$ \citep{2019ApJ...870...81T} and an inclination angle of 130$^{\circ}$ \citep{2017ApJ...835..259Y}.  The opposite kinematics would occur where the cavity walls appear at the opposite side to the majority of the outflow cavities with respect to the plane of sky (i.e., the northern red-shifted outflow has parts of its cavity wall in front of the plane of sky).}
	\label{fig:blobs_outflow}
\end{figure}

We propose two scenarios for those features, which have kinematics opposite to the outflows and small ($\sim$1\,\kms) velocity shifts.  The HCN features may be part of the outflows.  The emission of the red-shifted outflow can appear as blue-shifted if the outflow opening angle is large enough to have parts of the outflow at the front side of the envelope.  With an outflow opening angle of 22.5$^{\circ}$ measured at 10\arcsec\ from the protostar using CO observations \citep{2019ApJ...870...81T} and an inclination angle of 130$^{\circ}$ \citep{2017ApJ...835..259Y} from the Y17 model, the projected positions of the HCN features are consistent with the region where the outflow extends in front of the plane of sky in high inclination, supporting this scenario (Figure\,\ref{fig:blobs_outflow}).  The outflow opening angle is defined as the angle between the rotation axis to the edge of the outflow cavity at 10\,000\,au \citep{2017ApJ...835..259Y} and the outflow cavity is assumed as $z \propto \varpi^{1.5}$.  However, if shocks lead to the excitation of those HCN features, we expect to observe a broad spectral feature from the shocked gas that is blue-shifted for N1, N2, and N3 and red-shifted for S1.  Only N2 and S1 have broad features in their spectra, but the features appear at the opposite velocity for what we expect.

The blue-shifted infalling gas behind the outflow cavity can also result in the small blue-shifted emission in the red-shifted outflow.  The best-fitting model of the infall profile has similar velocities as the velocity offset of those HCN features; however, the model also predicts more red-shifted emission, which disagrees with the observation, and the model does not include the kinematics of the outflows.  In this case, we expect to see a broad spectral feature at the red-shifted velocity for N1, N2, and N3, which is consistent with the spectra of N2.  Future high resolution observations for probing the excitation mechanism of these HCN features, such as SiO tracing the shocked gas, may reveal the nature of the peculiar kinematics.

\begin{figure*}
	\includegraphics[width=\textwidth]{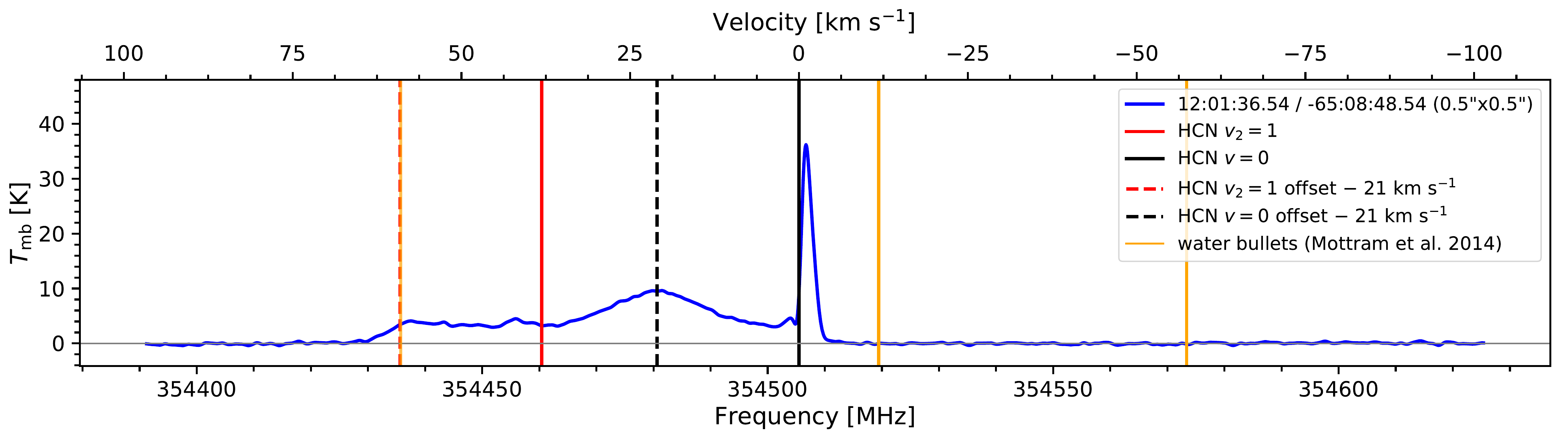}
	\caption{The spectrum of the high velocity compact feature in the spectral cube centered on the HCN\,\jj{4}{3} line extracted from a 0\farcs{5}$\times$0\farcs{5} region.  This feature is identified at $\sim$354480\,MHz.  The black solid and dashed lines indicate the rest frequencies of the HCN\,\jj{4}{3} $v=0$ and $v_{2}=1$ lines, respectively, while the red solid and dashed lines show the same two transitions shifted by $-$21\,\kms\ to match the broad peak at 21\,\kms.  The orange lines indicate the velocities of the water bullets identified by \citet{2012A&A...542A...8K} and \citet{2014A&A...572A..21M}.}
	\label{fig:offset_high_v}
\end{figure*}

A compact feature also appears at high velocity.  An offset compact emission peaks at $\sim$354480\,MHz ($\sim$22\,\kms), if the source velocity of $-$4.45\,\kms is assumed, in the spectrum centered on the HCN\,\jj{4}{3} line, extending over a wide range of frequencies (Figure\,\ref{fig:offset_high_v}).  Measured by the CASA \texttt{imfit} task, this compact high velocity emission peaks at 12\h01\m36\fs{53} and $-$65\dd08\m48\fs{42} with a size of 0\farcs{54}$\times$0\farcs{47}.  Only seen offset from the center, the broad line profile is unlikely to be emitted by COMs.  Shock-excited HCN\,\jj{4}{3} $v=0$ and $v_{2}=1$ lines may contribute to the broad features extending toward high velocities.  A velocity offset of 22\,\kms\ to the HCN\,\jj{4}{3} $v=0$ and $v_{2}=1$ lines would agree with the observations; and shocks would broaden the linewidth.  Comparing to the HCN features defined at low blue-shifted velocities, this high red-shifted velocity compact feature peaks in between N2 and N3 and much closer to N2, suggesting that the same physical structure may contribute to the emission at both the low blueshift velocity and the high redshift velocity.

\begin{figure*}[htbp!]
	\centering
	\includegraphics[width=\textwidth]{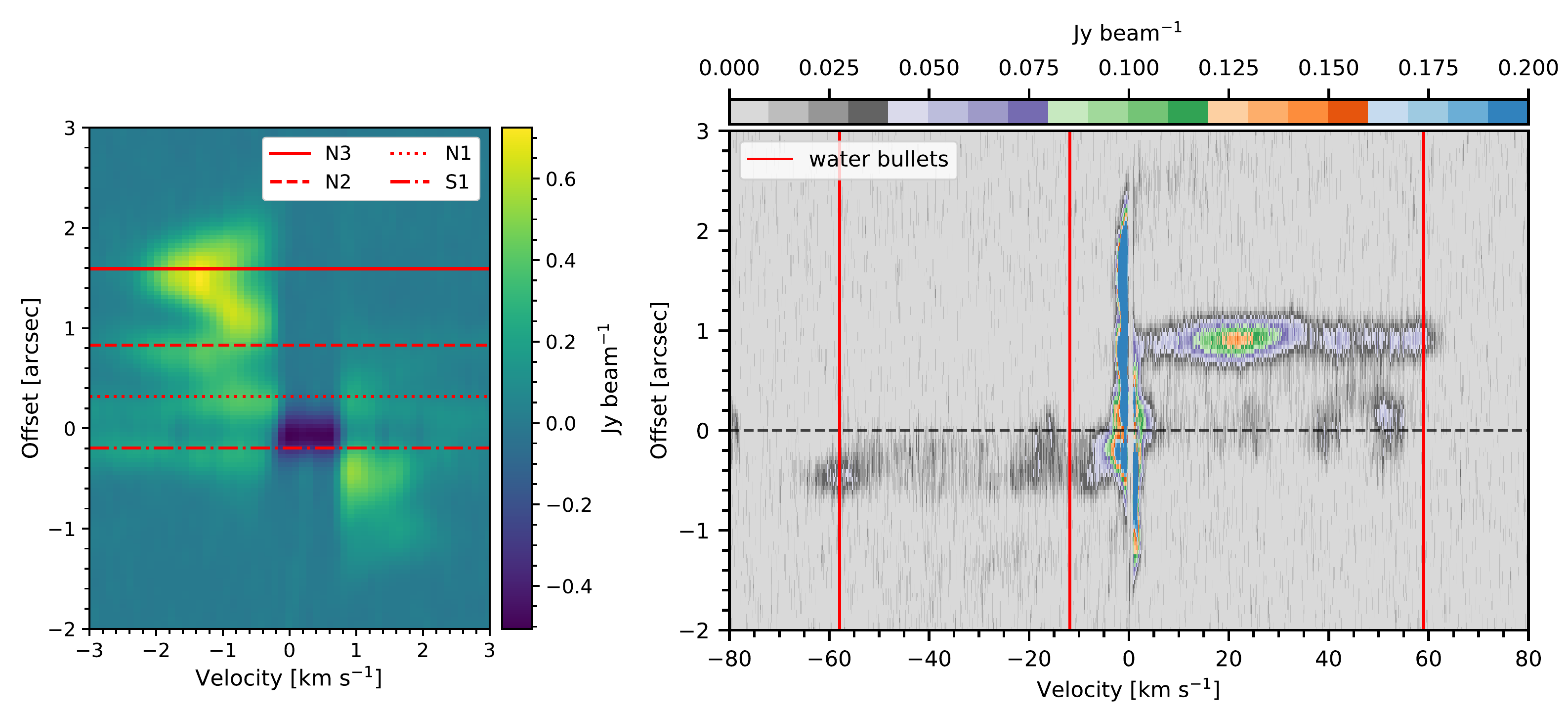}
	\caption{\textbf{Left:} The position-velocity diagram extracted along the HCN features.  The horizontal lines indicate the locations of the HCN features. \textbf{Right:} The position-velocity diagram extracted from the same slice as the left figure shown in the full velocity range.  The vertical lines indicate the velocities of the water bullets identified by \citet{2012A&A...542A...8K} and \citet{2014A&A...572A..21M}.}
	\label{fig:blob_pv}
\end{figure*}

Across the entire velocity range, high velocity broad emission appears at both red-shifted and blue-shifted velocities, occurring at $+0.9^{\prime\prime}$ and $-0.4^{\prime\prime}$ to the continuum source (Figure\,\ref{fig:blob_pv}, right).  \citet{2012A&A...542A...8K} identify high velocity ``bullets,'' narrow offset features, from the water 1$_{10}\rightarrow$1$_{01}$ emission, where \citet{2014A&A...572A..21M} isolate the bullets at $-$57.4, $-$11.8, and 59.0\,\kms.  The spectra of S1 and N3 as well as the PV diagrams along the outflows show emission peaks around $-$57.4\,\kms, but the observations cannot confirm other bullets.  However, the spectra of the HCN compact features show broad emission over the velocities of the water bullets; thus, the high velocity bullets may also present in our observations.

\section{Conclusions}
\label{sec:conclusion}

Using a 0\farcs{3} beam of ALMA, we successfully observe red-shifted absorption against the continuum toward BHR\,71 from the spectra of the \hcop\,\jj{4}{3} and HCN\,\jj{4}{3} lines, which is an unambiguous signature of infall.  The spectra of CS\,\jj{7}{6} and \htcn\,\jj{4}{3} also show red-shifted self-absorption.

To constrain the underlying infall kinematics, we test the 2D axisymmetric TSC infalling envelope model \citep{1984ApJ...286..529T} with radiative transfer calculations in 3D.  The comparison between the synthetic line profiles and the ALMA observations optimizes the envelope model.  We iterate the parameterized chemical abundance profiles to find the best-fitting models that reproduce the observations.  However, the fitting of the CS\,\jj{7}{6} line demands two evaporation zones, consistent with the results of the chemical modeling by \citet{2004ApJ...617..360L}.  The best-fitting model has an model-derived age of 12\,000 years since the initial collapse, younger than the age fitted with the continuum SED using the same model prescription.  The shape of the observed absorption features suggests a lower turbulent velocity of 0.25\,\kms\ than the $v_\text{turb}$ of 0.34\,\kms\ derived in Y17, but this discrepancy is comparable to the spectral resolution of the observations, 0.1\,\kms.

While the synthetic spectra reproduce the key features of the \hcop\ \jj{4}{3}, HCN \jj{4}{3}, and CS \jj{7}{6} lines, the best-fitting models cannot emulate all the features in the observations.  Our model of the TSC infalling envelope underestimates the red-shifted emission at the \hcop\ \jj{4}{3} and HCN \jj{4}{3} lines, and fails to reproduce the broad high velocity emission seen in the HCN \jj{4}{3} line.  Keplerian rotating disks could contribute to the high velocity emission in the HCN and \htcn\ lines; however, our observations cannot constrain the properties of the disk.  The models also overestimate low velocity emission at offset positions along the midplane of the envelope.  The infall signatures across the midplane of the envelope suggests a higher rotation at the inner 100\,au region, which would lead to a larger disk and may mitigate the shortcomings of our model, such as the lack of asymmetry in the PV diagrams (Figure\,\ref{fig:infall_pv}).

The ALMA observations serendipitously detect the emission of eight species of COMs along with SO$_{2}$, H$_{2}$CS, and HCN\,$v_{2}=1$ at the central $\sim$100\,au region, suggesting that BHR\,71 has a hot corino.  For \dmethanol, \methylformate, \methylformatev, \acetone, and \ethanol, which have multiple transitions detected in our observations, we estimate their excitation temperatures and column densities with the LTE radiative trasnfer calculations using \textsc{xclass}.  Their excitation temperatures range from 57\,K to 263\,K with a typical temperature of $\sim$100\,K.

The emission of \tmethanol\ and \methylformate\ shows a bar-shape morphology from their PV diagrams, probing the kinematics at the inner $\sim$100\,au region.  A Keplerian rotating disk extending from 15\,au to 50\,au produces a similar PV diagrams as the observations after considering the telescope beam size; however, the resolution of the observations prevents us from constraining the disk properties.  A 50\,au disk is significantly larger than the centrifugal radius of the best-fitting envelope model, 0.6\,au, suggesting a faster rotation in the inner region.

\acknowledgements
We thank Nami Sakai, Benny Tsang, Lars Kristensen, Sebastien Manigand, Nagayoshi Ohashi, Chang Won Lee, James Di\,Francesco, Kazuya Saigo, and Munan Gong for insightful discussions.  Y.-L. Yang acknowledges the supports of University Continuing Graduate Fellowship from The University of Texas at Austin and the JSPS Postdoctoral Fellowship from Japan Society for the Promotion of Science.  \added{AS acknowledges support for Program number HST-HF2-51421.001-A was provided by NASA through a grant from the Space Telescope Science Institute, which is operated by the Association of Universities for Research in Astronomy, Incorporated, under NASA contract NAS5-26555.  This research was supported by the Basic Science Research Program through the National Research Foundation of Korea (grant No. NRF-2018R1A2B6003423) and the Korea Astronomy and Space Science Institute under the R\&D program supervised by the Ministry of Science, ICT and Future Planning.}  The research of JKJ is supported by the European Research Council (ERC) under the European Union’s Horizon 2020 research and innovation programme through the ERC Consolidator Grant ``S4F'' (grant agreement No 646908).  This paper makes use of the following ALMA data: ADS/JAO.ALMA\#2016.0.00391.S. ALMA is a partnership of ESO (representing its member states), NSF (USA) and NINS (Japan), together with NRC (Canada), MOST and ASIAA (Taiwan), and KASI (Republic of Korea), in cooperation with the Republic of Chile. The Joint ALMA Observatory is operated by ESO, AUI/NRAO and NAOJ.  The National Radio Astronomy Observatory is a facility of the National Science Foundation operated under cooperative agreement by Associated Universities, Inc.

\facilities{ALMA}

\software{astropy \citep{2013A&A...558A..33A,2018AJ....156..123A}, CLASS \citep{2013ascl.soft05010G}, XCLASS \citep{2017A&A...598A...7M}, spectral-cube \citep{2016ascl.soft09017R}, LIME \citep{2010A&A...523A..25B}, Hyperion \citep{2011A&A...536A..79R}, CASA \citep{2007ASPC..376..127M}}

\appendix
\section{The Ray-tracing Method of \textsc{lime-aid}}
\label{sec:limeaid}
In steady state conditions without scattering, the specific intensity $I_\nu$ obeys the radiative transfer equation:
\begin{equation} \label{eq:RTE}
	\frac{\text{d}I_\nu}{\text{d}s} = j_\nu - k_\nu I_\nu \, ,
\end{equation}
where d$s$ is the traversed distance along the LOS, $j_\nu$ denotes the volume emissivity (in $\text{erg\,s}^{-1}\text{cm}^{-3}\text{Hz}^{-1}$), and $k_\nu$ denotes the absorption coefficient (in $\text{cm}^{-1}$). Equation\,\ref{eq:RTE} may be solved for individual sightlines by integrating along the corresponding rays, starting (finishing) at the far (near) end of a sphere of radius $R_\text{max}$, with zero intensity initial conditions. Assuming piecewise constant conditions within each Voronoi cell, the observed intensity is modified according to the solution in homogeneous media:

\begin{equation}
	\Delta I_\nu = \left(\frac{j_\nu}{k_\nu} - I_{\nu,0}\right) \left(1 - e^{-\Delta\tau_\nu} \right) \, ,
\end{equation}
where $I_{\nu,0}$ denotes the intensity carried over from the previous cell and $\Delta\tau_\nu = k_\nu \Delta \ell$ is the optical depth accrued by traversing a LOS distance $\Delta\ell$ within the cell.
We assume a Gaussian profile for line emission.
To account for gas motion, we also need to Doppler-shift into the frame of the gas such that the coefficients $j_\nu$ and $k_\nu$ are effectively evaluated at the normalized frequency in the absolute frame:
$\Delta u = (\Delta v + \bm{n} \cdot \bm{v}_i) / b_i$, where $\Delta v \equiv -c\Delta \nu / \nu_0$ with $\nu_0$ the frequency at line center, $\bm{n}$ is a unit vector in the direction of the observer,
$\bm{v}_i$ is the cell velocity, and $b_i \equiv (2 k_\text{B} T_i / m + v_\text{turb}^2)^{1/2}$ is the line broadening parameter in the cell with $k_\text{B}$ the Boltzmann constant, $T_i$ the gas temperature, $m$ the mass of the molecule, and $v_\text{turb}$ the microturbulent velocity. Assuming thermal and turbulent broadening are dominant, under this prescription we evaluate the line coefficients as
\begin{equation}
	j_{\nu,\text{gas}} = j_{\nu,0} e^{-\Delta u^2} \quad \text{and} \quad k_{\nu,\text{gas}} = k_{\nu,0} e^{-\Delta u^2} \, ,
\end{equation}
with the precomputed values at line center given by
\begin{equation}
	j_{\nu,0} = \frac{hc}{4\pi} \frac{n_2 A_{21}}{\sqrt{\pi} b_i} \;\, \text{and} \;\,
	k_{\nu,0} = \frac{hc}{4\pi} \frac{(n_1 B_{12} - n_2 B_{21})}{\sqrt{\pi} b_i} \, ,
\end{equation}
where the Einstein coefficients are related by the degeneracies of the two transition levels $g_1 B_{12} = g_2 B_{21}$ and $B_{21} = A_{21} c^2 / (2 h \nu_0^3)$. The emissivity and absorption coefficients for the dust are
\begin{equation}
	j_{\nu,\text{dust}} = k_{\nu,\text{dust}} B_\nu(T_\text{dust}) \;\;\, \text{and} \;\;\, k_{\nu,\text{dust}} = \kappa \rho_\text{dust} \, ,
\end{equation}
where $B_\nu$ denotes the Planck
function for a given dust temperature, while $\kappa$ and $\rho_\text{dust}$ are the dust opacity per gram of dust and density, respectively.
Finally, for this study when the ray intersects the LOS midplane we also add a continuum source, whose brightness follows a 2D Gaussian profile in the plane parallel to the image plane.  The fitted continuum parameters from the ALMA observation (Table\,\ref{tbl:continuum}) determine the input 2D Gaussian profile.
For completeness and concreteness, we list the basic data for the modeled transitions in Table\,\ref{tbl:line_data}.

To ensure convergence of the image cubes in space and frequency we employ an adaptive convergence scheme. Specifically, we specify a base image resolution and perform ray tracing at each of the pixel corner locations, saving each value. Then for each pixel we trace additional rays for the center and midpoints of the edges. We then compare the average pixel intensity based on the trapezoidal (linear) and Simpson's (quadratic) integration schemes, which use four and nine point stencils, respectively. This allows us to accurately estimate the error caused by a sparse point sampling. If the relative error is above a specified tolerance level, e.g., $|1 - I_{\nu,\text{trap}} / I_{\nu,\text{Simp}}| > \varepsilon \sim 10^{-6}$ for any frequency, then we refine the pixel into four sub-pixels. Each of these are tested and adaptively refined until the convergence criterion is met for all frequencies, with the hierarchy of rays tracked maintained with a quadtree data structure, which is eventually aggregated from the bottom up for an efficient and robust determination of the area-averaged intensity for each pixel. Our radiative transfer calculations produce line profiles that are nearly indistinguishable from the results obtained by \textsc{lime}, when we do not include a central continuum source.  With \textsc{lime-aid}, we can efficiently and realistically test various models for the abundance profile and viewing angles in a rapid post-processing fashion.

\section{Effects of Cavities and the Inclination of the Envelope}
\label{sec:infall_effect}
BHR 71 drives outflows in the north-south direction, which carve out cavities within the envelope.  Traditional 1D or 2D radiative transfer calculations fail to fully test the effect of the cavity on the infall profile due to the lack of geometric dimensions.  With \textsc{lime} and \textsc{lime-aid}, we investigate the effect of the outflow cavities and inclination angles on the infall profile.

\begin{figure}[htbp!]
	\centering
	\includegraphics[width=0.48\textwidth]{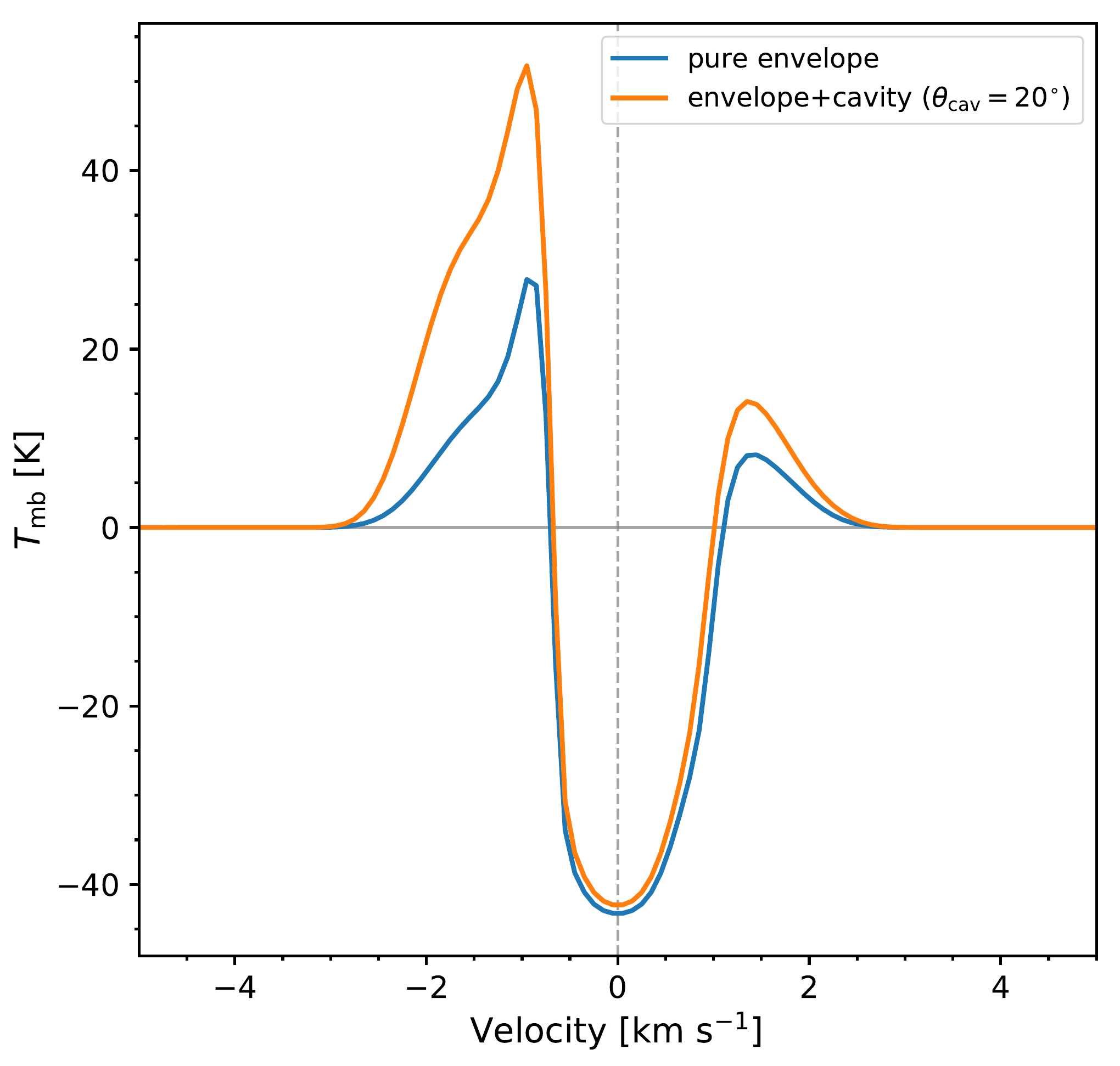}
	\caption{The synthetic continuum-subtracted \hcop\,\jj{4}{3} line profiles calculated with the envelopes with and without the outflow cavities.}
	\label{fig:model_cavity}
\end{figure}

Figure\,\ref{fig:model_cavity} illustrates the effect of cavities on the infall profile.  The model with cavities has more intensity at low velocity (from $-2$ to 2\,\kms), while the high velocity emission remains the same.  Although the LOS at an inclination angle of 130$^{\circ}$ passes through the cavity when $r \lesssim 1000$\,au, where the abundance becomes zero, the finite size of the telescope beam still includes parts of the outflow cavity wall at small radii.  Thus, with a cavity, the overall intensity increases because of the higher temperature near the cavity walls.

\begin{figure}[htbp!]
	\centering
	\includegraphics[width=0.48\textwidth]{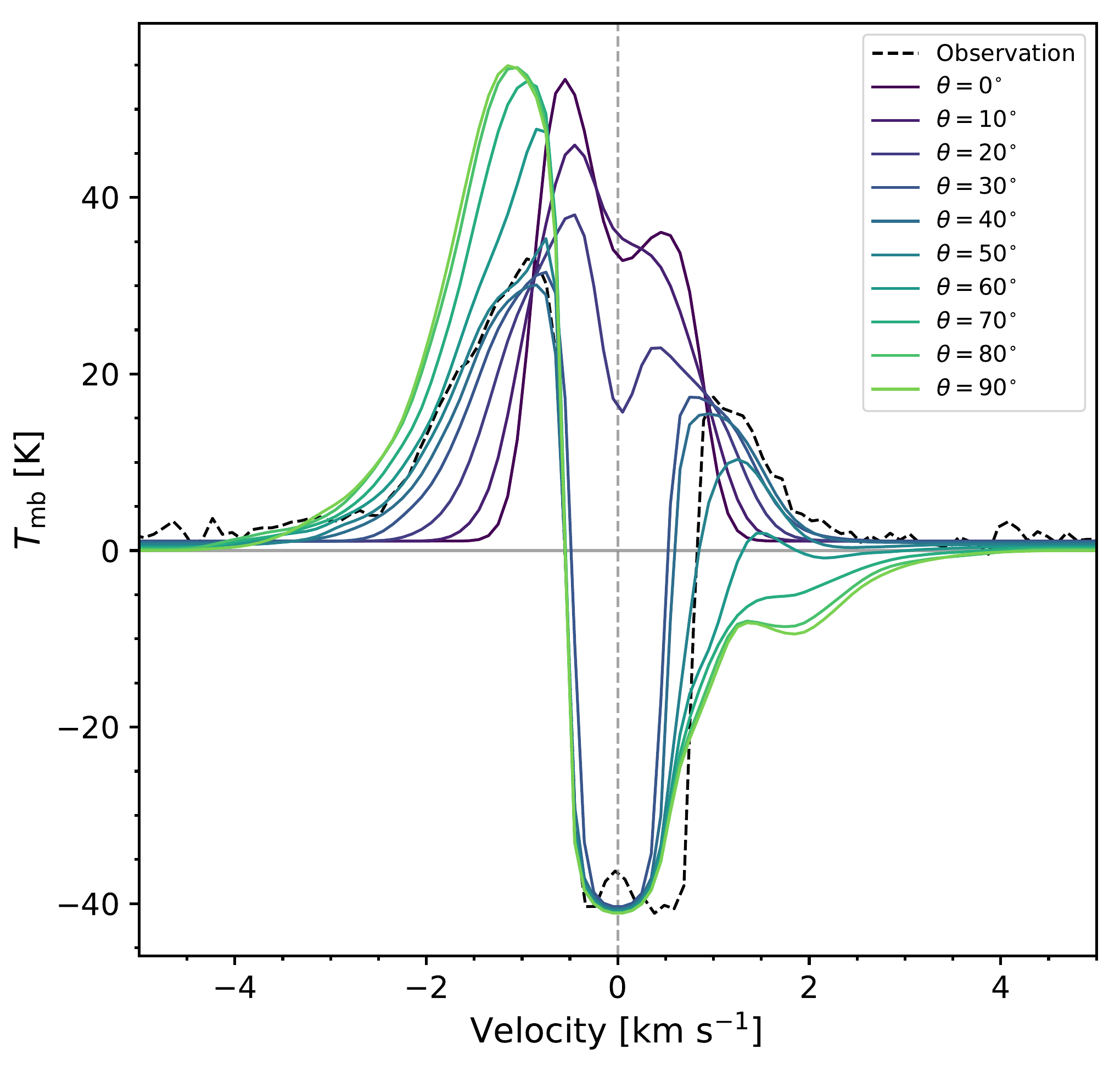}
	\caption{The synthetic continuum-subtracted infall profiles of the \hcop\,\jj{4}{3} line with inclination angles ranging from 0$^{\circ}$ to 90$^{\circ}$. The model has a bipolar outflow with an opening angle of 20$^{\circ}$.}
	\label{fig:infall_inclination}
\end{figure}

The inclination angle is another critical parameter for the synthetic infall profile.  Here we only consider the geometric effect of the inclination angle as the model does not include the outflowing gas.  For the TSC envelope with cavities, the infall profile strongly depends on the inclination angle, which is defined in Section\,\ref{sec:intro} (Figure\,\ref{fig:infall_inclination}).  The line profiles are extracted with the size of the observed continuum source, 0\farcs{52}$\times$0\farcs{39} with PA$=$131\fdg{25}.  Along the LOS close to the outflow axis, the infall profile becomes narrower due to the zero abundance in the outflow cavity, which eliminates the emission at high velocity.  Along the face-on view, the absorption disappears as the molecular abundance becomes zero in the cavities, and the emission from the inner edge of the envelope dominates the line profile.  The inner radius of the evaporation zone is 50\,au, which is 0\farcs{25} at 200 pc.

As the inclination increases, the infall profile becomes dominated by the envelope, which consists of low-velocity warm gas and high-velocity hot gas.  Thus, the intensity increases as more warm and hot gas enters the LOS, and the high velocity emission increases from the hot gas in the evaporation zone.  At high inclination, the second absorption feature appears at 2.5\,\kms\ as the molecules in the evaporation zone become optically thick to the continuum source.
%
% Find a place to show the LOS beam-averaged velocity as a function of radius, and the panels of intensity maps at different inclination.

\begin{figure*}
	\includegraphics[width=\textwidth]{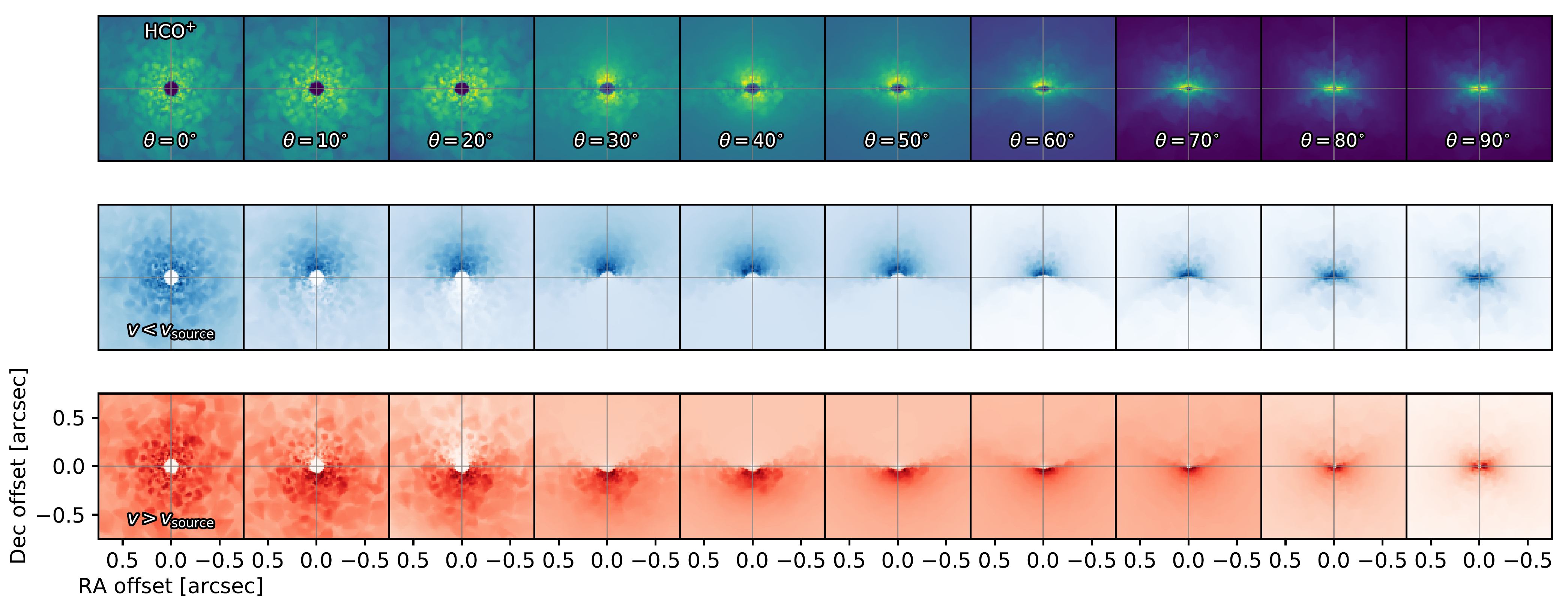}
	\caption{The intensity maps of the best-fitting \hcop\ model viewed from different inclination angles.  The second row shows the morphology at the blue-shifted velocities from 0\,\kms$>v-v_\text{source}>-5$\,\kms; the third row shows the morphology at the red-shifted velocities from 5\,\kms$>v-v_\text{source}>0$\,\kms.  This model is ray-traced without the central continuum source to highlight the effects of cavities and inclinations.  The inclination angles shown here follow the same definition in Section\,\ref{sec:intro}.  An inclination of 50$^{\circ}$ shown here is equivalent to the inclination of BHR\,71 (130$^\circ$) but flipped upside down.}
	\label{fig:inclination_grid}
\end{figure*}

\section{Line Fitting Results of the ALMA Spectra}
\label{sec:line_fitting}
Table\,\ref{tbl:line_fitting} shows the line fitting results using \textsc{class}\footnote{\href{http://www.iram.fr/IRAMFR/GILDAS}{http://www.iram.fr/IRAMFR/GILDAS}}, a data reduction and analyses package for radio observations.  The line fitting starts with a single Gaussian profile.  If a single Gaussian profile fails to reproduce the spectral feature, we tried a double or triple Gaussian profile to achieve a better fit.  Also the line width is sometimes fixed to 3.5\,\kms\ to fit blended lines.  Table\,\ref{tbl:line_fitting} also lists the adopted methods other than the default single Gaussian profile, and the source size fitted with 2D Gaussian profiles, using the CASA \texttt{imfit} task.

\startlongtable
    \begin{deluxetable*}{r l l l l l l}
    \tabletypesize{\scriptsize}
    \tablecaption{Line Fitting \label{tbl:line_fitting}}
    \tablewidth{\textwidth}
    \tablehead{\colhead{Line No.} & \colhead{Frequency} & \colhead{Line Width} & \colhead{Integrated Flux} & \colhead{Deconvolved FWHMs} & \colhead{PA} & \colhead{Method\tablenotemark{a}} \\
               \colhead{} & \colhead{[MHz]} & \colhead{[\kms]} & \colhead{[K~\kms]} & \colhead{[arcsec]} & \colhead{[deg.]} & \colhead{}}
    \startdata
    \multicolumn{5}{c}{\hcop~\jj{4}{3}} \\
    \hline
    1 & 354442.15 (0.07) & 3.5 (n/a) & 10.5 (0.3) & $0\farcs{61}\pm0\farcs{06}\times0\farcs{19}\pm0\farcs{04}$ & $160\fdg{76}\pm4\fdg{17}$ &  DG/FW \\
    2 & 354445.94 (0.07) & 3.5 (n/a) & 11.3 (0.3) & $0\farcs{75}\pm0\farcs{06}\times0\farcs{17}\pm0\farcs{04}$ & $163\fdg{93}\pm2\fdg{37}$ & DG/FW \\
    3 & 354459.44 (0.06) & 6.6 (0.1) & 20.9 (0.3) & $0\farcs{48}\pm0\farcs{07}\times0\farcs{15}\pm0\farcs{09}$ & $163\fdg{4}\pm8\fdg{68}$ &  \\
    4 & 354476.57 (0.08) & 7.7 (0.2) & 16.3 (0.3) & $0\farcs{55}\pm0\farcs{11}\times0\farcs{18}\pm0\farcs{13}$ & $164\fdg{71}\pm10\fdg{74}$ &  \\
    5 & 354525.27 (0.13) & 9.1 (0.4) & 28.4 (0.9) & $0\farcs{67}\pm0\farcs{07}\times0\farcs{27}\pm0\farcs{07}$ & $16\fdg{21}\pm6\fdg{54}$ &  \\
    6 & 354600.95 (0.12) & 5.9 (0.2) & 15.1 (0.5) & $0\farcs{45}\pm0\farcs{11}\times0\farcs{22}\pm0\farcs{12}$ & $159\fdg{46}\pm23\fdg{35}$ &  \\
    7 & 354607.97 (0.02) & 3.8 (0.0) & 16.6 (0.2) & $0\farcs{53}\pm0\farcs{06}\times0\farcs{21}\pm0\farcs{06}$ & $166\fdg{72}\pm7\fdg{83}$ &  \\
    8 & 354622.82 (0.05) & 3.3 (0.1) & 8.0 (0.2) & $0\farcs{55}\pm0\farcs{14}\times0\farcs{23}\pm0\farcs{19}$ & $176\fdg{44}\pm19\fdg{62}$ &  \\
    \hline
    \multicolumn{5}{c}{HCN~\jj{4}{3}} \\
    \hline
    9 & 356626.83 (0.03) & 3.6 (0.1) & 10.1 (0.1) & $0\farcs{63}\pm0\farcs{10}\times0\farcs{25}\pm0\farcs{06}$ & $152\fdg{83}\pm8\fdg{26}$ &  \\
    10 & 356661.53 (0.12) & 7.5 (0.3) & 7.6 (0.2) & $0\farcs{62}\pm0\farcs{28}\times0\farcs{18}\pm0\farcs{14}$ & $161\fdg{76}\pm40\fdg{95}$ &  \\
    11 & 356675.94 (0.05) & 2.6 (0.1) & 2.1 (0.1) & $0\farcs{54}\pm0\farcs{27}\times0\farcs{17}\pm0\farcs{02}$\tablenotemark{b} & $132\fdg{84}\pm3\fdg{98}$\tablenotemark{b} &  \\
    12 & 356686.39 (0.08) & 3.0 (0.2) & 2.5 (0.1) & $0\farcs{45}\pm0\farcs{19}\times0\farcs{25}\pm0\farcs{06}$\tablenotemark{b} & $111\fdg{21}\pm14\fdg{71}$\tablenotemark{b} &  \\
    13 & 356697.01 (0.07) & 2.3 (0.1) & 3.2 (0.1) & $0\farcs{50}\pm0\farcs{26}\times0\farcs{26}\pm0\farcs{19}$ & $8\fdg{34}\pm37\fdg{52}$ &  \\
    14 & 356705.68 (0.05) & 2.4 (0.1) & 3.3 (0.1) & $0\farcs{34}\pm0\farcs{21}\times0\farcs{28}\pm0\farcs{19}$ & $175\fdg{46}\pm85\fdg{37}$ & DG \\
    15 & 356712.81 (0.05) & 6.1 (0.1) & 14.2 (0.2) & $0\farcs{39}\pm0\farcs{13}\times0\farcs{29}\pm0\farcs{14}$ & $164\fdg{22}\pm58\fdg{56}$ & DG \\
    16 & 356723.39 (0.12) & 2.9 (0.1) & 7.6 (0.1) & $0\farcs{58}\pm0\farcs{14}\times0\farcs{46}\pm0\farcs{14}$ & $23\fdg{16}\pm62\fdg{77}$ & TG \\
    17 & 356726.06 (0.12) & 2.0 (0.1) & 4.4 (0.1) & $0\farcs{51}\pm0\farcs{11}\times0\farcs{22}\pm0\farcs{17}$ & $5\fdg{11}\pm17\fdg{43}$ & TG \\
    18 & 356729.28 (0.12) & 1.9 (0.1) & 3.2 (0.1) & $0\farcs{46}\pm0\farcs{14}\times0\farcs{31}\pm0\farcs{19}$ & $42\fdg{73}\pm135\fdg{22}$ & TG \\
    19 & 356755.67 (0.07) & 6.0 (0.2) & 15.7 (0.4) & $0\farcs{92}\pm0\farcs{11}\times0\farcs{27}\pm0\farcs{08}$ & $16\fdg{96}\pm4\fdg{35}$ & TG \\
    20 & 356761.86 (0.14) & 3.1 (0.2) & 3.9 (0.4) & $0\farcs{47}\pm0\farcs{12}\times0\farcs{34}\pm0\farcs{15}$ & $30\fdg{08}\pm84\fdg{25}$ & TG \\
    21 & 356767.25 (0.13) & 3.5 (0.3) & 4.4 (0.3) & $0\farcs{68}\pm0\farcs{25}\times0\farcs{28}\pm0\farcs{13}$ & $171\fdg{58}\pm30\fdg{66}$ & TG \\
    22 & 356796.28 (0.08) & 4.9 (0.2) & 17.5 (0.6) & $0\farcs{66}\pm0\farcs{08}\times0\farcs{21}\pm0\farcs{07}$ & $175\fdg{06}\pm5\fdg{86}$ &  \\
    23 & 356831.65 (0.04) & 1.6 (0.1) & 1.5 (0.1) & $0\farcs{37}\pm0\farcs{24}\times0\farcs{21}\pm0\farcs{07}$\tablenotemark{b} & $53\fdg{03}\pm22\fdg{59}$\tablenotemark{b} & DG \\
    24 & 356834.87 (0.05) & 3.2 (0.1) & 3.9 (0.1) & $0\farcs{43}\pm0\farcs{34}\times0\farcs{06}\pm0\farcs{26}$ & $170\fdg{54}\pm31\fdg{33}$ & DG \\
    \hline
    \multicolumn{5}{c}{CS~\jj{7}{6}} \\
    \hline
    25 & 342782.23 (0.12) & 9.4 (0.3) & 13.3 (0.3) & $0\farcs{39}\pm0\farcs{17}\times0\farcs{06}\pm0\farcs{09}$ & $135\fdg{37}\pm25\fdg{55}$ &  \\
    26 & 342935.06 (0.12) & 3.7 (0.4) & 10.7 (0.7) & $0\farcs{54}\pm0\farcs{11}\times0\farcs{20}\pm0\farcs{14}$ & $167\fdg{08}\pm13\fdg{99}$ &  \\
    27 & 342946.84 (0.08) & 4.5 (0.2) & 11.8 (0.4) & $0\farcs{79}\pm0\farcs{14}\times0\farcs{40}\pm0\farcs{10}$ & $173\fdg{42}\pm12\fdg{93}$ &  \\
    28 & 342997.61 (0.07) & 6.8 (0.2) & 11.3 (0.3) & $0\farcs{28}\pm0\farcs{18}\times0\farcs{12}\pm0\farcs{09}$ & $142\fdg{82}\pm24\fdg{05}$ &  \\
    \hline
    \multicolumn{5}{c}{\htcn~\jj{4}{3}} \\
    \hline
    29 & 345132.49 (0.07) & 3.5 (0.1) & 11.2 (0.4) & $0\farcs{52}\pm0\farcs{06}\times0\farcs{26}\pm0\farcs{05}$ & $164\fdg{94}\pm10\fdg{2}$ &  \\
    30 & 345143.86 (0.11) & 3.5 (n/a) & 3.5 (0.1) & $0\farcs{47}\pm0\farcs{23}\times0\farcs{08}\pm0\farcs{11}$ & $145\fdg{08}\pm30\fdg{91}$ & DG/FW \\
    31 & 345149.14 (0.09) & 3.5 (n/a) & 4.0 (0.1) & $0\farcs{65}\pm0\farcs{19}\times0\farcs{03}\pm0\farcs{13}$ & $154\fdg{56}\pm9\fdg{46}$ & DG/FW \\
    32 & 345162.02 (0.16) & 8.7 (0.4) & 7.7 (0.3) & $0\farcs{31}\pm0\farcs{17}\times0\farcs{07}\pm0\farcs{27}$ & $16\fdg{09}\pm62\fdg{29}$ &  \\
    33 & 345179.92 (0.31) & 12.1 (0.6) & 9.8 (0.4) & $0\farcs{38}\pm0\farcs{08}\times0\farcs{26}\pm0\farcs{03}$\tablenotemark{b} & $160\fdg{0}\pm12\fdg{82}$\tablenotemark{b} &  \\
    34 & 345209.92 (0.24) & 5.4 (0.2) & 7.5 (0.2) & $0\farcs{46}\pm0\farcs{12}\times0\farcs{14}\pm0\farcs{09}$ & $166\fdg{95}\pm17\fdg{18}$ & TG \\
    35 & 345219.31 (0.24) & 4.7 (0.2) & 3.7 (0.2) & $0\farcs{56}\pm0\farcs{12}\times0\farcs{30}\pm0\farcs{04}$\tablenotemark{b} & $10\fdg{47}\pm7\fdg{2}$\tablenotemark{b} & TG \\
    36 & 345227.49 (0.24) & 5.1 (0.2) & 5.8 (0.2) & $0\farcs{51}\pm0\farcs{14}\times0\farcs{24}\pm0\farcs{10}$ & $142\fdg{58}\pm22\fdg{79}$ & TG \\
    37 & 345235.81 (0.24) & 1.8 (0.2) & 1.9 (0.2) & $0\farcs{36}\pm0\farcs{22}\times0\farcs{27}\pm0\farcs{18}$ & $89\fdg{92}\pm86\fdg{65}$ & DG \\
    38 & 345238.35 (0.24) & 1.6 (0.2) & 1.6 (0.2) & $0\farcs{55}\pm0\farcs{24}\times0\farcs{25}\pm0\farcs{18}$ & $125\fdg{19}\pm79\fdg{34}$ & DG \\
    39 & 345246.53 (0.23) & 10.8 (0.7) & 9.6 (0.5) & $0\farcs{45}\pm0\farcs{19}\times0\farcs{37}\pm0\farcs{31}$ & $80\fdg{18}\pm84\fdg{75}$ &  \\
    40 & 345257.27 (0.17) & 5.5 (0.5) & 5.4 (0.3) & $0\farcs{78}\pm0\farcs{15}\times0\farcs{32}\pm0\farcs{03}$\tablenotemark{b} & $164\fdg{72}\pm4\fdg{37}$\tablenotemark{b} &  \\
    41 & 345283.15 (0.08) & 4.6 (0.2) & 9.2 (0.3) & $0\farcs{45}\pm0\farcs{09}\times0\farcs{24}\pm0\farcs{08}$ & $159\fdg{98}\pm20\fdg{63}$ & DG \\
    42 & 345286.34 (0.05) & 1.1 (0.1) & 1.2 (0.2) & $0\farcs{58}\pm0\farcs{09}\times0\farcs{33}\pm0\farcs{03}$\tablenotemark{b} & $164\fdg{57}\pm6\fdg{85}$\tablenotemark{b} & DG \\
    43 & 345295.18 (0.03) & 3.6 (0.1) & 6.7 (0.1) & $0\farcs{56}\pm0\farcs{11}\times0\farcs{14}\pm0\farcs{06}$ & $157\fdg{86}\pm8\fdg{35}$ & DG \\
    44 & 345300.73 (0.05) & 2.9 (0.1) & 3.2 (0.1) & $0\farcs{54}\pm0\farcs{10}\times0\farcs{29}\pm0\farcs{03}$\tablenotemark{b} & $157\fdg{73}\pm6\fdg{49}$\tablenotemark{b} & DG \\
    45 & 345306.00 (0.33) & 3.5 (n/a) & 3.7 (0.4) & $0\farcs{57}\pm0\farcs{21}\times0\farcs{24}\pm0\farcs{16}$ & $143\fdg{29}\pm33\fdg{65}$ & FW \\
    46 & 345319.30 (0.11) & 3.5 (n/a) & 12.8 (0.5) & $0\farcs{52}\pm0\farcs{05}\times0\farcs{18}\pm0\farcs{06}$ & $168\fdg{3}\pm6\fdg{29}$ & DG/FW \\
    47 & 345324.74 (0.34) & 3.5 (n/a) & 4.8 (0.5) & $0\farcs{44}\pm0\farcs{06}\times0\farcs{30}\pm0\farcs{03}$\tablenotemark{b} & $147\fdg{78}\pm9\fdg{86}$\tablenotemark{b} & DG/FW \\
    48 & 345398.61 (0.06) & 3.6 (0.1) & 8.0 (0.2) & $0\farcs{48}\pm0\farcs{09}\times0\farcs{18}\pm0\farcs{08}$ & $156\fdg{98}\pm11\fdg{65}$ &  \\
    49 & 345406.47 (0.00) & 3.5 (n/a) & 4.2 (0.2) & $0\farcs{48}\pm0\farcs{07}\times0\farcs{34}\pm0\farcs{03}$\tablenotemark{b} & $158\fdg{65}\pm11\fdg{42}$\tablenotemark{b} & DG/FW/FC \\
    50 & 345410.51 (0.06) & 3.5 (n/a) & 6.1 (0.2) & $0\farcs{39}\pm0\farcs{11}\times0\farcs{21}\pm0\farcs{14}$ & $165\fdg{46}\pm28\fdg{93}$ & DG/FW \\
    51 & 345444.94 (0.07) & 2.1 (0.2) & 4.2 (0.5) & $0\farcs{51}\pm0\farcs{09}\times0\farcs{21}\pm0\farcs{10}$ & $172\fdg{77}\pm13\fdg{21}$ & DG \\
    52 & 345448.79 (0.41) & 6.9 (0.6) & 7.3 (0.8) & $0\farcs{44}\pm0\farcs{17}\times0\farcs{27}\pm0\farcs{19}$ & $108\fdg{81}\pm67\fdg{36}$ & DG \\
    53 & 345459.96 (0.13) & 5.1 (0.3) & 6.8 (0.3) & $0\farcs{43}\pm0\farcs{05}\times0\farcs{39}\pm0\farcs{05}$\tablenotemark{b} & $53\fdg{47}\pm54\fdg{24}$\tablenotemark{b} &  \\
    54 & 345466.71 (0.15) & 3.5 (n/a) & 6.9 (0.3) & $0\farcs{42}\pm0\farcs{10}\times0\farcs{25}\pm0\farcs{16}$ & $7\fdg{53}\pm29\fdg{13}$ & FW \\
    55 & 345473.52 (0.09) & 3.5 (n/a) & 4.0 (0.1) & $0\farcs{30}\pm0\farcs{13}\times0\farcs{22}\pm0\farcs{11}$ & $18\fdg{0}\pm85\fdg{98}$ & DG/FW \\
    56 & 345477.96 (0.13) & 3.5 (n/a) & 2.6 (0.1) & $0\farcs{50}\pm0\farcs{11}\times0\farcs{24}\pm0\farcs{02}$\tablenotemark{b} & $115\fdg{56}\pm4\fdg{91}$\tablenotemark{b} & DG/FW \\
    57 & 345485.49 (0.42) & 2.7 (0.7) & 4.0 (1.2) & $0\farcs{33}\pm0\farcs{10}\times0\farcs{11}\pm0\farcs{17}$ & $16\fdg{0}\pm35\fdg{58}$ & TG \\
    58 & 345488.34 (0.38) & 2.2 (0.8) & 2.6 (1.4) & $0\farcs{43}\pm0\farcs{15}\times0\farcs{21}\pm0\farcs{16}$ & $17\fdg{49}\pm33\fdg{47}$ & TG \\
    59 & 345492.06 (0.34) & 3.0 (0.6) & 2.9 (0.6) & $0\farcs{51}\pm0\farcs{10}\times0\farcs{41}\pm0\farcs{07}$\tablenotemark{b} & $52\fdg{49}\pm30\fdg{88}$\tablenotemark{b} & TG \\
    60 & 345509.46 (0.07) & 4.9 (0.1) & 10.0 (0.2) & $0\farcs{29}\pm0\farcs{09}\times0\farcs{15}\pm0\farcs{09}$ & $1\fdg{13}\pm33\fdg{02}$ &  \\
    61 & 345546.66 (0.06) & 3.9 (0.1) & 3.8 (0.1) & $0\farcs{47}\pm0\farcs{21}\times0\farcs{05}\pm0\farcs{16}$ & $159\fdg{85}\pm33\fdg{31}$ &  \\
    62 & 345570.54 (0.10) & 3.3 (0.2) & 3.9 (0.2) & $0\farcs{48}\pm0\farcs{09}\times0\farcs{31}\pm0\farcs{04}$\tablenotemark{b} & $150\fdg{65}\pm11\fdg{41}$\tablenotemark{b} &  \\
    \enddata
    \tablenotetext{a}{DG: Double Gaussian; TG: Triple Gaussian; FW: Fixed line widths; FC: Fixed line centroids.}
    \tablenotetext{b}{One (or both) of the fitted FWHM is smaller than the beam size.
                      The fitted convolved sizes are shown instead.  Such situation is typical for weak lines close to 3$\sigma$}
    \tablecomments{
                   The uncertainty of line width are shown inside the prentice when the line width remains flexible for the fitting.}
\end{deluxetable*}

\section{Identified COMs}
\label{sec:coms_identified}
\noindent \textbf{\newline O-bearing molecules} \\
\noindent \textit{Methanol (\methanol):} We identified seven emission lines of methanol and its isotopologs, including CH$_{3}$OH, $^{13}$CH$_{3}$OH, CH$_{2}$DOH, and CH$_{3}$OD.  Accroding to \citet{1988ApJS...67..135A}, our observations only cover one line of CH$_{3}$OD, which is detected as Line 46. \citet{2006A&A...454L..79P}
Previous observations \citep{1998ApJ...509..768G,2006A&A...454L..79P} have detected the emission of methanol (\methanol), supporting the identification of methanol.  The emission of methanol (Line 9) exhibits absorption at the line center, suggesting that the emission of methanol is optically thick so that the fitted column density serves as a lower limit.  If we exclude Lines 2 and 46, which are either blended with an unidentified line or has its line width fixed during the line fitting, the mean width of methanol and its isotopologs is 3.7$\pm$0.2\,\kms.

\begin{figure*}[htbp!]
	\includegraphics[width=\textwidth]{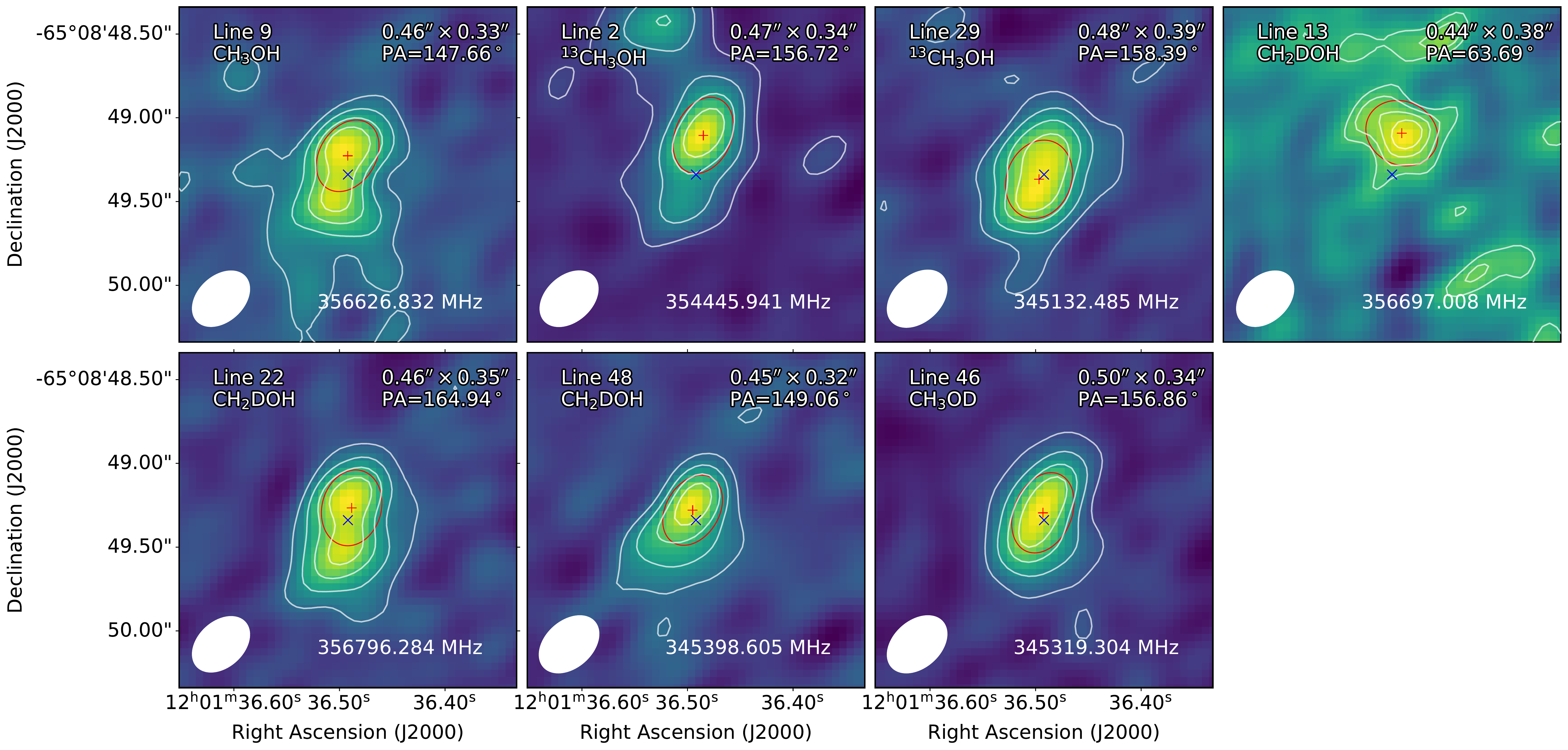}
	\caption{The moment 0 maps of methanol and its isotopologs ($^{13}$CH$_{3}$OH, CH$_{2}$DOH, and CH$_{3}$OD), calculated with a line width of 3.5\,\kms.  The white contours show the emission from the maximum to 3$\sigma$ level.  The red contour indicates the fitted size of the half power beam width (HPBW).  The red cross shows the center of the fitted profile, while the blue ``$\times$'' indicates the peak of the continuum emission.}
	\label{fig:com_morphology_methanol}
\end{figure*}

Figure\,\ref{fig:com_morphology_methanol} shows the morphology of methanol and its isotopologs.  Excluding the offset emission at the same frequency as Line 2 (see Section\,\ref{sec:blobs}), all the emission of methanol and its isotopologs extends in the north-south direction, except for one of the \dmethanol\ lines (Line 13).

\noindent \textit{Methyl formate (\methylformate):}  The ALMA observations detect methyl formate in both the ground state and its first torsional excited state.  The ground state rotational transitions correspond the emission of Lines 15, 53, 54, and 57, while the emission of the vibrationally excited methyl formate matches Lines 12, 31, 39, 55, 60, and 62.  The complex features of Lines 54 and 57/58 hint that the emission of methyl formate may be optically thick.  A more detailed model dedicated to methyl formate may better constrain its properties.

\begin{figure*}[htbp!]
	\includegraphics[width=\textwidth]{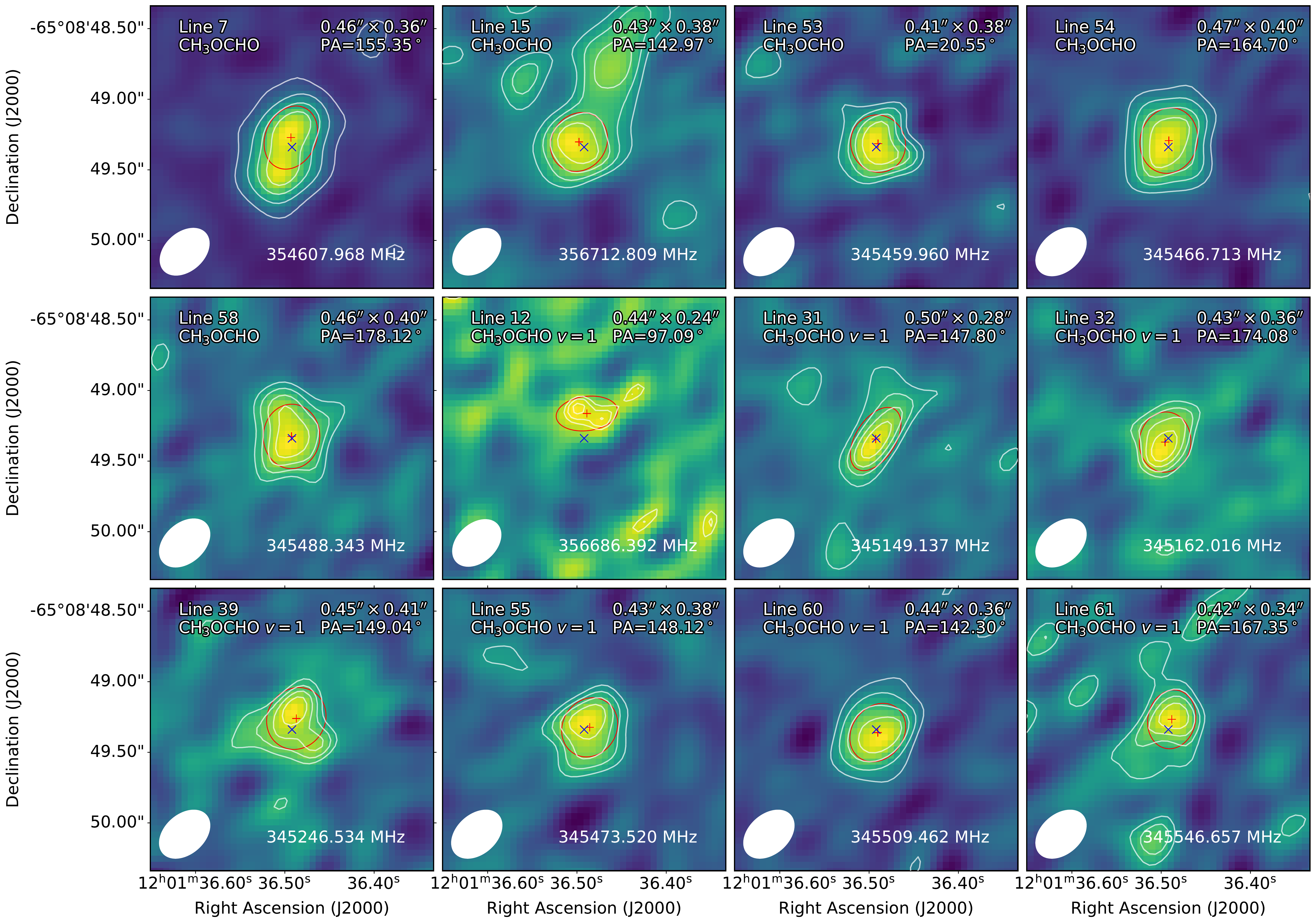}
	\caption{Similar to Figure\,\ref{fig:com_morphology_methanol} but for methyl formate (\methylformate).}
	\label{fig:com_morphology_methylformate}
\end{figure*}

Figure\,\ref{fig:com_morphology_methylformate} shows the morphology of methyl formate in both ground state and the vibrational excited state.  One of the ground state transitions (Line 7) peaks both north and south of the continuum source, while another ground state transition (Line 15) peaks at the center with extended emission toward north.  The rest of the methyl formate transitions (10 of 12) show a compact emission peaked at the center.

\begin{figure*}[htbp!]
	\includegraphics[width=\textwidth]{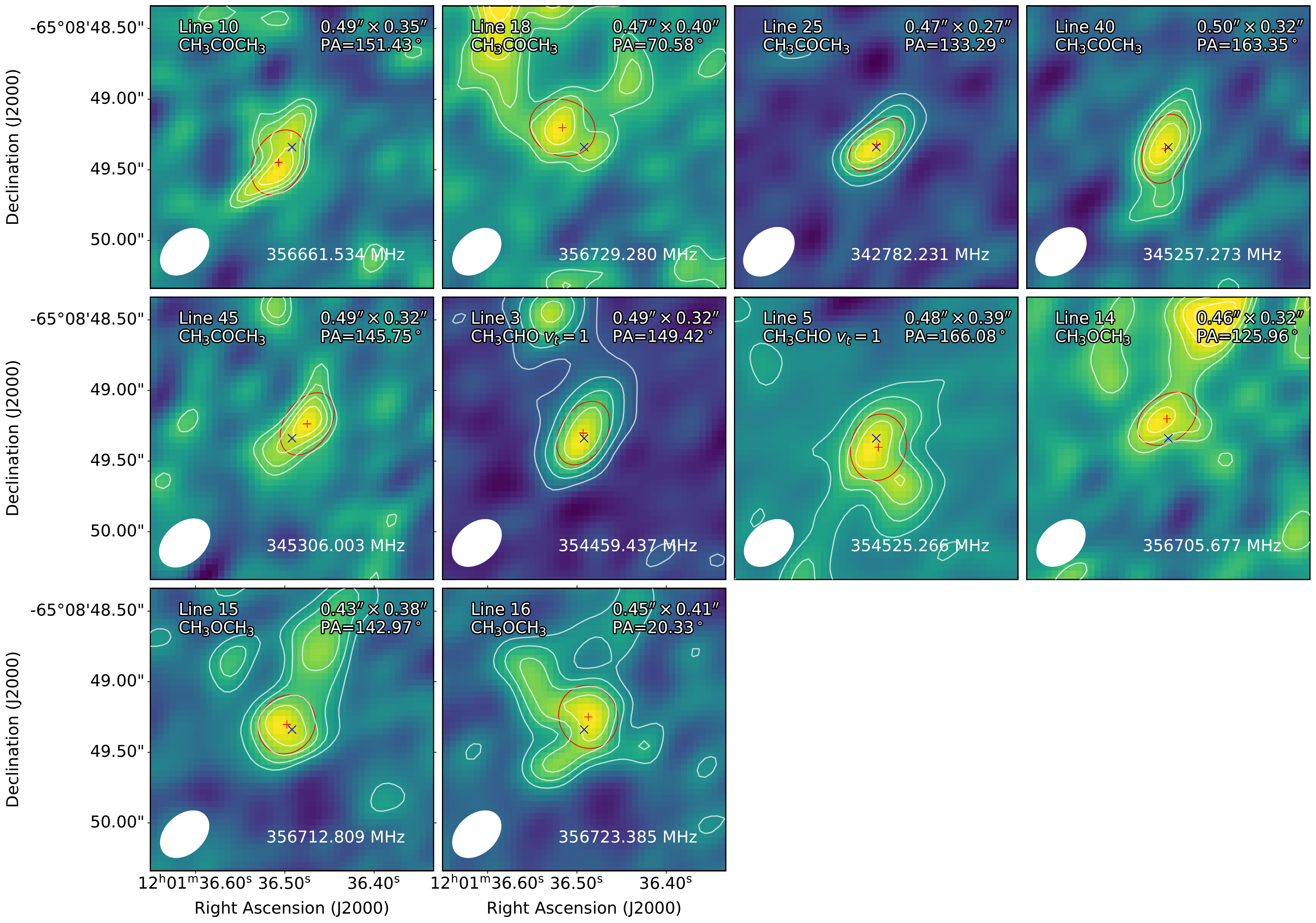}
	\caption{Similar to Figure\,\ref{fig:com_morphology_methanol} but for acetone (\acetone), acetaldehyde (\acetaldehyde), and dimethyl ether (\dimethylether).}
	\label{fig:com_morphology_acetone_acetaldehyde_dimethylether}
\end{figure*}

\noindent \textit{Acetone (\acetone):} Line 25 provides the strongest indication of acetone, which also matches Lines 11, 18, 40, and 45.  The emission of acetone also appears at Lines 10, 23 and 24, but our simple model underestimates the strengths of these lines.  Acetone emits from a compact region with a size of the continuum at the center (Figure\,\ref{fig:com_morphology_acetone_acetaldehyde_dimethylether}).

\noindent \textit{Dimethyl ether (\dimethylether):} The emission of dimethyl ether matches Lines 14, 15, and 16.  The apparent absorption at the center of Line 16 suggests that this transition of dimethyl ether may be optically thick.  All transitions of dimethyl ether peak at the center with an irregular morphology (Figure\,\ref{fig:com_morphology_acetone_acetaldehyde_dimethylether}).

\begin{figure*}[htbp!]
	\includegraphics[width=\textwidth]{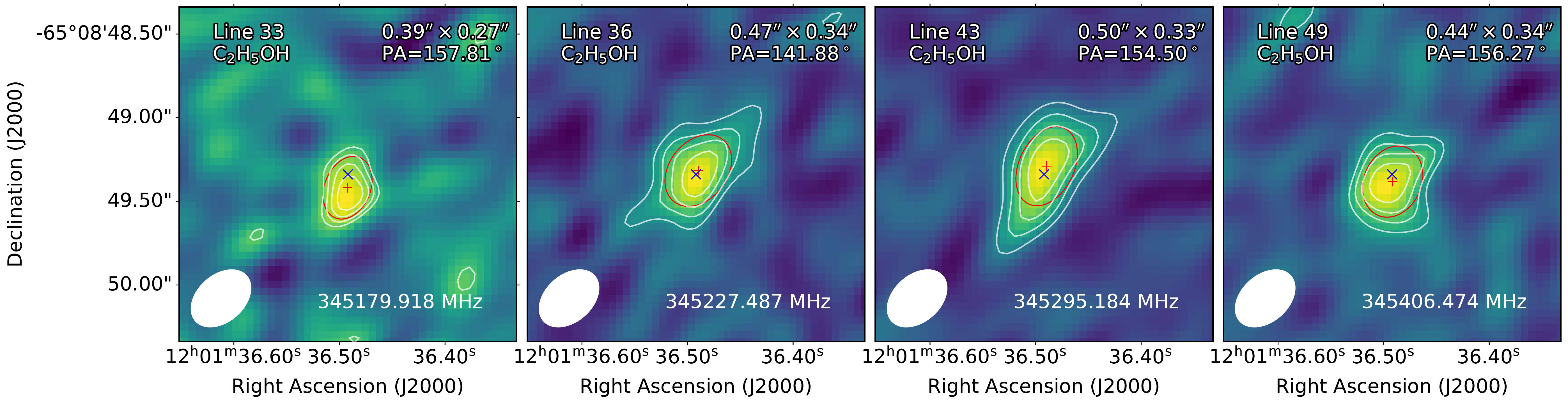}
	\caption{Similar to Figure\,\ref{fig:com_morphology_methanol} but for gauche-ethanol (gauche-\ethanol).}
	\label{fig:com_morphology_gaucheethanol}
\end{figure*}

\noindent \textit{Ethanol (\ethanol):} The window centered on the \htcn\ \jj{4}{3} line detects the emission of gauche-ethanol, a conformer of ethanol, where the hydrogen of the OH group points toward hydrogens of the methyl group.  The gauche form dominates in the gas phase because of its twofold degeneracy \citep{Scheiner2009}.  Moreover, the transitions of gauche-ethanol has lower upper energies, compared to $\sim$1000\,K for trans-ethanol.  Our model of gauche-ethanol matches Lines 33, 36, 43, 49/50, as well as the red wing of the \htcn\ \jj{4}{3} line.  Lines 49 and 50 heavily blend with each other, and the peak frequency of gauche-ethanol lies between the two identified peaks.  The identified gauche-ethanol emission shows a compact feature peaked at the center (Figure\,\ref{fig:com_morphology_gaucheethanol}).

\noindent \textbf{S-bearing molecules} \\
\noindent \textit{Sulfur dioxide (SO$_{2}$):} Line 19 agrees with the emission of sulfur dioxide, which blends with another line, possibly CH$_{3}$C$^{15}$N, in its blue wing.  The emission of SO$_{2}$ extends over a 1\arcsec$\times$1\arcsec\ region, more than other identified complex molecules, peaking both north and south sides of the continuum source (Figure\,\ref{fig:com_morphology_sulfurmol_cnmol}).

\noindent \textit{Thioformaldehyde (H$_{2}$CS):} The emission of thioformaldehyde agrees with Line 27.  The emission of H$_{2}$CS has a morphology similar to that of SO$_{2}$, which is more extended than other molecules (Figure\,\ref{fig:com_morphology_sulfurmol_cnmol}).

\section{Tentatively Identified COMs}
\label{sec:coms_tentative}
The species discussed here are catagorized as tentatively identified because our observations do not cover enough unblended lines to support a confirmed identification.  In particular, the three lines of \methylamine\ covered by our observations yield a conflicting result; thus, we only discuss \methylamine\ as a possible identification.

\noindent \textbf{O-bearing molecules} \\
\noindent \textit{Acetaldehyde (\acetaldehyde):} The torsionally excited acetaldehyde contributes to Lines 3 and 5; however, both lines have complex line profiles due to blending with other lines.  Together with the emission of HCN\,$v_{2}=1$, the model of acetaldehyde reproduces the width of Line 3, but it only reproduces part of the complex profile of Line 5.  Line 3 appears as a compact source centered on the continuum source, while Line 5 has more extended emission toward south with a peak at the center (Figure\,\ref{fig:com_morphology_acetone_acetaldehyde_dimethylether}).

\noindent \textbf{N-bearing molecules} \\
\noindent \textit{Hydrogen cyanide (HCN $v_{2}=1$):} Together with the torsionally excited acetaldehyde, the emission of hydrogen cyanide from a vibrational excited state ($v_{2}=1$) successfully reproduces Line 3, which is significantly broader than the other lines.
While hydrogen cyanide is abundant in protostellar environments, a vibrational transition of HCN ($v_{2}=1\rightarrow0$) has a much greater Einstein-A coefficient (1.3\,s$^{-1}$, \citealt{1991ApJ...383..674E}) than that of the rotational transition of vibrationally excited HCN (A$=1.9\times10^{-3}$\,s$^{-1}$ for $J=4\rightarrow3, v_{2}=1$).  Thus, the detection of the rotational emission in the vibrational excited state suggests a mechanism like radiative pumping to keep HCN vibrationally excited.

\begin{figure*}[htbp!]
	\includegraphics[width=\textwidth]{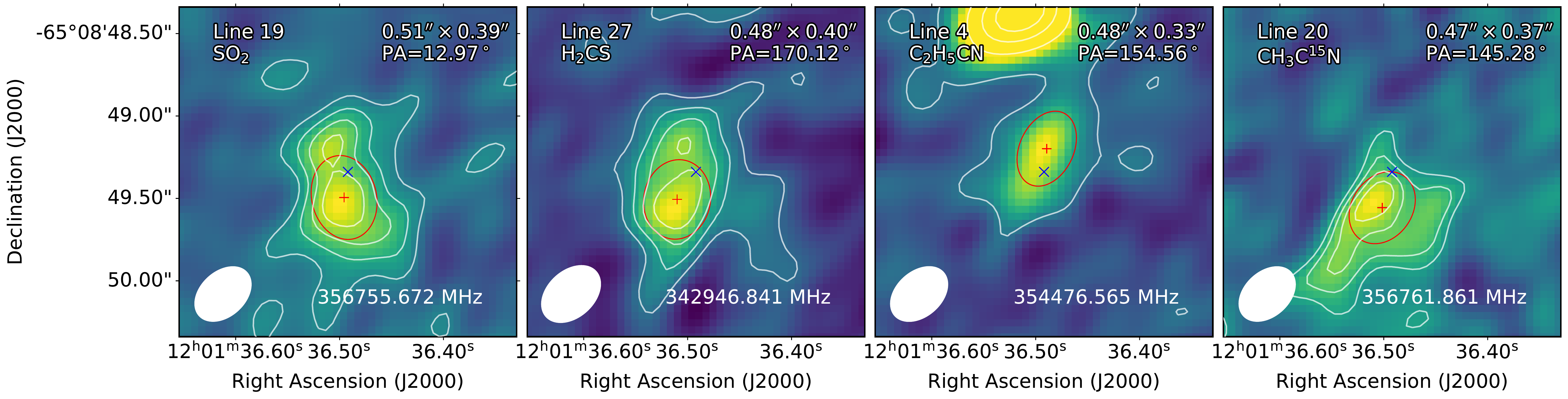}
	\caption{Similar to Figure\,\ref{fig:com_morphology_methanol} but for sulfur-bearing molecules, SO$_{2}$ and H$_{2}$CS, and complex cyanides, CH$_{3}$C$^{15}$N and C$_{2}$H$_{5}$CN.}
	\label{fig:com_morphology_sulfurmol_cnmol}
\end{figure*}

\noindent \textit{Methyl cyanide (\methylcyanideFT)}: Two lines of \methylcyanideFT\ fall in the range of the observations, which are consistent with Line 20 and the blue wing of the \hcop\ \jj{4}{3} line.  The extended emission of \methylcyanideFT\ peaks south of the continuum source (Figure\,\ref{fig:com_morphology_sulfurmol_cnmol}).
% The low isotopic ratio of \methylcyanideFT in the solar system, $\sim$440 \citep{2012A&ARv..20...56C}, questions the identification of \methylcyanideFT; however, our observations do not cover the frequencies of CH$_{3}$CN to investigate the ratio of $^{15}$N/$^{14}$N.

\noindent \textit{Ethyl cyanide (\ethylcyanide):} The emission of ethyl cyanide contributes to Line 4 as well as the unlabelled feature at the lower frequency end of Line 22.  At Line 4, the emission of ethyl cyanide appears to be compact around the center (Figure\,\ref{fig:com_morphology_sulfurmol_cnmol}).

\noindent \textit{Methylamine (\methylamine):} Three lines of methylamine fall in the range of our observations, while two of them are detected as the blue wing of Line 5 and Line 6.  The line at 345520\,MHz is not detected.  However, lower the excitation temperature to $\sim$70\,K would make the line at 345520\,MHz as weak as the the noise in our observations.  Thus, methylamine is the best candidate for these lines; however, it cannot be unambiguously confirmed.

\noindent \textbf{Cl-bearing molecules} \\
\noindent \textit{Methyl Chloride (CH$_{3}$Cl):}  Gauche-ethanol alone cannot fully describe the strength of the blended Line 49 and 50.  Our observations cover four lines of methyl chloride, which are detected with low S/N as Line 49/50, 52, 53, and the emission at $\sim$345435\,MHz.  Due to the low S/N of these lines, our spectra cannot unambiguously confirm the identification of methyl chloride.

\bibliography{research,fixed}

\end{document}